\documentclass[10pt,twocolumn,twoside]{IEEEtran}    

\usepackage{amsmath}

\usepackage{amsfonts}

\usepackage{mathtools}

\usepackage{bbold}

\usepackage{upgreek}

\usepackage{graphicx}

\usepackage{subfigure}

\usepackage{bm}

\usepackage{cite}

\usepackage{color}

\usepackage{CJK}

\usepackage{verbatim}

\usepackage{url}

\usepackage{multirow}

\usepackage{array}

\usepackage{booktabs}

\usepackage{colortbl}

\usepackage{algorithm}

\usepackage{algorithmicx,algpseudocode}

\usepackage{diagbox}

\usepackage{makecell}

\usepackage{enumerate}

\usepackage{cases}

\usepackage{lipsum,cuted,arydshln}

\usepackage{BOONDOX-cal}

\usepackage{threeparttable}

\usepackage[colorlinks,
            citecolor=black, 
            linkcolor=black, 
            urlcolor=black 
            ]{hyperref}

\usepackage{varioref}

\usepackage[misc]{ifsym}

\usepackage{stfloats}

\usepackage{balance}

\usepackage{caption}

\usepackage[table]{xcolor}

\usepackage{setspace}

\usepackage{tikz}

\usepackage{pifont}

\def\QEDclosed{\mbox{\rule[0pt]{1.3ex}{1.3ex}}} 

\usepackage[thmmarks]{ntheorem}

\usepackage{nomencl}

\usepackage{wasysym}

\usepackage{hyperref}

\usepackage{leading}

\usepackage{setspace}

\makeatletter
\def\UrlAlphabet{%
      \do\a\do\b\do\c\do\d\do\e\do\f\do\g\do\h\do\i\do\j%
      \do\k\do\l\do\m\do\n\do\o\do\p\do\q\do\r\do\s\do\t%
      \do\u\do\v\do\w\do\x\do\y\do\z\do\A\do\B\do\C\do\D%
      \do\E\do\F\do\G\do\H\do\I\do\J\do\K\do\L\do\M\do\N%
      \do\O\do\P\do\Q\do\R\do\S\do\T\do\U\do\V\do\W\do\X%
      \do\Y\do\Z}
\def\UrlDigits{\do\1\do\2\do\3\do\4\do\5\do\6\do\7\do\8\do\9\do\0}
\g@addto@macro{\UrlBreaks}{\UrlOrds}
\g@addto@macro{\UrlBreaks}{\UrlAlphabet}
\g@addto@macro{\UrlBreaks}{\UrlDigits}
\makeatother

{
  \theoremstyle{nonumberplain}
  \theoremheaderfont{\bfseries\em}
  \theorembodyfont{\normalfont}
  \theoremseparator{:}
  \theoremsymbol{\mbox{$\QEDclosed$}}
  
}

\theoremheaderfont{\bf\em}
\theorembodyfont{\normalfont}
\theoremseparator{:}

\definecolor{gray1}{rgb}{0.9,0.9,0.9}
\definecolor{gray2}{rgb}{0.8,0.8,0.8}
\definecolor{gray3}{rgb}{0.7,0.7,0.7}
\definecolor{gray4}{rgb}{0.5,0.5,0.5}

\definecolor{blue1}{rgb}{0.578,0.66,0.844}
\definecolor{red1}{rgb}{0.844,0.676,0.719}

\definecolor{white}{rgb}{1,1,1}

\usepackage[acronym]{glossaries}
\makenoidxglossaries

\newacronym{der}{DER}{Distributed Energy Resource}
\newacronym{pv}{PV}{Photovalitic}
\newacronym{wt}{WT}{Wind Turbine}
\newacronym{ev}{EV}{Electric Vehicle}
\newacronym{ict}{ICT}{Information Communications Technology}
\newacronym{sdn}{SDN}{Software-defined network}
\newacronym{eiot}{EIoT}{Energy Internet-of-Thing}
\newacronym{hilp}{HILP}{High Impact and Low Probability}
\newacronym{aic}{AIC}{Availability, Integrity, Confidentiality}
\newacronym{cre}{CRE}{Cyber-resiliency Enhancement}
\newacronym{ids}{IDS}{Intrusion Detection System}
\newacronym{ims}{IMS}{Intrusion Mitigation System}
\newacronym{rep}{REP}{Retail Energy Provider}
\newacronym{vpp}{VPP}{Virtual Power Plant}
\newacronym{scada}{SCADA}{Supervisory Control and Data Acquisition}
\newacronym{derms}{DERMS}{DER Management System}
\newacronym{dms}{DMS}{Distribution Management System}
\newacronym{agc}{AGC}{Automatic Generation Control}
\newacronym{avc}{AVC}{Automatic Voltage Control}
\newacronym{sced}{SCED}{Security-Constrained Economic Dispatch}
\newacronym{scopf}{SCOPF}{Security-Constrained Optimal Power Flow}
\newacronym{iso}{ISO}{Independent System Operator}
\newacronym{rto}{RTO}{Regional Transmission Organization}
\newacronym{ami}{AMI}{Advanced Metering Infrastructure}
\newacronym{it}{IT}{Information Technology}
\newacronym{ot}{OT}{Operation Technology}
\newacronym{opf}{OPF}{Optimal Power Flow}
\newacronym{ml}{ML}{Machine Learning}
\newacronym{ics}{ICS}{Industrial Control System}
\newacronym{p2p}{P2P}{Peer-to-Peer}
\newacronym{pll}{PLL}{Phase Lock Loop}
\newacronym{dos}{DoS}{Denial-of-Service}
\newacronym{fdi}{FDI}{False Data Injection}
\newacronym{mtd}{MTD}{Moving Target Defense}
\newacronym{dmz}{DMZ}{Demilitarized Zone}
\newacronym{vlan}{VLAN}{Virtual Local Area Network}
\newacronym{rbac}{RBAC}{Role-based Access Control}
\newacronym{ldap}{LDAP}{Lightweight Directory Access Protocol}
\newacronym{tls}{TLS}{Transport Layer Security}
\newacronym{pki}{PKI}{Public Key Infrastructure}
\newacronym{hids}{HIDS}{Host-based Intrusion Detection System}
\newacronym{nids}{NIDS}{Network-based Intrusion Detection System}
\newacronym{pids}{PIDS}{Physics-based Intrusion Detection System}
\newacronym{ied}{IED}{Intelligent Electronic Device}
\newacronym{hpc}{HPC}{Hardware Performance Counter}
\newacronym{svr}{SVR}{Support Vector Regression}
\newacronym{lstm}{LSTM}{Long Short-Term Memory}
\newacronym{cnn}{CNN}{Convolutional Neural Network}
\newacronym{stl}{STL}{Signal Temporal Logic}
\newacronym{cvf}{CVF}{Cooperative Vulnerability Factor}
\newacronym{hss}{HSS}{Harmonic-State-Space}
\newacronym{uio}{UIO}{Unknown Input Observer}
\newacronym{lfc}{LFC}{Load Frequency Control}
\newacronym{dt}{DT}{Digital Twin}
\newacronym{smo}{SMO}{Sliding Mode Observe}
\newacronym{hod}{HOD}{High-order Differentiator}
\newacronym{han}{HAN}{Home Area Network}
\newacronym{ann}{ANN}{Artificial Neural Network}
\newacronym{ubb}{UBB}{Uniformly Ultimately Bounded}
\newacronym{wmsr}{WMSR}{Weighted Mean Subsequence Reduced}
\newacronym{ess}{ESS}{Energy Storage System}
\newacronym{ndn}{NDN}{Named Data Networking}
\newacronym{fdems}{FDEMS}{Facilities DER Energy Management System}
\newacronym{milp}{MILP}{Mixed Integer Linear Programming}
\newacronym{drl}{DRL}{Deep Reinforcement Learning}
\newacronym{iot}{IoT}{Internet of Things}
\newacronym{hmi}{HMI}{Human Machine Interface}
\newacronym{oem}{OEM}{Original Equipment Manufacturer}
\newacronym{pmu}{PMU}{Phasor Measurement Unit}
\newacronym{lihp}{LIHP}{Low Impact and High Probability}
\newacronym{mimp}{MIMP}{Medium Impact and Medium Probability}

\begin{document}


\title{Enhancing Cyber-Resiliency of DER-based Smart Grid: A Survey}

\author{
Mengxiang Liu, Fei, Teng,~\IEEEmembership{Senior~Member,~IEEE}, Zhenyong Zhang,~\IEEEmembership{Member,~IEEE}, \\Pudong Ge,~\IEEEmembership{Student~Member,~IEEE},
Mingyang Sun,~\IEEEmembership{Senior~Member,~IEEE}, \\Ruilong Deng,~\IEEEmembership{Senior~Member,~IEEE},  Peng Cheng,~\IEEEmembership{Member,~IEEE},
and Jiming Chen,~\IEEEmembership{Fellow,~IEEE}

\thanks{{\color{black}This work was supported in part by the National Natural Science Foundation of China under Grants 62293503, 62293502, 62293500, 62073285, 62303126, 52161135201, U20A20159, 62103371, 62362008, in part by the Engineering and Physical Sciences Research Council of U.K. (EPSRC) under Award EP/T021780/1, in part by the Natural Science Foundation of Zhejiang Province under Grants LR23F030001 and LZ23F030009, and in part by the Newton Advanced Fellowships NAF/R1/201101.}}
\thanks{M. Liu, P. Ge, and F. Teng are with the Department of Electrical and Electronic Engineering, Imperial College London, London, UK.}
\thanks{Z. Zhang, M. Sun, P. Cheng, and J. Chen are with the College of Control Science and Engineering, Zhejiang University, Hangzhou, China.}
\thanks{R. Deng is with the State Key Laboratory of Industrial Control Technology and the College of Control Science and Engineering, Zhejiang University, Hangzhou 310027, China, and also with the Huzhou Institute of Industrial Control Technology, Huzhou 313000, China.}
\thanks{{Corresponding Authors: Ruilong Deng (dengruilong@zju.edu.cn); Fei Teng (f.teng@imperial.ac.uk).}}

}
\maketitle

\begin{spacing}{1}
\begin{abstract}
The rapid development of information and communications technology has enabled the use of digital-controlled and software-driven distributed energy resources (DERs) to improve the flexibility and efficiency of power supply, and support grid operations. However, this evolution also exposes geographically-dispersed DERs to cyber threats, including hardware and software vulnerabilities, communication issues, and personnel errors, etc. Therefore, enhancing the cyber-resiliency of DER-based smart grid - the ability to survive successful cyber intrusions - is becoming increasingly vital and has garnered significant attention from both industry and academia. In this survey, we aim to provide a comprehensive review regarding the cyber-resiliency enhancement (CRE) developments of the DER-based smart grid, present a holistic CRE framework, and thoroughly discuss the research directions of the next-generation CRE methods. Firstly, an integrated threat modeling method is tailored for the hierarchical DER-based smart grid with special emphasises on vulnerability identification and impact analysis. Then, the defense-in-depth strategies encompassing prevention, detection, mitigation, and recovery are comprehensively surveyed, systematically classified, and rigorously summarized. A holistic CRE framework is subsequently proposed to incorporate the five key resiliency enablers. Finally, challenges and future directions are discussed in details. The overall aim of this survey is to illustrates  the recent development of CRE methods and motivate further efforts to improve the cyber-resiliency of DER-based smart grid.
\end{abstract}






{\color{black}
\leading{8pt}
\printnoidxglossary[type=\acronymtype,title=Acronyms, nonumberlist, style=index]}

\printnomenclature
\section{Introduction}
The power system is rapidly transitioning to address the ever-increasing power demand, energy crisis, and climate challenges, through decentralization and digitization. \acrshort{der}s including both small-scale conventional generators and inverter-based resource (such as \acrshort{pv}s, \acrshort{wt}s, \acrshort{ev}s, and storage units) are driving this transition from the traditional large spinning generation to the decarbonized \acrshort{der}-dominated generation \cite{DoE2022report, LTRA2020report, DoEOE2021whitepaper}. The utilization of digital-controlled and software-driven \acrshort{der}s can greatly enhance the flexibility and efficiency of power supply to customers. Moreover, IEEE Stdandard 1547-2018 has been put on the table to formalize the interconnection and interoperability of \acrshort{der}s with associated power system interfaces, such as frequency disturbance ride-through capability, to support grid operations \cite{IEEStd1547_2018}. Along with the transition towards the low-carbon future, there is an increasing demand for advanced \acrshort{ict} like 5G, \acrshort{eiot}, and \acrshort{sdn} technologies. These technologies, together with smart inverter devices, offer numerous benefits for the transition. However, they also pose various cyber threats \cite{emmanuel2016communication,nafees2022smart, zheyuanIEMag_2018}.

\begin{figure*}[hb]
  \centering
  \includegraphics[width=18cm]{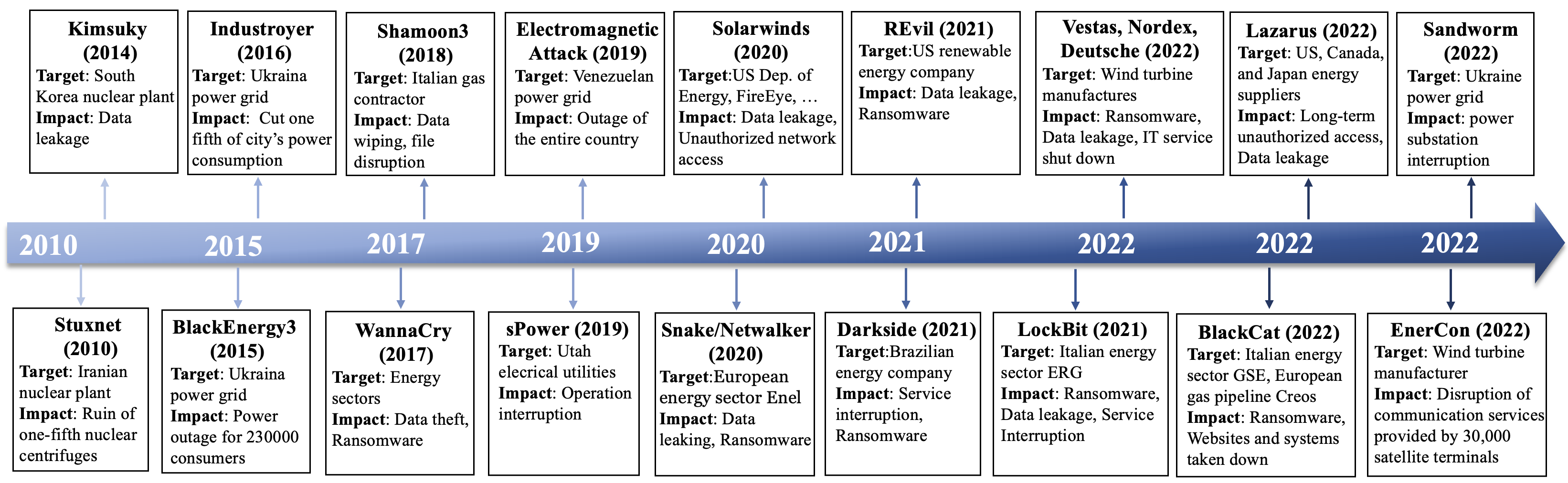}
  \caption{\color{black}Timeline of the cyberattacks targeting at the smart grid from 2010 to 2022 with an emphasis on the recent three years.}\label{fig:AttackTimeline}
\end{figure*}

{\color{black} A timeline documenting the major cyberattacks against power grid between 2010 and 2022 with a focus on the last three years is included in Fig. \ref{fig:AttackTimeline}.} The power grid, being a critical infrastructure of a country, has been a prime target for state-sponsored or profit-driven attackers since the infamous 2015 Ukraine Blackout event \cite{7752958}. Recent cyberattack incidents, such as the REvil \cite{Revil} and EnerCon \cite{EnerCon2022} events, indicate that renewable energy resources are frequently targeted by the adversaries seeking to extort ransom or disrupt communication links. Furthermore, as \acrshort{der}s are physically connected to the power grid and increasingly involved in grid operations, attackers can maliciously control their behaviors to cause system-wide impact, such as frequency/voltage instability, line overloading, and power outages. Given the unique characteristics of \acrshort{der}-based smart grid, several exclusive cybersecurity challenges can be summarized as follows: i) Utility operators do not have complete access and awareness to \acrshort{der}s installed and maintained by individuals and third parties; ii) Geographically dispersed \acrshort{der} systems lack industrial-graded security mechanisms to prevent physical intrusion; and iii) Numerous private and public network access points do not have sufficient security measures in place. In this context, there is a growing consensus in the community that 100\% secure cyber network is unlikely achievable in the future power grids.

To address these challenges, cyber-resiliency - the ability to survive successful cyber intrusions - must be developed and integrated into the planning, control, and management processes of \acrshort{der} hardware, software, and communication networks. This integration will ensure continuous electricity flow to meet the critical load of customers, even during major cyberattacks. Resiliency, which was first defined by Holling in 1973 as a system's ability to maintain its functionality and behavior after a disturbance \cite{holling1973resilience}, was initially proposed to address growing natural disasters in the power grid. However, given the increasing threat of cyberattacks, the concept of cyber-resiliency is recently introduced and is defined as a system's ability to limit the impact, duration, and extent of physcial degradation caused by cyberattacks \cite{9913670,7579185}. Enhancing the cyber-resiliency is particularly crucial to pave the way towards the large-scale deployment of \acrshort{der}s.

\begin{figure*}[ht]
  \centering
  \includegraphics[width=18cm]{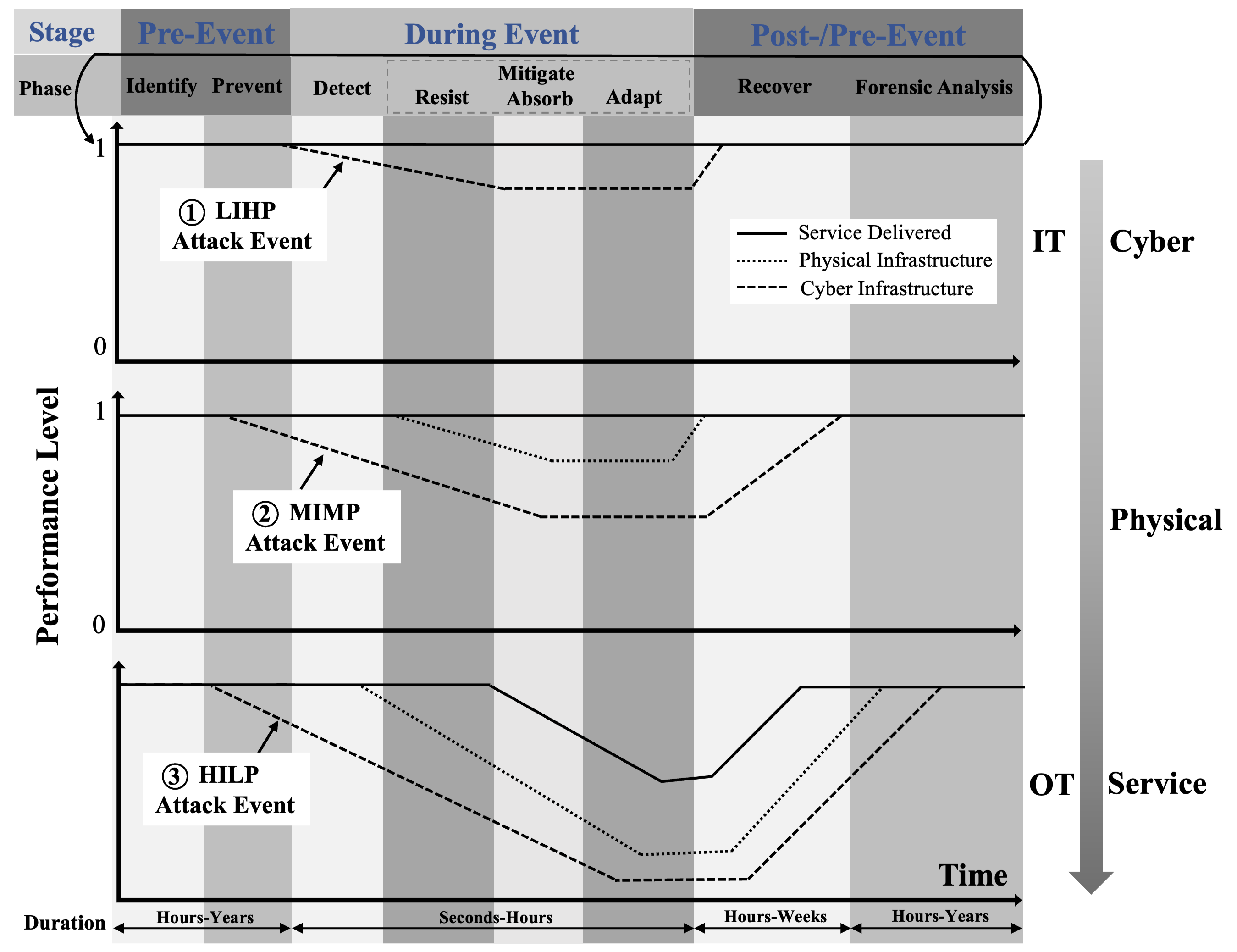}
  \caption{Cyber-resiliency stages and phases for the DER-based smart grid.}\label{fig:CyberResiliencyOrientedStagesPhases}
\end{figure*}

The cyber-resiliency consideration of DER-based smart grid can be classified into three stages and five phases based on the occurrence time of attack events as shown in Fig. \ref{fig:CyberResiliencyOrientedStagesPhases}. The three stages are pre-event, during event, and post-event, while the five phases are identification, prevention, detection, mitigation, and recovery. In the pre-event stage (hours to years), threat identification \cite{EPRI2013report} as well as prevention technologies \cite{onunkwo2020recommendations} are employed to identify possible vulnerabilities and provide preventative capabilities against common and naive cyberattacks. The identification and prevention stages are mainly targeted at known and common attacks while at the same time prepare the response plans for the unknown and sophisticated attacks. Given undisclosed zero-day vulnerabilities and inappropriate configurations or management of prevention technologies, they may be bypassed and invalidated by powerful and persistent adversaries. After a successful cyber intrusion event that bypasses preventive defenses, detection modules \cite{lai2021review, jones2020implementation,kavousi2020machine} and mitigation strategies \cite{appiah2020decentralized,liu2021robust,ge2022cyber} work sequentially or parallelly to respond to cyber contingencies. {\color{black}Since the propagation from cyber degradation to physical consequences usually occurs quickly such that the operator would not have enough time to make response strategies, the resistance and absorption capabilities provided by the smart grid's inherent $N-k$ robustness can be leveraged in the first two mitigation sub-phases.} In the post-event stage, when the system under attacks is maintained stable, recovery plans are made to recover power supply services, repair power infrastructures, and remove cyber malware sequentially \cite{wang2022cyber,liu2022towards,9881595}, after which forensic analysis will be conducted for further guideline development \cite{erol2013smart}. 

Different from the traditional resilience curve under extreme HILP natural disasters as discussed in the recent literature \cite{9913670}, in addition to the power supply service, the infrastructure capabilities are further classified into the cyber aspect in maintaining data \acrshort{aic} and the physical part of providing power generation, transmission, and distribution functionalities to illustrate the interdependence between them in the presence of attack events. {Three resilience curves are depicted in Fig. \ref{fig:CyberResiliencyOrientedStagesPhases} to illustrate the performance levels under \acrshort{lihp}, \acrshort{mimp}, and HILP attack events summarised from three mainstream real attack incidents. Under the LIHP attack event (\ding{172}) that aims to encrypt critical operating data or steal sensitive customer data and then possibly ask for a ransom like the Vestas \cite{Vestas} and REvil \cite{Revil} events, it will not propagate to the OT network to disrupt the physical infrastructure and power supply service and its impact on the cyber infrastructure can be detected, mitigated, and recovered in a timely manner. The MIMP attack event (\ding{173}) can further intrude into the OT network but has limited impacts on the field devices like the EnerCon event \cite{EnerCon2022} that interrupted the onsite WTs' remote monitoring and maintenance services, where massive WTs were forced offline. Although this kind of attack will not directly affect the power supply service due to the smart grid's inherent $N-k$ robustness, it can result in severe cascading failures and blackouts when coordinated with other attacks or occurring concurrently with emergency faults. Compared with the LIHP attack event, the performance level will drop to a lower value under the MIMP event, from which the required recovery time is much longer due to the compromised OT network and onsite physical devices. The HILP attack event (\ding{174}) such as the BlackEnergy \cite{7752958} event is usually launched by powerful adversaries to induce wide-area and long-duration blackouts by manipulating the field devices' operating statuses. Under the highly coordinated HILP event, the power supply service will be affected when the cyber and physical infrastructure performances decrease to significantly low values. The unique cyber resilience feature lies in the gap between the performance curves of cyber and physical infrastructures, implying that coordinated efforts from both \acrshort{it} and \acrshort{ot} areas are required to ensure the smart grid's survivability under HILP attack events.}

\begin{figure*}[ht]
  \centering
  \includegraphics[width=18cm]{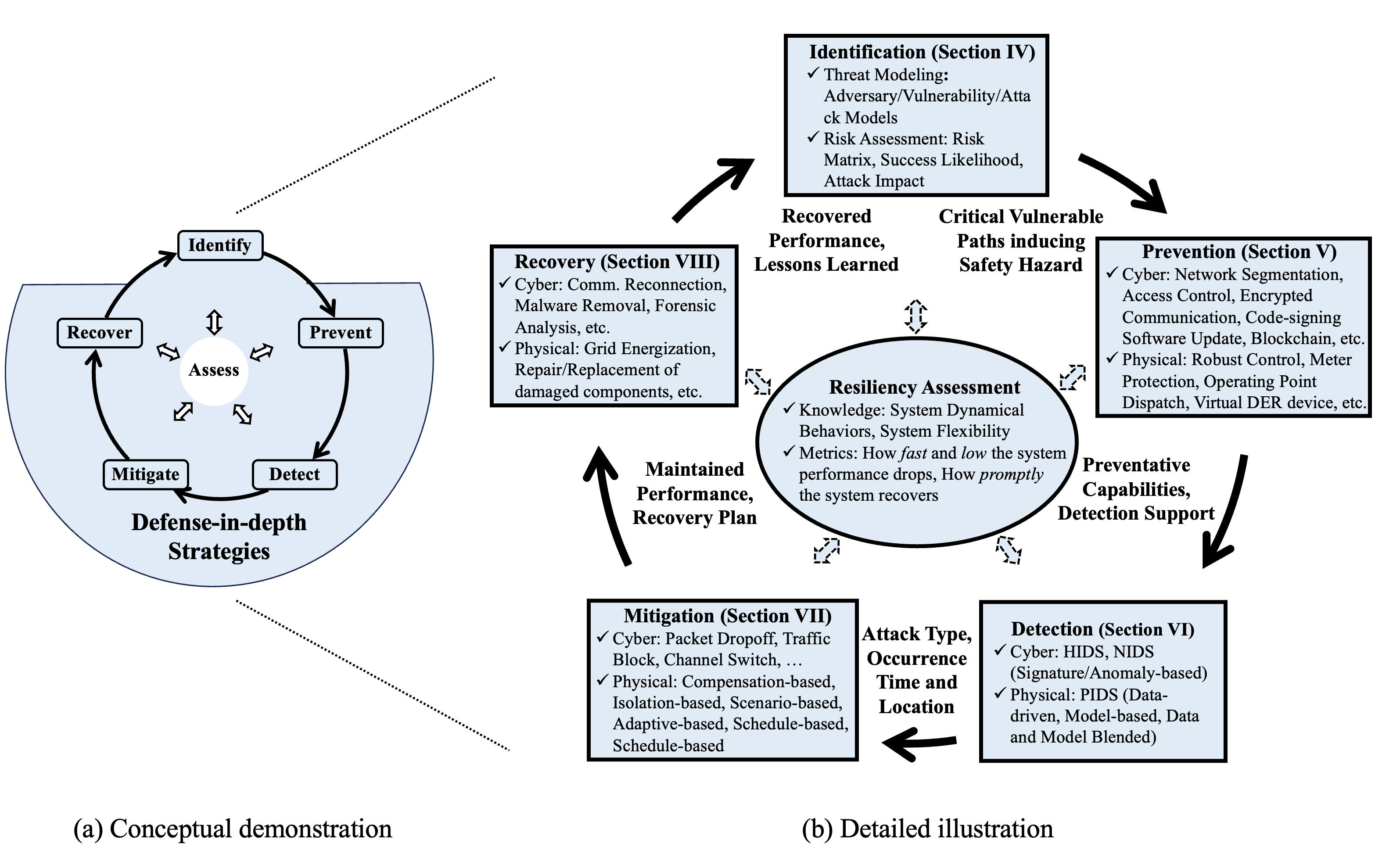}
  \caption{\color{black}The cyber-resiliency enhancement framework for the \acrshort{der}-based smart grid.}\label{fig:CREProcess}
\end{figure*}

Drawing inspirations from the NIST cybersecurity improvement framework \cite{NISTCybersecurityFramework}, which provides a high-level and strategical view of the life-cycle of an organization's cybersecurity risk management, we propose a holistic \acrshort{cre} framework tailored for the \acrshort{der}-based smart grid as shown in Fig. \ref{fig:CREProcess}. In addition to the risk-based approaches for cybersecurity management, the \acrshort{cre} framework specifies the detection, mitigation, and recover capabilities by utilizing the characteristic, controllability, and flexibility of field physical devices. Furthermore, short- and long-term resiliency assessment schemes are included to measure how quickly and to what extent the system performance drops, as well as how promptly the performance recovers, based on knowledge of system dynamics and flexibility \cite{panteli2017metrics,arghandeh2016definition,qiu2023virtual}. To improve the system's resiliency, all five phases should be considered in a holistic approach, as the resiliency level is determined by the phase with the worst performance, akin to the “Buckets effect”.
This requires a global understanding of the \acrshort{cre} process. To this end, we aim to provide a comprehensive survey of the state-of-the-art CRE developments in the DER-based smart grid, present a holistic CRE framework, and thoroughly discuss the research directions of the next-generation CRE methods. The detailed contributions of this survey are listed as follows:

1) The hierarchical architecture of \acrshort{der}-based smart grid is presented to illustrate the participating actors and the corresponding functionalities.
    
2) An integrated threat modeling method is tailored for the hierarchical \acrshort{der}-based smart grid to clarify the adversary model, asset/vulnerability model and attack model, after which a general risk assessment matrix is established to rank the attack scenarios' risk levels considering both their occurrence likelihoods and consequence severity.
    
3) The state-of-the-art developments of prevention, detection, and mitigation technologies are comprehensively reviewed, systematically classified according to their work principles, and rigorously summarized to highlight their implementation guidelines and respective cons and pros. Besides, the necessity and focus of the cyber-recovery under \acrshort{hilp} cyberattack events are clarified for the first time.

4) A holistic \acrshort{cre} framework that incorporates the five key enablers of resiliency is proposed, with their challenges and future directions being discussed in details.

{\color{black}The remaining sections are organized as follows. Section \ref{section II} introduces the differences between our survey and related works. Section \ref{Section III} presents the hierarchical architecture of DER-based smart grid, threat modeling and risk assessment methods. The comprehensive review of existing defense-in-depth strategies encompassing prevention, detection, mitigation, and recovery is provided in Sections \ref{Section V}-\ref{section: VIII}, respectively. Challenges and future directions are discussed in Section \ref{Section IX}, and Section \ref{Section X} concludes this survey.}


\begin{table*}[!ht]
{\color{black}
\footnotesize
\centering
\begin{threeparttable}
\caption{\color{black}Comparisons between this survey and existing ones}\label{Table:SurveyCompariosn}
\begin{tabular}{p{1.5cm}<{\centering}|p{3cm}<{\centering}p{0.3cm}<{\centering}p{0.3cm}<{\centering}p{0.3cm}<{\centering}p{0.3cm}<{\centering}p{0.3cm}<{\centering}p{0.3cm}<{\centering}p{0.3cm}<{\centering}p{0.3cm}<{\centering}p{0.3cm}<{\centering}p{0.3cm}<{\centering}p{0.3cm}<{\centering}p{0.3cm}<{\centering}p{0.3cm}<{\centering}p{0.3cm}
<{\centering}p{0.3cm}<{\centering}p{1.2cm}<{\centering}}
\Xhline{1.2pt}
\multicolumn{2}{c}{\textbf{Resilience Enhancement Phases}} & [31] & [32] & [33] & [34] & [35] & [36] & [37] & [38] & [39] & [40] & [41] & [42] & [43] & [44] & [45] 
& { Our work} \\ \Xhline{0.5pt}
\multirow{3}{1.5cm}{Threat Identification} & Adversary Model & \cellcolor{gray1}N & 
\cellcolor{gray1}N & \cellcolor{gray1}N & \cellcolor{gray2}P & \cellcolor{gray1}N & \cellcolor{gray2}P & \cellcolor{gray1}N & \cellcolor{gray2}P & \cellcolor{gray4}C & \cellcolor{gray1}N & \cellcolor{gray1}N & \cellcolor{gray1}N & \cellcolor{gray1}N & \cellcolor{gray1}N & \cellcolor{gray1}N & \cellcolor{gray4}C \\
   & Vulnerability Coverage & \cellcolor{gray2}P 
   & \cellcolor{gray2}P & \cellcolor{gray3}M & \cellcolor{gray2}P & \cellcolor{gray2}P & \cellcolor{gray3}M & \cellcolor{gray2}P & \cellcolor{gray2}P & \cellcolor{gray3}M & \cellcolor{gray2}P & \cellcolor{gray2}P & \cellcolor{gray3}M & \cellcolor{gray3}M & \cellcolor{gray3}M & \cellcolor{gray4}C & \cellcolor{gray4}C \\
   & Risk Assessment & \cellcolor{gray2}P & 
   \cellcolor{gray3}M & \cellcolor{gray2}P & \cellcolor{gray2}P & \cellcolor{gray3}M & \cellcolor{gray2}P & \cellcolor{gray3}M & \cellcolor{gray3}M & \cellcolor{gray2}P & \cellcolor{gray4}C & \cellcolor{gray2}P & \cellcolor{gray4}C & \cellcolor{gray4}C & \cellcolor{gray2}P & \cellcolor{gray2}P & \cellcolor{gray4}C \\
\multirow{4}{1.5cm}{Defense-in-Depth Strategies} & Prevention & \cellcolor{gray3}M & \cellcolor{gray4}C & \cellcolor{gray3}M & \cellcolor{gray2}P & \cellcolor{gray4}C & \cellcolor{gray3}M & \cellcolor{gray2}P & \cellcolor{gray2}P & \cellcolor{gray3}M & \cellcolor{gray1}N & \cellcolor{gray3}M & \cellcolor{gray1}N & \cellcolor{gray3}M & \cellcolor{gray2}P & \cellcolor{gray1}N & \cellcolor{gray4}C \\
   & Detection & \cellcolor{gray4}C & 
   \cellcolor{gray3}M & \cellcolor{gray4}C & \cellcolor{gray2}P & \cellcolor{gray3}M & \cellcolor{gray4}C & \cellcolor{gray3}M & \cellcolor{gray2}P & \cellcolor{gray2}P & \cellcolor{gray1}N & \cellcolor{gray1}N & \cellcolor{gray3}M & \cellcolor{gray3}M & \cellcolor{gray3}M & \cellcolor{gray3} M & \cellcolor{gray4}C \\
   & Mitigation & \cellcolor{gray1}N & 
   \cellcolor{gray1}N & \cellcolor{gray2}P & \cellcolor{gray1}N & \cellcolor{gray2}P & \cellcolor{gray1}N & \cellcolor{gray2}P & \cellcolor{gray2}P & \cellcolor{gray2}P & \cellcolor{gray2}P & \cellcolor{gray2}P & \cellcolor{gray2}P & 
\cellcolor{gray2} P & \cellcolor{gray3} M & \cellcolor{gray2} P & \cellcolor{gray4} C\\
   & Recovery & \cellcolor{gray1}N & 
   \cellcolor{gray1}N & \cellcolor{gray1}N & \cellcolor{gray1}N & \cellcolor{gray1}N & \cellcolor{gray1}N & \cellcolor{gray1}N & \cellcolor{gray1}N & \cellcolor{gray1}N & \cellcolor{gray1}N & \cellcolor{gray1}N & \cellcolor{gray1}N & \cellcolor{gray1}N & \cellcolor{gray1} N & \cellcolor{gray1} N & \cellcolor{gray4}C \\ \Xhline{1.2pt}

\end{tabular}
\begin{tablenotes}
  \footnotesize
  \item[] {\colorbox{gray4}C}: Covered, {\colorbox{gray3}M}: Mostly Covered, {\colorbox{gray2}P}: Partially Covered, {\colorbox{gray1}N}: Not Covered
\end{tablenotes}
\end{threeparttable}}
\end{table*}

\section{Related Surveys}\label{section II}

{\color{black} Regarding the cyber security of the general smart grid, there have been many prominent survey papers that discuss, classify, summarise related research and industry developments. 
Yan \textit{et al.} \cite{yan2012survey} comprehensively reviewed and discussed security requirement, network vulnerabilities, prevention and defense countermeasures, and secure communication protocols and architectures in the smart grid. Sun \textit{et al.} \cite{sun2018cyber} provided a state-of-the-art of smart grid's cyber security R\&D including vulnerabilities, industry practices and standards, and anomaly detection methods, and demonstrated the feasibility of detection methods in a University hardware-in-the-loop testbed. Liu \textit{et al.} \cite{liu2012cyber} presented a thorough survey regarding the cyber security issues from device, network, dispatch and management, and anomaly detection aspects, and a brief overview of the privacy issue in the smart grid. Tan \textit{et al.} \cite{tan2016survey} conducted a survey of recent security advances in smart grid, via the data perspective, and classified the security vulnerabilities and solutions into data generation, data acquisition, data storage, and data processing.

Komninos \textit{et al.} \cite{komninos2014survey} emphasised on the cyber security issues of the smart home environment that is interacted with the smart grid, and presented a systematical survey compromising vulnerability identification, impact assessment, detection and prevention countermeasures, as well as standardisation efforts. Gunduz \textit{et al.} \cite{gunduz2020cyber} comprehensively analyzed the cyber threats and potential solutions of the IoT-based smart grid, and provided in-depth discussion and examination of network vulnerabilities, attack countermeasures, and security requirements. Liang \textit{et al.} \cite{liang2016review} presented a holistic review regarding a typical cyber threat to the smart grid, the FDI attack, where the theoretical analysis of attack construction, potentially induced physical and economic impacts, and corresponding defense strategies have been thoroughly discussed and summarised. Qin \textit{et al.} \cite{huang2018survey} provided a comprehensive study for the evolutionary of cyberattacks from the initial intrusion to inducing serious consequences in the smart grid, where the root causes of attacks are clearly identified by analysing the vulnerabilities of communication protocols and the system-level consequences are assessed thoroughly based on a multi-stage model.}

There also exist several surveys regarding the cybersecurity of \acrshort{der}-based smart grid \cite{zografopoulos2022distributed,sahoo2019cyber,vosughi2022cyber,9534741,qi2016cybersecurity,li2022cybersecurity,tuyen2022comprehensive}. Zografopoulos \textit{et al.} \cite{zografopoulos2022distributed} provided a \acrshort{der} cybersecurity outlook covering {\color{black}the device - and communication - levels vulnerabilities, attacks, impacts, and mitigation schemes.} Sahoo \textit{et al.} \cite{sahoo2019cyber} presented a brief review of the vulnerabilities in the control and cyber layer of the voltage source converters both in the grid-connected and standalone modes. Vosughi \textit{et al.} \cite{vosughi2022cyber} discussed the latest trends in the \acrshort{der} control schemes along with the cyber-physical vulnerabilities, standard communication protocols, and key security mechanisms. Ye \textit{et al.} \cite{9534741} discussed the challenges and future visions of the cyber-physical security of \acrshort{pv} systems from firmware, network, PV converter control, and grid security perspectives. Qi \textit{et al.} \cite{qi2016cybersecurity} proposed a holistic attack-resilient framework compromising threat modeling and defensive actions (attack prevention, detection, and response) to help ensure the secure integration of \acrshort{der} without harming the grid reliability and stability. Li \textit{et al.} \cite{li2022cybersecurity} presented a comprehensive review of critical attacks and defense strategies for smart inverters and inverter-based systems like microgrids. {\color{black}Tuyen \textit{et al.} \cite{tuyen2022comprehensive} presented a comprehensive review of the system structure and vulnerabilities of typical inverter-based power system with \acrshort{der} integration, nature of several types of cyberattacks, state-of-the-art defense strategies including detection and mitigation techniques.}

Nevertheless, the survey papers \cite{yan2012survey,sun2018cyber,liu2012cyber,tan2016survey,komninos2014survey,gunduz2020cyber,liang2016review,huang2018survey} mainly focus on the cyber security issues of the smart grid while not paying enough attentions to the increasingly exposed cyber vulnerabilities as the high penetration of DERs and the corresponding defense countermeasures. {\color{black}Besides, a holistic analysis regarding the cyber resilience of the DER-dominated smart grid is still lacking} and the DER-related surveys \cite{zografopoulos2022distributed,sahoo2019cyber,vosughi2022cyber,9534741,qi2016cybersecurity,li2022cybersecurity,tuyen2022comprehensive} either lack systematical threat modeling, risk assessment methods or neglect a comprehensive review of existing defense-in-depth strategies. In particular, for threat modeling and assessment, only \cite{zografopoulos2022distributed} detailed the adversary model, while \cite{sahoo2019cyber} and \cite{vosughi2022cyber} lack comprehensive vulnerability investigations. For defense-in-depth strategies, \cite{sahoo2019cyber,9534741,tuyen2022comprehensive} did not discuss prevention technologies, and \cite{sahoo2019cyber} and \cite{vosughi2022cyber} did not consider \acrshort{ids}s. All literature includes \acrshort{ims}s but only \cite{li2022cybersecurity} 
briefly classified and summarized them. Moreover, recovery scheduling is not covered in any of the literature. To fill the aforementioned gaps, this paper aims to provide a high-level threat modeling framework, specific risk assessment methods, and a systematical review of state-of-the-art defense-in-depth strategies.

\section{Identification: Hierarchical Framework, Threat Modeling, and Risk Assessment}\label{Section III}
In this paper, we adopt the bottom-up principle to identify potential threats arouse from hardware, software, communication, and personnel, and assess their risks considering the success probability and consequence severity. Before introducing the technical parts, a refined description of the hierarchical framework of \acrshort{der}-based smart grid is presented first.

\begin{figure}[ht]
  \centering
  \includegraphics[width=9cm]{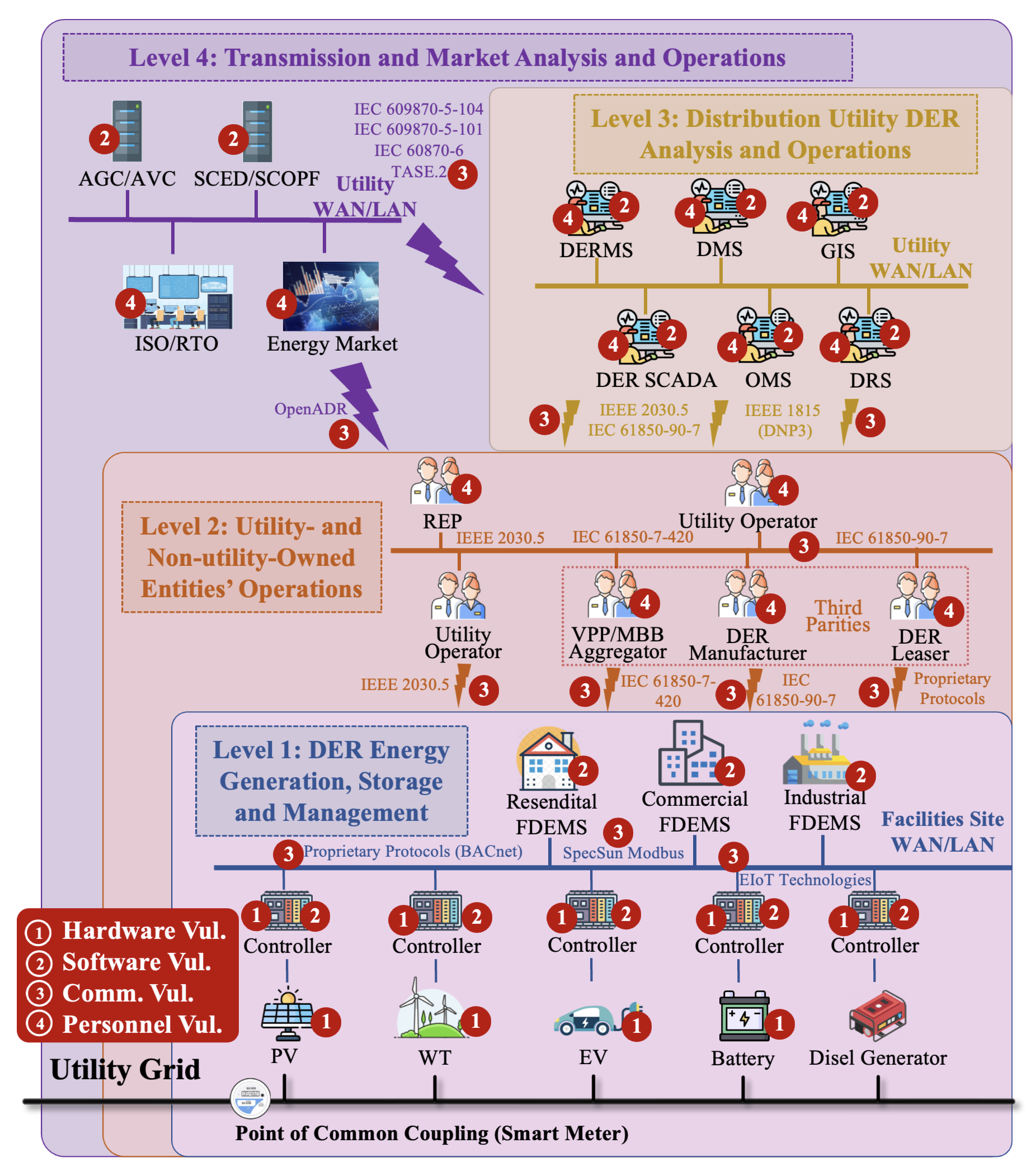}
  \captionsetup{font={small}}
  \caption{Hierarchical framework of the \acrshort{der}-based smart grid and the potential hardware, software, communication, and personnel vulnerabilities.}\label{fig:HierarchicalArchitecturewithVulnerabilities}
\end{figure}

\subsection{Hierarchical Framework of DER-based Smart Grid}
Given the large and increasing amount of geographically dispersed \acrshort{der}s, it is difficult for utility operators and stakeholders to directly control and manage their operations, and a generic hierarchical architecture is needed to coordinate them. As shown in Fig. \ref{fig:HierarchicalArchitecturewithVulnerabilities}, according to the functionalities and corresponding properties of actors, they are divided into four levels: 
{\color{black} 1) Level 1 - \acrshort{der} energy generation, storage, and management; 2) Level 2 - Utility and third parities' operations; 3) Level 3 - Distribution utility \acrshort{der} analysis and operations; and 4) Level 4 - Transmission and market analysis and operations. }

Level 1 collects the basic \acrshort{der} units compromising renewable energy source (\acrshort{pv}, \acrshort{wt}, \acrshort{ev}), non-renewable energy source (disel generator), and storage systems (battery). 
{\color{black}Open standard communication protocols (SunSpec Modbus \cite{alliance2021sunspec}), proprietary protocols (BACnet \cite{bushby2002bacnet}), and emerging \acrshort{iot} technologies (ZigBee, WiFi, and 5G) are widely adopted to enable the real-time interaction among \acrshort{der} units and facilities DER energy management systems (\acrshort{fdems}s) and thus provide \acrshort{der}'s autonomous response capabilities and ancillary services \cite{vosughi2022cyber}. }
{Level 2 includes the actors beyond local sites like utility operators, \acrshort{rep}s, as well as the non-utility-owned  third parities like \acrshort{vpp}s \cite{pudjianto2007virtual}, microgrids \cite{DoEOE2021whitepaper}, \acrshort{der} manufacturers, and \acrshort{der} leasers. For the non-utility-owned DER facilities, the utility operator does not have full accesses to read/write all facility information. To enable the DER functionalities like reactive power and voltage control, the facility's elements including 1) Nameplate information (read), 2) Configuration information (read and write), 3) Monitoring information (read), and 4) Management information (read and write) are required to be interacted with the utility operator \cite{IEEStd1547_2018}. These information interactions have to be implemented through unified information models and non-proprietary protocols such as IEEE 2030.5 (Smart Energy Profile 2.0) \cite{8608044}, IEC 61850-90-7 and IEC 61850-7-420 \cite{7953513} to guarantee the interoperability. Nevertheless, these standard communication protocols will expose the DER facilities' management and maintenance information flows to a wider area of cyber vulnerabilities if not equipped with strong security mechanisms as discussed in the subsequent subsection. Besides, proprietary protocols from manufactures can be also utilised to achieve interactions between DER facilities and third parties.}




Level 3 is responsible for the state analysis and operation determination of \acrshort{der} units in the region of the distribution power system. Many utility actors including \acrshort{scada}, \acrshort{derms}, and \acrshort{dms} are employed to ensure the safe, efficient, and reliable operation and scheduling of wide-area dispersed \acrshort{der} units. 
{\color{black} The involved communication protocols include IEEE 2030.5, IEC 61850-90-7, IEEE 1815 (DNP3) \cite{6327578}, and proprietary protocols of utilities. }
Level 4 is responsible for the analysis and operation of wide-area dispersed transmission system and related energy trading market. Applications including \acrshort{agc}, \acrshort{avc}, \acrshort{sced}, \acrshort{scopf}, and \acrshort{iso} and \acrshort{rto} balancing authorities should be reconsidered given the uncertainty, variability, and market participation of geographically dispersed \acrshort{der} units. 
{\color{black}The advanced metering infrastructure (\acrshort{ami}) plays a fundamental role for two-way data exchange between remote \acrshort{der} units and the transmission control center \cite{bennett2008networking,emmanuel2016communication,international20131}.}

{\color{black}Compared with the existing \acrshort{der} system architectures \cite{EPRI2013report,qi2016cybersecurity}, the uniqueness of the proposed hierarchical framework is i) The four-level framework comprising the \acrshort{der} device, \acrshort{der} aggregator, distribution utility, and transmission operation is proposed for the first time. ii) Actors, functionalities, and communication protocols in each layer are specified to clarify potential vulnerabilities and possible consequences; iii) Newly emerging \acrshort{der}-related entities like VPP and MBB aggregators and the P2P energy trading mode are incorporated.}

\subsection{Threat Modeling}
Threat modeling aims to identify, classify and describe threats to highlight a campaign of attacks or attackers. 
A holistic threat modeling framework that integrates both \acrshort{it} and \acrshort{ot} perspectives has been tailored for the \acrshort{der}-based smart grid, comprising the adversary model, key vulnerability and attack model.

\subsubsection{Adversary Model}
The adversary model details the identity, motivation, knowledge, access, and resource of a threat.
The threat actors include state-sponsored actors, terrorists, cybercriminals, hacktivists, cyber fighters, and disgruntled employees.
The adversary motivation include ransomware, competitor discrediting, cyberwarfare, economic gain, and terrorism/political. The adversary knowledge includes both the cyber-domain operational information \cite{9910566}.
The adversary access includes the physical access through Fieldbus/Serial/USB/Ethernet interfaces \cite{staggs2017wind}, remote access through phishing emails \cite{holm2013cyber}, and close proximity access through wireless compromise \cite{WindfarmBlackHat}. The adversary resource consists of substantial and limited privileges. 

\subsubsection{Key Vulnerability}
The \acrshort{der}-based smart grid is a typical human-in-the-loop cyber-physical system, where the cyber vulnerabilities may come from hardware, software, communication, and personnel. The typical hardware vulnerability is the weak physical access control to \acrshort{der} assets, which directly exposes various communication interfaces to the adversary. The software vulnerabilities can exist in the firmware, user code, management software, etc, and allow the adversary to access the system illegally, steal sensitive data, and disrupt system services. The software-driven principle of \acrshort{der}s makes it particularly impressionable to this kind of vulnerability. 

{\color{black}The ingenious SolarWinds incident has been disclosed to infect three \acrshort{oem}s by installing malicious software through a routine security update \cite{zetter2020solarwinds}. The OEMs may have remote accesses to critical parts of customer networks, as well as privileges to make changes to those networks, install new software, or even control critical operations. If an OEM has bi-directional access to a DER network, it is possible to manipulate the DERs' configuration modes and operating points to induce disruptive consequences. Hence, the insecure supply chain of DER associated \acrshort{oem}s can expose the DER-based smart grid to severe cyber threats. Hence, the typical software vulnerabilities include 1) Insecure supply chain of DER \acrshort{oem}s \cite{zetter2020solarwinds}, 2) Insufficient test and validation on firmware and user code \cite{bayer2006dynamic}, and 3) Zero-day vulnerabilities \cite{bilge2012before}.}

The communication vulnerability is the most common type and can come from communication protocols, network component/participator, network services, etc. {\color{black}As an open communication standard, the SunSpec Modbus enables multi-vendor interoperability for solar inverters, energy storage devices, meters, and other devices incorporated into DER systems, and is semantically identical with IEEE 2030.5 and IEEE 1815 communication protocols, ensuring a high signal-to-noise ratio for the majority of DER networks that implement multiple protocols \cite{alliance2021sunspec}. But no encryption, node authentication, or key management features were included in Modbus until Oct. 2018 as the field Modbus protocol is usually used for the data exchange in an enclosed and controlled environment, and therefore contemporary SunSpec Modbus implementations of the standard lack over-the-wire security \cite{SANDIA2019REPORT}. 

{Under the Common Smart Inverter Profile implementation guideline, the encryption, authentication, and key management features of IEEE 2030.5 standard have been adopted. However, questions still exist in these techniques' scability in highly-distributed and resource-constrained DER environments. According to the technical report of Sandia National Laboratories \cite{SANDIA2019REPORT}, the scalability gaps include i) Non-expiring certificates and no certificate revocation methods; ii) No method to update the cryptographic algorithms for the lifetime of the DER devices; iii) No physical security requirements, etc. For example, the attacker can masquerade as an authenticated user and steal information, take unauthorized actions, and possibly “plant” malware, once the adversary maliciously controls a certificated device as certification revocation is not supported. The security requirements of IEC 61850-90-7 and IEC 61850-7-420 have been included in the IEC 62351 series standard, which contains provisions to ensure data confidentiality, integrity, and authenticity for different protocols used in power systems \cite{hussain2019review}. Nevertheless, according to a recent security assessment work on IEC 62351 \cite{schlegel2017security}, two weaknesses still exist: i) Some inaccurate (e.g., Cipher suite designations) and unconventional choices (e.g., RSA signatures for IEC 61850) are adopted in the standard; ii) The provided security level is constrained by the requirements related to backwards-compatibility.

}

DNP3-SA and DNP3Sec protocols employ encryption and authentication technologies to assure the integrity and confidentiality of exchanged data. However, the security of data/service's availability is not considered in the two protocols \cite{volkova2018security}. DoS attacks can be created by modifying the length ﬁeld of a DNP3 payload sent from slave IECs to the master device, under which the master device rejects the corresponding frame and consequently the required physical mechanism fails \cite{rodofile2015real}. In addition to this, by using formal modeling and analysis methods, Amoah \emph{et al.} revealed a previously unidentified flaw in the DNP3-SA protocol \cite{amoah2016formal}. The attacker can replay a previously authenticated command to an outstation with arbitrary parameters.} According to these latest technical reports and academia literature, the main communication protocol related vulnerabilities include 1) Insufficient security mechanisms in SunSpec Modbus \cite{alliance2021sunspec}, 2) Scalability gaps of IEEE 2030.5's security features \cite{SANDIA2019REPORT}, 
and 3) Inadequate security consideration in DNP3-SA and DNP3Sec \cite{volkova2018security}.

As the integration of third parties into the system operation, management, and maintenance, the network component/participator related vulnerabilities also appear: 1) Insufficient network segmentation between \acrshort{der} systems \cite{zheyuanIEMag_2018}, 2) Unknown trust level among multiple stakeholders \cite{jefferies1995proposed}, and 3) Multiple access points from external networks \cite{almutairi2022web}.
Based on the communication infrastructure, numerous network services can be provided to enable convenient device management and cost-efficient operation. These services also expose service oriented vulnerabilities, including 1) Insecure remote management services on \acrshort{der} systems \cite{bret2016all}, 2) Security challenges of \acrshort{p2p} energy trading \cite{aloul2012smart}, and 3) Vulnerable \acrshort{ml}-based applications \cite{jagielski2018manipulating}. The personnel vulnerability appears as a critical concern as the wide integration of human-involved control and management into the \acrshort{der}-based smart grid. However, it is hard to guarantee the security qualification of all stakeholder staffs especially as their scale increases. {\color{black}According to Fig. \ref{fig:HierarchicalArchitecturewithVulnerabilities}, the hardware vulnerability is mainly from \acrshort{der}s and field controllers, and the personnel vulnerability is among the operation and management staff in upper levels. Moreover, software and communication vulnerabilities spread throughout all the levels.}

\begin{table*}[]
{\color{black}
\footnotesize
\centering
\begin{threeparttable}
\caption{\color{black}Attack Techniques Summary and Classification}\label{Table:AttackTech}
\begin{tabular}{p{1.2cm}|p{3.3cm}p{0.5cm}p{11.5cm}}
\Xhline{1pt}
  \textbf{Types} &  \textbf{Attack Techniques} & \textbf{AIC} & \textbf{Description and Direct Impacts} \\ \Xhline{0.5pt}
  
  \multirow{5}{1cm}{Initial Access Acquisition} 
  & {Network service exploitation} & IC & Use cross-site scripting or SQL injection to illegally access the \acrshort{der} network \cite{ics2016ics}. \\ \cline{2-4}
  & Wireless compromise & C & Exploit wireless protocol vulnerabilities to obtain illegal remote accesses to the \acrshort{der} network \cite{osti_1761987}. \\ \cline{2-4}
  & Supply-chain compromise & IC & Gain control systems' accesses by compromising products before receipt by end consumers \cite{SolarWind2020}. \\ \cline{2-4}
  & Zero-day attack & C & Exploit zero-day vulnerabilities to get illegal accesses to the \acrshort{der} system \cite{case2016analysis}. \\ \cline{2-4}
  & Social engineering attack & C & Use personal information or subterfuge to learn a legal user’s password \cite{SANDIA2017REPORT}. \\ \cline{1-4}
  
  \multirow{3}{1cm}{Information Discovery}
  & Insider attack & C & Employ persons within the organization that have access to critical resources \cite{Hunker2011InsidersAI}. \\ \cline{2-4}
  & Side-channel attack & C & Analyze time/power/electromagnetic information to infer critical information \cite{standaert2010introduction}. \\ \cline{2-4}
  & Eavesdropping attack & C & Take screenshots of \acrshort{hmi}s and workstations or Listen to communicated credentials \cite{ScreenShotAttack}. \\ \cline{1-4}
  
  \multirow{8}{1cm}{Execution and Implication}
  & Firmware manipulation & I & Install malicious firmware into inverters/converters to execute illegal actions \cite{garcia2017hey}. \\ \cline{2-4}
  & Trojan attack & IC & A malware disguising itself as legitimate code or software to gain legitimate users' privileges \cite{konstantinou2016taxonomy}. \\ \cline{2-4}
  
  & Hall spoofing attack & I & Mislead hall sensor’s measurements by placing a camouflaged attack tool near the inverter \cite{barua2020hall}. \\ \cline{2-4}
  & Control logic modification & AI & Modify control logic of the \acrshort{der} controller to manipulate outputs or trigger overflow bug \cite{tychalas2021icsfuzz}. \\ \cline{2-4}
  & Brute force attack & AI & Repetitively change I/O point values to affect the process function associated with that point \cite{cherepanov2017win32}. \\ \cline{2-4}
  & \acrshort{dos} attack & A & Deliberately overload a \acrshort{der} stakeholder and prevent it from performing normal functions \cite{sPowerDoSAttack}. \\ \cline{2-4}
  & \acrshort{fdi} attack & I & Modify and inject data streams exchanged in the \acrshort{der} network \cite{JohnDERSecurity,deng2018false,deng2017ccpa,deng2018false1,8625609}. \\ \cline{2-4}
  & Replay attack & I & Replace current transmission data with previously recorded data in the \acrshort{der} network \cite{gallo2018distributed}. \\ \cline{2-4}
  & \acrshort{p2p} energy market attack &  I & Submit fake contracts and modifications of transactions to gain illegal profits \cite{aloul2012smart,aitzhan2016security,shuaib2016cognitive}.  \\ \cline{2-4}
  & \acrshort{ml} adversarial attack & I & Create adversarial examples with imperceptible perturbations to mislead the \acrshort{ml} outputs \cite{yuan2019adversarial}. \\ \Xhline{1pt}

\end{tabular}
\end{threeparttable}}
\end{table*}

\begin{figure}
    \centering
    \includegraphics[width=9cm]{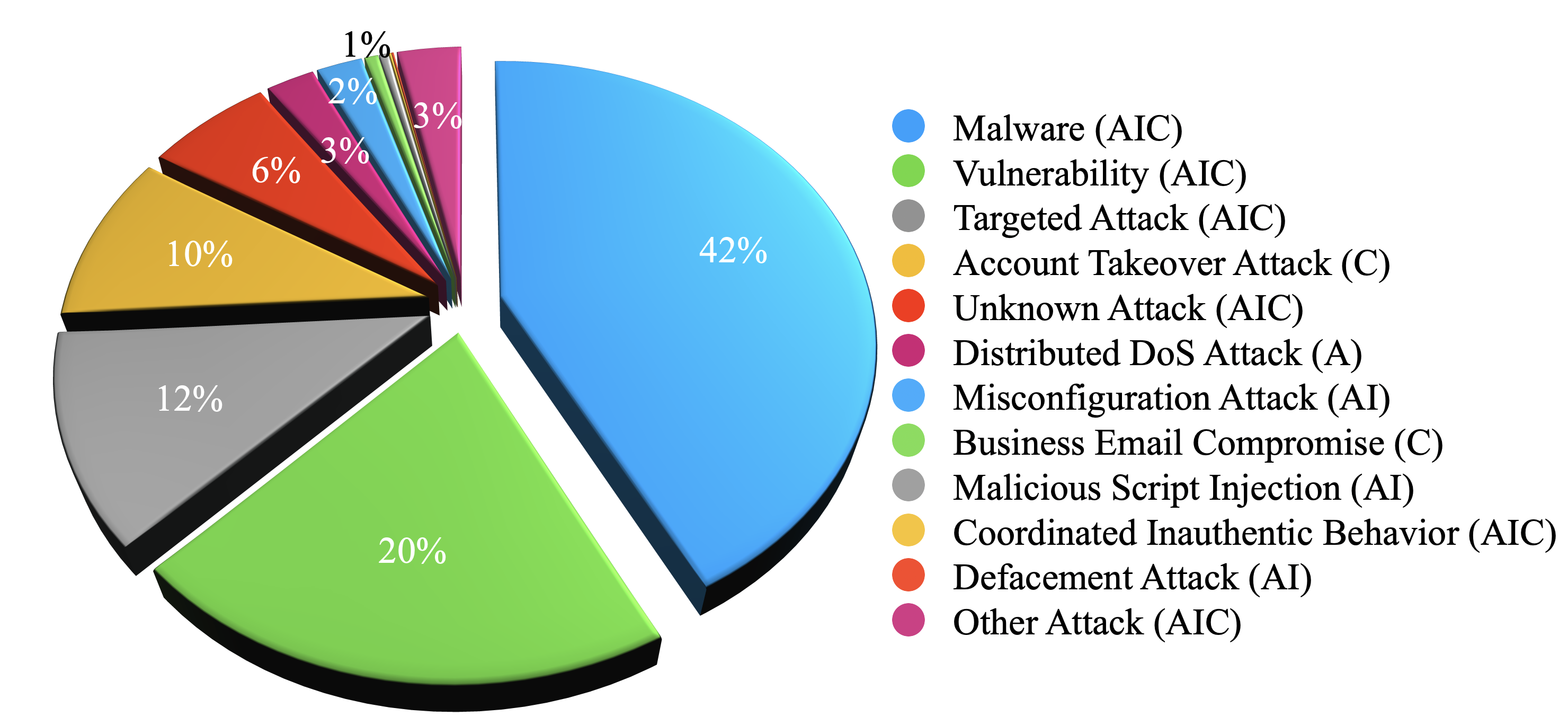}
    \caption{\color{black}Statistics of attack techniques adopted by the cyberattack events against \acrshort{ics}s during 2022.}
    \label{fig:2DPieChartAttacktechniques}
\end{figure}

{\color{black}\subsubsection{Attack Model} The attack model specifies the attack techniques by exploiting those vulnerabilities and potential attack impacts in the context of \acrshort{der}-based smart grid. Inspired by the MITRE ATT\&CK Matrix for \acrshort{ics}s, the attack techniques are divided into initial access acquisition, information discovery, and execution and implication according to the adversary's intrusion and execution phases \cite{assante2015industrial}. {\color{black}As shown in TABLE \ref{Table:AttackTech}, the attack techniques' descriptions and their impacts on \acrshort{aic} are clearly illustrated.} 
According to the statistical data published on HACKMAGEDDON\footnote{\url{https://www.hackmageddon.com/}}, the top attack techniques adopted by the cyber attack events against \acrshort{ics}s during 2022 are depicted in Fig. \ref{fig:2DPieChartAttacktechniques} to provide a high-level understanding of different attack techniques' risks and occurrence. In particular, the malware from supply-chain compromise, malicious firmware installation, and Trojan attacks are the most commonly adopted attack techniques, followed by known/zero-day vulnerabilities, targeted attacks, and account takeover attacks. The take home message from these attack statistics is that the observed cyberattack events are becoming more and more mature, implying the adversary's increasing intelligence. Moreover, the human-involved threats like insiders and social engineering attacks are gaining increasing attentions. Note that TABLE \ref{Table:AttackTech} merely lists the attack techniques against the smart grid, without including all attack techniques highlighted in Fig. \ref{fig:2DPieChartAttacktechniques}. }

{As the rapid development of ML-based applications in the smart grid, the ML adversarial attack that generates adversarial examples with imperceptible perturbations to mislead the output of ML models \cite{zhang2023vulnerability,10100624} has attracted widespread attentions. The credibility of the input data to ML models determines essentially the application's performance, and the tricky designed perturbation on inputs can possibly degraded the performance significantly. For the ML-enabled renewable energy output forecasting method, when the adversary modifies the meteorological input data with a 10\% error, the mean-absolute-error of PV's output forecasting will be 3 times larger than that in the normal case, with the average economic loss more than 700\$/10min \cite{ruan2023vulnerability}. In the ML-based voltage and transient stability assessment schemes, the assessment accuracy can be decreased by more than 20\% under appropriate adversarial perturbation strategies \cite{9693255,9642059}. The ML-based AC state estimation's root-mean-square-error can be enlarged by 10 times when the active and reactive power flows are perturbed strategically \cite{tian2022adversarial}. Moreover, for the ML-based load forecasting, the adversarial attack by manipulating the temperature and historical data can increase the mean-absolute-percentage-error from 2\% to more than 10\% \cite{chen2019exploiting}. There exist studies that investigate the impacts of adversarial attacks on other ML-based applications like inertial forecasting \cite{9895140}, grid event classification \cite{niazazari2020attack}, power flow control \cite{zeng2022physics}, etc. Interested readers are sugested to check the latest review paper on the vulnerability of ML approaches applied in IoT-based smart grid \cite{zhang2023vulnerability}.
}

{Compared with compromising single point, the coordinated attack that modifies multiple points in the mean time can induce longer and severer attack impacts \cite{rahman2022challenges,xiang2017coordinated}. Two representative type of coordinated attacks are analysed in the subsequent part from the perspectives of attack stealthiness and impact severity. 1) \textbf{Coordinated FDI attacks against microgrid communication links}: Due to the adoption of the consensus-based secondary controller, arbitrary bias injection into one communication link can induce ramping voltage deviations \cite{liu2022false}, which can quickly enter the unsafe area and trigger the islanding of DERs. When the adversary is able to compromise multiple communication links and appropriately design the bias injections to meet the zero-sum criteria, controllable and stable voltage deviations can be induced \cite{sahoo2018stealth}. Hence, the coordination of multiple FDI attacks on communication links can improve their stealthiness and enhance the controllability of attack impacts. 2) \textbf{Coordination of multiple load alternating attacks through EIoT botnet attack}: Considering the inherent $N-k$ robustness of smart grid, the impacts of load alternations resulted from compromising single EIoT device or a small portion of EIoT devices can be almost neglected. Nevertheless, it has been demonstrated by Soltan \textit{et al.} \cite{soltan2018blackiot} that the EIoT botnet attack manipulating a large number of high-wattage devices will induce frequency instability, cascading failures, and generation cost increase by increasing $30\%$, $1\%$, and $5\%$ of the total load demand, respectively, at the same time. Hence, the coordination of multiple load alternations through the EIoT botnet attack can significantly enlarge the severity of attack impacts.
}

{\color{black}Following the perspective of \acrshort{aic}, the potential attack impacts on the \acrshort{der}-based smart grid are divided into security- and privacy-related. The security-related impact focuses on how can the cyber-physical attacks impact/disrupt the data availability and integrity, consequently affecting the device-level functionalities and grid-level process and operation.

\noindent{{\textbf{Security-Related Transmission-Level}}} - {\emph{Energy Price/Load Manipulation:}}
        Utility, VPP, and third parties participate the biding process and determine the energy price and amount of energy to be traded \cite{zhang2018peer,dasgupta2021cyber}. When most of the participates are controlled by the adversary, the energy price may be manipulated. The consumers' electricity consumption can be also controlled according to the demand-response mechanism if the real-time price is manipulated \cite{liu2022real}.
        
        - {\emph{Generator Trip and Load Shedding:}}
        When the loads including EV and batteries at one or multiple critical locations are modulated with the purpose of exciting an existing inter-area \emph{oscillation} mode on the power system, the generator may be tripped and load shedding may happen. It was reported that the distributed load modulation event (500MW) can result in minor load shedding (22.2MW) due to the composite load model's internal protection \cite{johnson2022cybersecurity}.
        
        - {\emph{Load-Generation Imbalance:}}
        The adversary can deliberately trigger mandatory tripping operations of inverters by forging over-/under- voltage/frequency, unintentional islanding, and short-circuit/open-phase signals \cite{IEEStd1547_2018,shilay2017catching}, and increase steep EV and battery load increase simultaneously, such that traditional generators cannot supply the loads. Besides, when the parameters of voltage/frequency ride-through capabilities are misconfigured, the active/reactive power injected by DERs may vary in a malicious trend, i.e., decreasing power injection in peak-load period, causing load-generation imbalance. Actually, the load-generation imbalance can possibly lead to generator trip and load shedding, and also induce line overflow, cascading failure, and blackout \cite{soltan2018blackiot,10124159}.

\noindent{{{\textbf{Security-Related Distribution-Level}}}} - {\emph{Consumer Expense Increase in Residential Units:}} In the residential unit, the Home Management System (HMS) can monitor and control the battery such that minimal power utilization from the grid is achieved. After penetrating into the HMS network, the battery may be programmed to charge from the grid when excessive load presents and discharge when surplus power from PV exists, inducing extra expenses for the consumer \cite{sahoo2019cyber}. In addition, by compromising the control algorithm and control parameters (e.g., changing the power factor), the adversary could tune the maximum power output of a PV inverter to a certain percentage of the current available solar power, namely power curtailment, decreasing the efficiency of energy usage and thereby increasing the expense of buying power from the grid \cite{shilay2017catching}.

 - {\emph{Frequency/Voltage Deviation and Power Sharing Failure in Microgrids:}} The essential objectives of islanded microgrids are to control the PCC voltages and frequency to expected values, and achieve proper load sharing. When targeted by malicious adversary, the measurement inputs of primary controllers and transmitted data between secondary controllers may be compromised \cite{9580468,liu2022false}, disrupting the control objectives.
 
 - {\emph{Poor Power Quality:}} The IEEE Standard 1547-2018 limits the DC current, voltage fluctuations, current distortion, and overvoltage over a cumulative duration injected by DER to guarantee the power quality delivered to customers \cite{IEEStd1547_2018}. Since the settings are allowed to be adjusted locally and/or remotely as specified by the APS operator, the power quality can be impacted if the attacker manipulates the setting parameters locally or remotely.
 
 - {\emph{Intentional Islanding Failure:}} The support of intentional islanding scheduled by the APS operator requires inverters to have black-start and isochronous regulation capabilities \cite{IEEStd1547_2018}. The attacker can mislead the APS operator to include only the inverters that do not have islanding-support capabilities in a intentional islanding, causing local load-generation imbalance together with voltage/frequency instability, and eventually failing the intentional islanding.
        
 - {\emph{Increased Power Loss:}}
        The inverters are required to respond to voltage variations within the normal operating range, and different modes including constant power factor, voltage-reactive power, active power-reactive power, and voltage-active power are listed \cite{IEEStd1547_2018}. The attacker can manipulate the feeder voltage by tampering with the parameters utilized in the voltage-regulation functions (e.g., the power factor) such that the power loss is largely increased.
        
       - {\emph{Aggravated Equipment Wear:}}
        Traditional voltage regulation equipment such as tap changing transformers, voltage regulators, and shunt capacitor exist in the distribution system. The cost of these equipment is huge and their lifetime is related to the action frequency. The attacker can manipulate the feeder voltage near the action bound to cause frequent actions on these equipment, aggravating the equipment wear.
        
       - {\emph{Voltage Violation:}}
        The most severe impact is to cause voltage violation, i.e., manipulating the feeder voltage out of the operational bound (e.g., 0.95p.u. - 1.05p.u.). Besides affecting the active/reactive power injections by compromising the voltage-regulation functions of inverters \cite{ustun2019cybersecurity}, the adversary can also manipulate the on/off of EV and batteries to cause a steep and large load increase/drop when the DER generation output is low/high, possibly inducing regional reverse power flow, generation-load imbalance and voltage violation \cite{johnson2022cybersecurity,shilay2017catching}.
    
The privacy-related impact concerns the customer information leak caused by data confidentiality violation, including location information, personal behavior patterns and activities inside home, and real-time surveillance information.
    
      \textbf{Privacy-Related} - {\emph{Location Information:}} Historical EV data can be used to determine range of use since last recharge. Location of active EV charging activities can be used to determine the location of drivers;

        - {\emph{Personal Behavior Patterns and Activities inside Home:}} Electricity usage and DER generation patterns and appliance use can be used to infer behavioral patterns, habits, and activities taking place inside the home;

        - {\emph{Real-Time Surveillance Information:}} Real-time energy use data can be used to determine if anyone is home, potentially what they are doing, and where they are located in the home.}



\begin{figure}
    \centering
    \includegraphics[width=9cm]{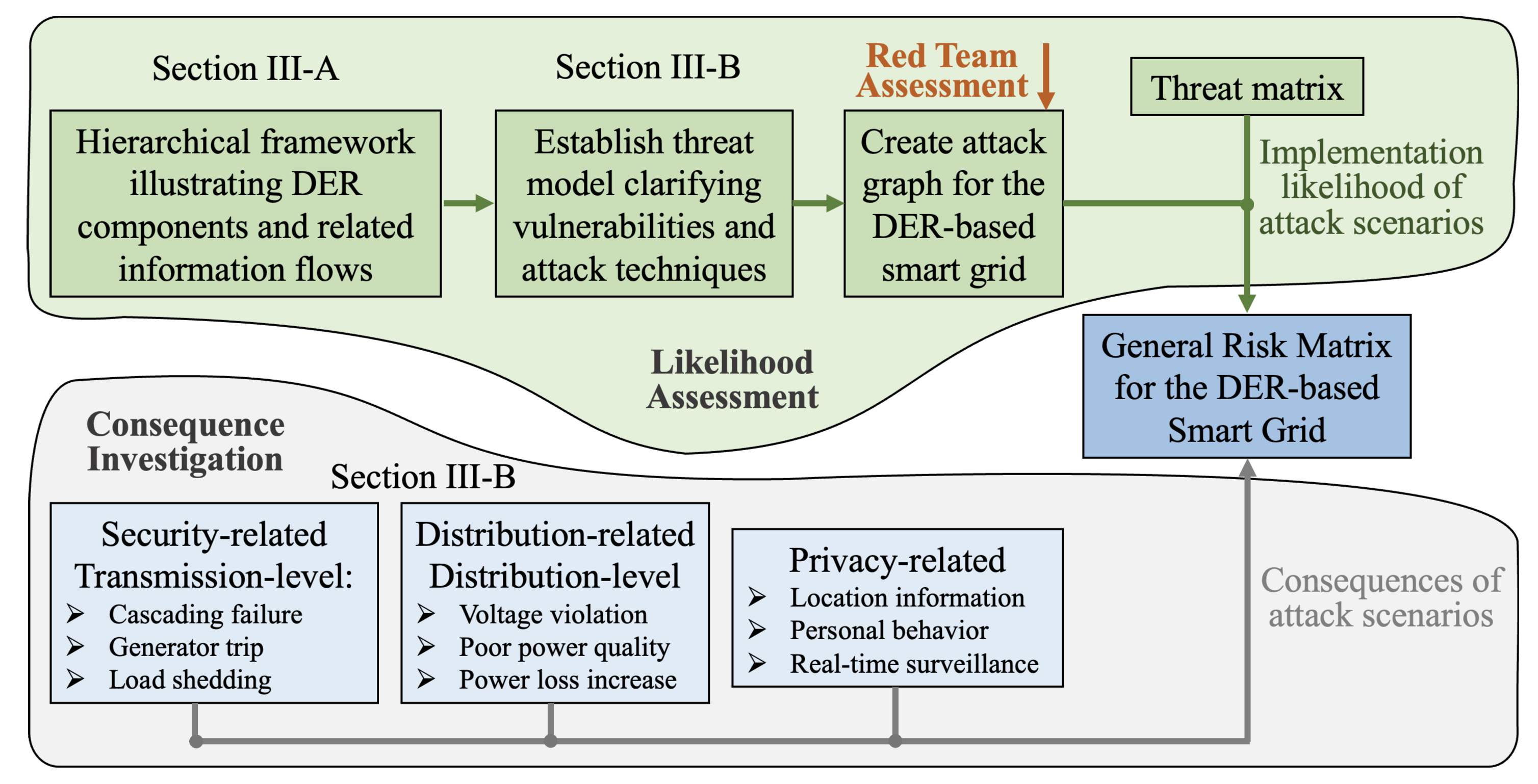}
    \caption{\color{black}Generation scheme of the risk matrix for the \acrshort{der}-based smart grid.}
    \label{fig:RiskMatrixGeneration}
\end{figure}

\subsection{Risk Assessment Matrix}
\begin{figure}[ht]
  \centering
  \includegraphics[width=9cm]{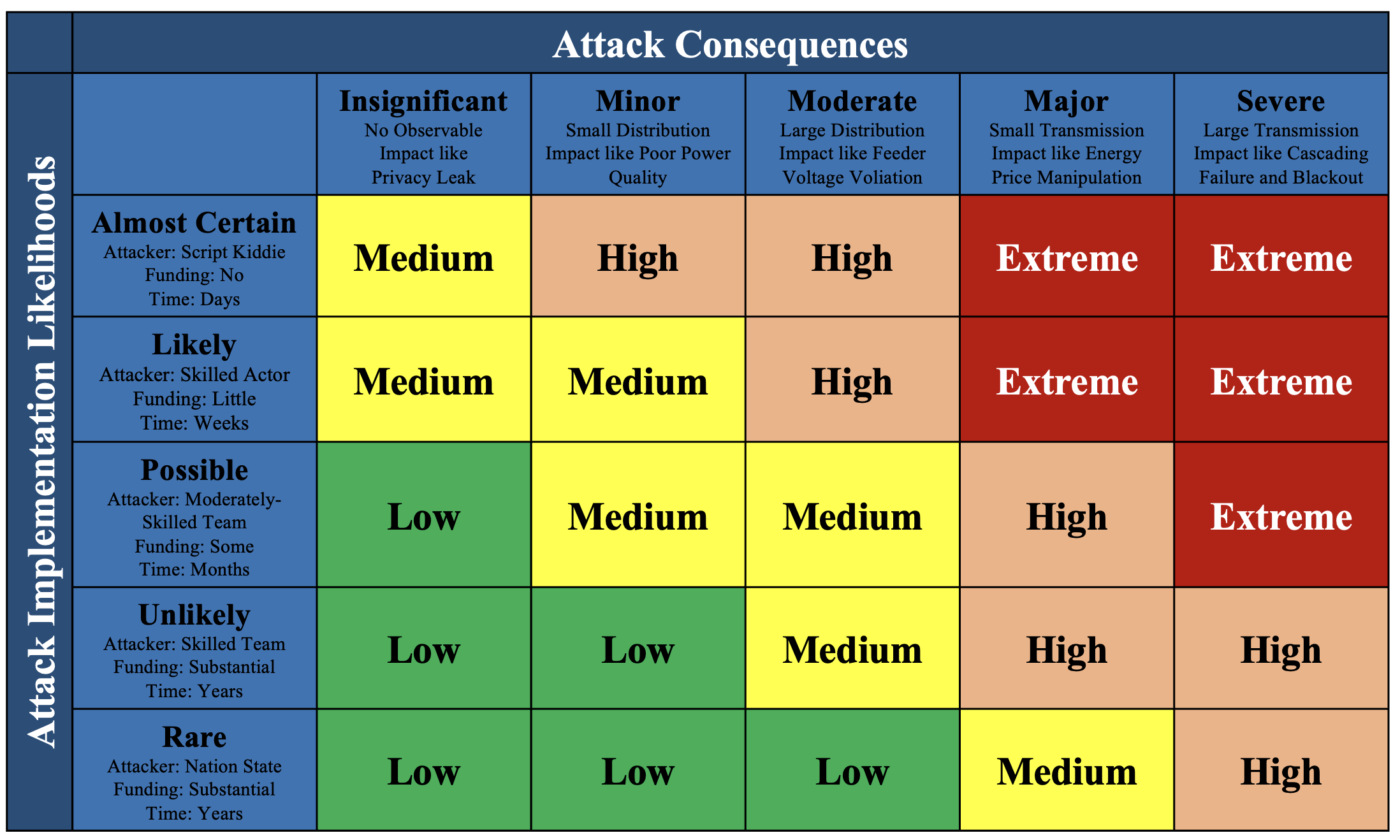}
  \caption{\color{black}Risk matrix reference for the \acrshort{der}-based smart grid.}\label{fig:RiskAssessmentMatrix}
\end{figure}

{\color{black}After identifying the potential vulnerabilities and associated attack techniques in the \acrshort{der}-based smart grid, it is crucial to assess the risk of each attack scenario. A risk matrix has been proposed that takes the inputs of attack implementation likelihoods and attack consequences as illustrated in Fig. \ref{fig:RiskMatrixGeneration}. 

{\color{black}In the attack implementation likelihood assessment phase, based on the established threat model, the red team will first conduct multiple assessment activities comprising visits to manufacturing facilities, development and testing labs, and assessments of fielded \acrshort{der} systems. The team mainly assesses the cybersecurity posture of state-of-the-art \acrshort{der} equipment using authorized, adversary-based assessment techniques, often in close collaboration with the vendors. 
Then, attack graphs will be created to show the steps an adversary must take to move from a system/network access point to a consequence or objective. 
A demonstrative example that illustrates the deployment of malicious firmware against \acrshort{ev}s \cite{johnson2022cybersecurity} is shown in Fig. \ref{fig:AttGraphMaliciousFirmware}. The first step in this attack graph is to craft the payload that will be delivered to the deployed \acrshort{ev} supply equipment. Afterwards, the adversary will gain access to the business network using either a malicious insider or using remote attack techniques, followed by pivoting through the business network until getting access to the firmware repository. Different methods will be chosen to insert malicious firmware depending on if the update requires code signing. Finally, by triggering shutdown signals following specific strategics, expected consequences can be induced.}

{\color{black}The attack graphs will then be utilised to estimate the skill and time requirements to execute different attack scenarios. Combined with the general threat matrix, which enables government entities and intelligence organizations to categorize threat into a common vocabulary \cite{mateski2012cyber}, the attack implementation likelihoods are qualitatively classified as Almost Certain, Likely, Possible, Unlikely, and Rare according to the adversary's knowledge, funding, and time. Note that other threat attributes like intents and targets and more granular classification levels can be incorporated into the risk matrix, and here the simplified version is shown only for demonstrative purpose.}

\begin{figure}
    \centering
    \includegraphics[width=9cm]{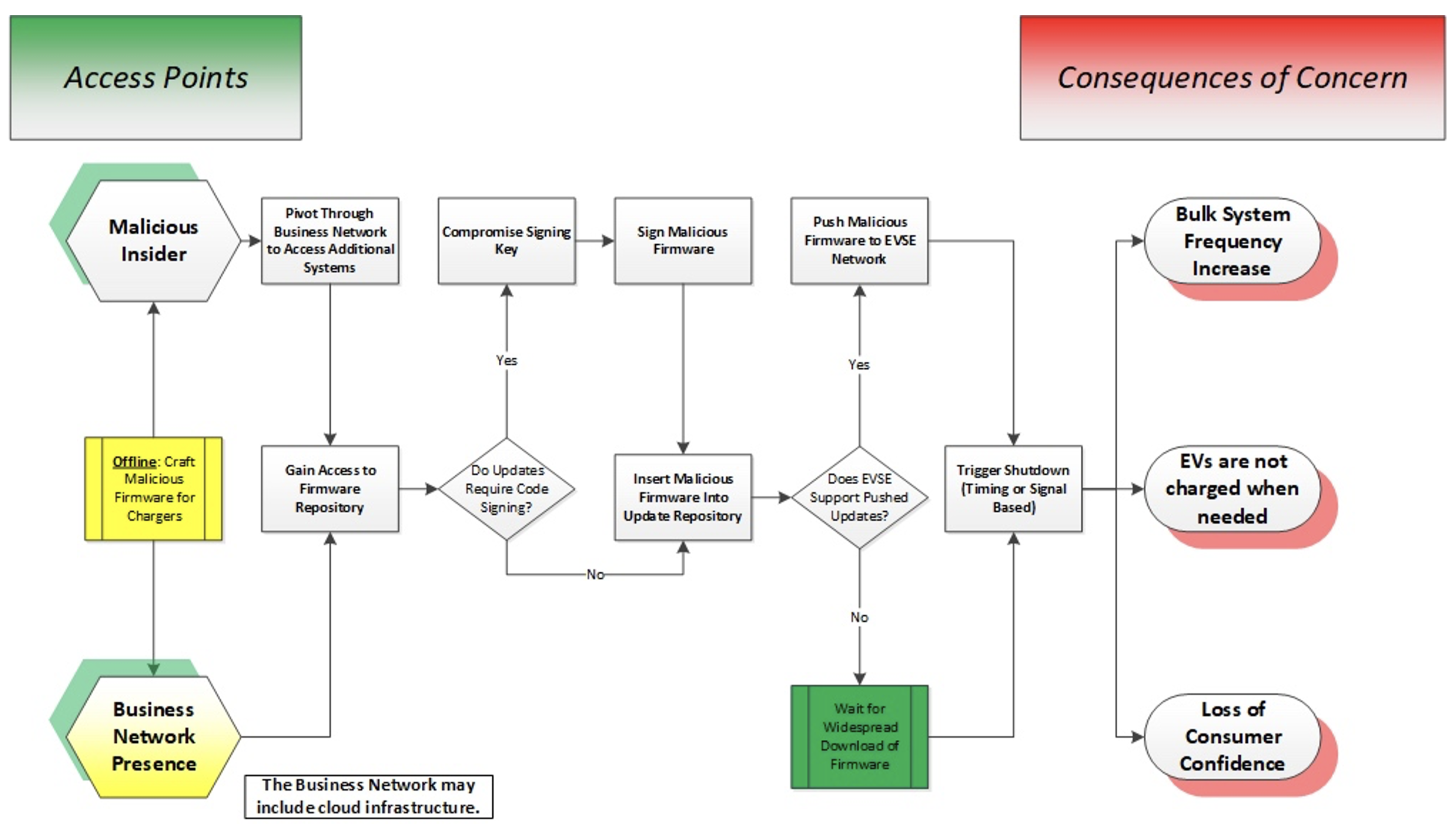}
    \caption{\color{black}Attack graph illustrating the malicious firmware deployment against \acrshort{ev}s \cite{johnson2022cybersecurity}.}
    \label{fig:AttGraphMaliciousFirmware}
\end{figure}

In the consequence investigation phase, the impacts of attack scenarios on smart grid are observed from the experimental results obtained using high-fidelity smart grid simulator like OPAL-RT, RTDS, and Typhoon HIL \cite{9484453}. According to the impact scale and severity, the attack consequences are qualitatively classified as Severe, Major, Moderate, Minor, and Insignificant. For example, the privacy leak normally not affects the power system operation and thus is deemed as insignificant, while the large-scale transmission level cascading failure and blackout will severely affect the power supply and hence is assumed as severe. Finally, the generated risk matrix is presented in Fig. \ref{fig:RiskAssessmentMatrix} with the columns being the consequence levels and rows being the likelihood levels. As indicated by the colors of entries, the qualitatively classified risk levels include Extreme, High, Medium, and Low, and help the operator determine the weakest system points and choose appropriate defense strategies. 

A thorough risk assessment process against EV supply equipment has been conducted coordinately and is detailed in \cite{johnson2022cybersecurity}. } {\color{black}Some insights can be synthesised from this practice as follows.
\begin{itemize}
    \item The attack with almost certain probability cannot currently be achieved as no public scripts and tools that can indeed impact the power system exist.
    \item The skilled actor/team or nation state can cause insignificant and minor impact on the \acrshort{der}-based smart grid. For example, the personal behavior pattern may be inferred after eavesdropping the energy usage and \acrshort{der} generation data from smart meters/\acrshort{pmu}s and data servers \cite{pillitteri2014guidelines}, and frequency/voltage deviations can appear in isolated microgrids when multiple primary/secondary controllers are compromised by a skilled team \cite{9580468,liu2022false}.
    \item Since the \acrshort{der} penetration is not high, moderate, major, and severe attack impact cannot be caused by purely manipulate the \acrshort{der} actions. It has been pointed out that approximately 30\% of \acrshort{der} deployment relative to peak load begins to show infrequent but potential grid-level consequences \cite{DoE2022report,osti_1505553}. Hence, attention should be paid although this threat that is currently impossible, but is likely to be possible under the global trend towards the low-carbon power system \cite{foxon2010developing}.
\end{itemize}}

{\color{black} Besides the risk matrix based assessment method, the probabilistic risk-based assessment methods have also been paid enough attentions to address the uncertainties from the adversary and smart grid. To describe the relations between uncertain causes (threats and vulnerabilities) and their possible consequences (cascading events), Ciapessoni \textit{et al.} proposed a probabilistic risk-based security assessment method based on an efficient contingency scenario generator and extended risk indicators \cite{ciapessoni2016probabilistic}. Considering that the false data injected into smart meters propagates randomly in the \acrshort{ami} network, Liang \textit{et al.} established an impact assessment framework to investigate their impacts on the smart grid control and operations based on the state machine approach \cite{9866019}.}

\section{Defense-in-Depth Strategies: Prevention}\label{Section V}
{\color{black}Preventive technologies are divided into cyber- and physics-based according to their application scenarios. Cyber-based technologies are collected from the \acrshort{it} domain like encryption and authentication, and they can be deployed at host, protocol, system, and network levels to \textit{prevent the adversary from intruding into the system network}. Physics-based methods aim to exploit the robustness of control and operation algorithms or deploy extra protection devices in the \acrshort{ot} environment to \textit{prevent the attack from inducing hazardous consequences on the system operation}.} 

\begin{figure}
    \centering
    \includegraphics[width=9cm]{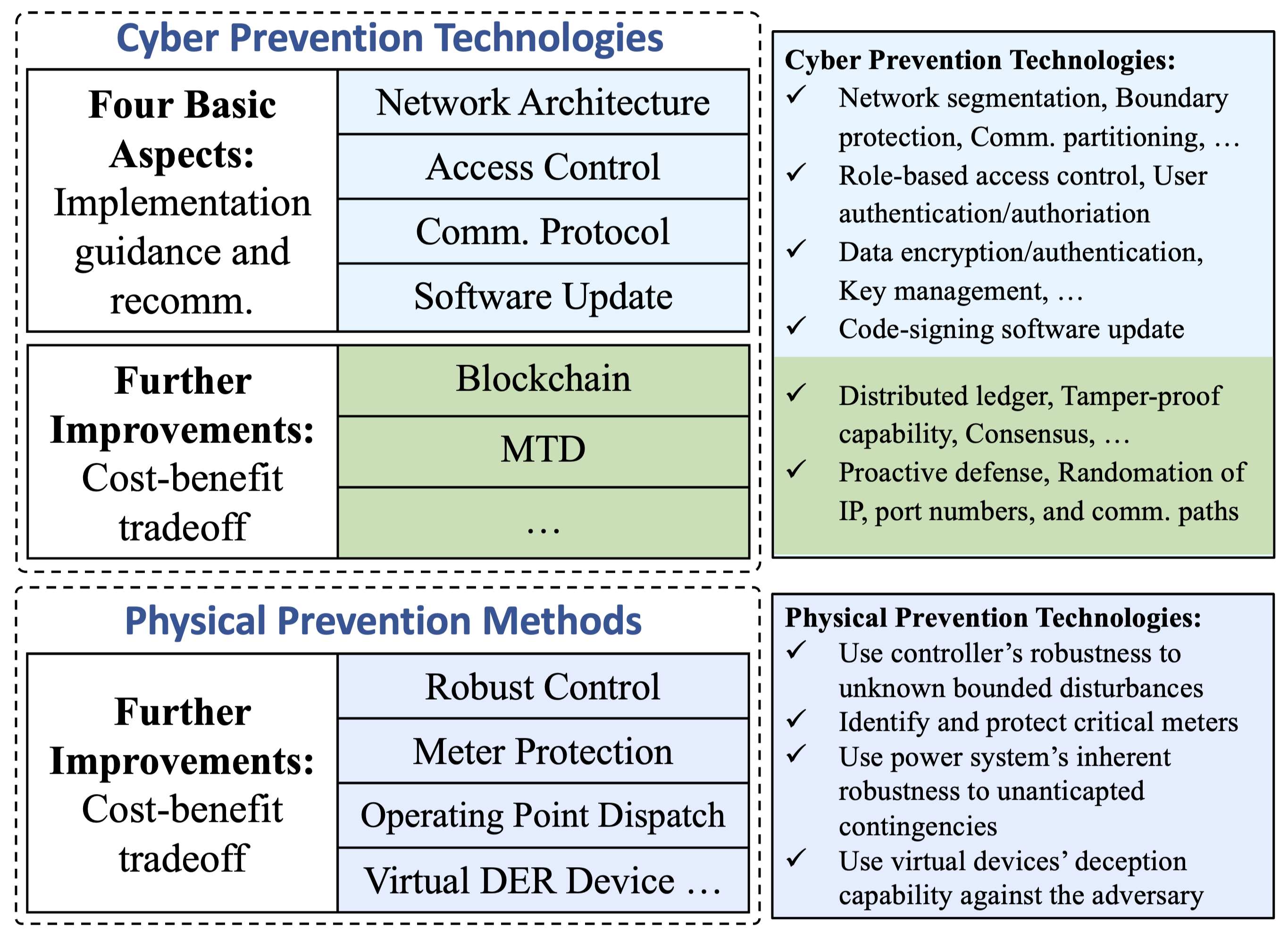}
    \caption{\color{black}Summary of Prevention Technologies and Methods}
    \label{fig:prevention technologies}
\end{figure}

\subsection{Cyber Prevention Technologies}
Various cyber preventative technologies can be found from the \acrshort{it} domain, and here we mainly summarize the results of the SunSpec/Sandia \acrshort{der} Cybersecurity Workgroup \cite{johnson2022distributed}, whose primary objective is to advance cybersecurity in the \acrshort{der} community by creating consensus around \acrshort{der} cybersecurity standards, guidelines, and best practice documents. As shown in Fig. \ref{fig:prevention technologies}, the subsequent parts will be expanded following the four basic aspects including network architectures guidelines, access control requirements, communication requirements, patching requirement, and two further improving technologies comprising blockchain and moving target defense (\acrshort{mtd}).

\begin{figure}
    \centering
    \includegraphics[width=9cm]{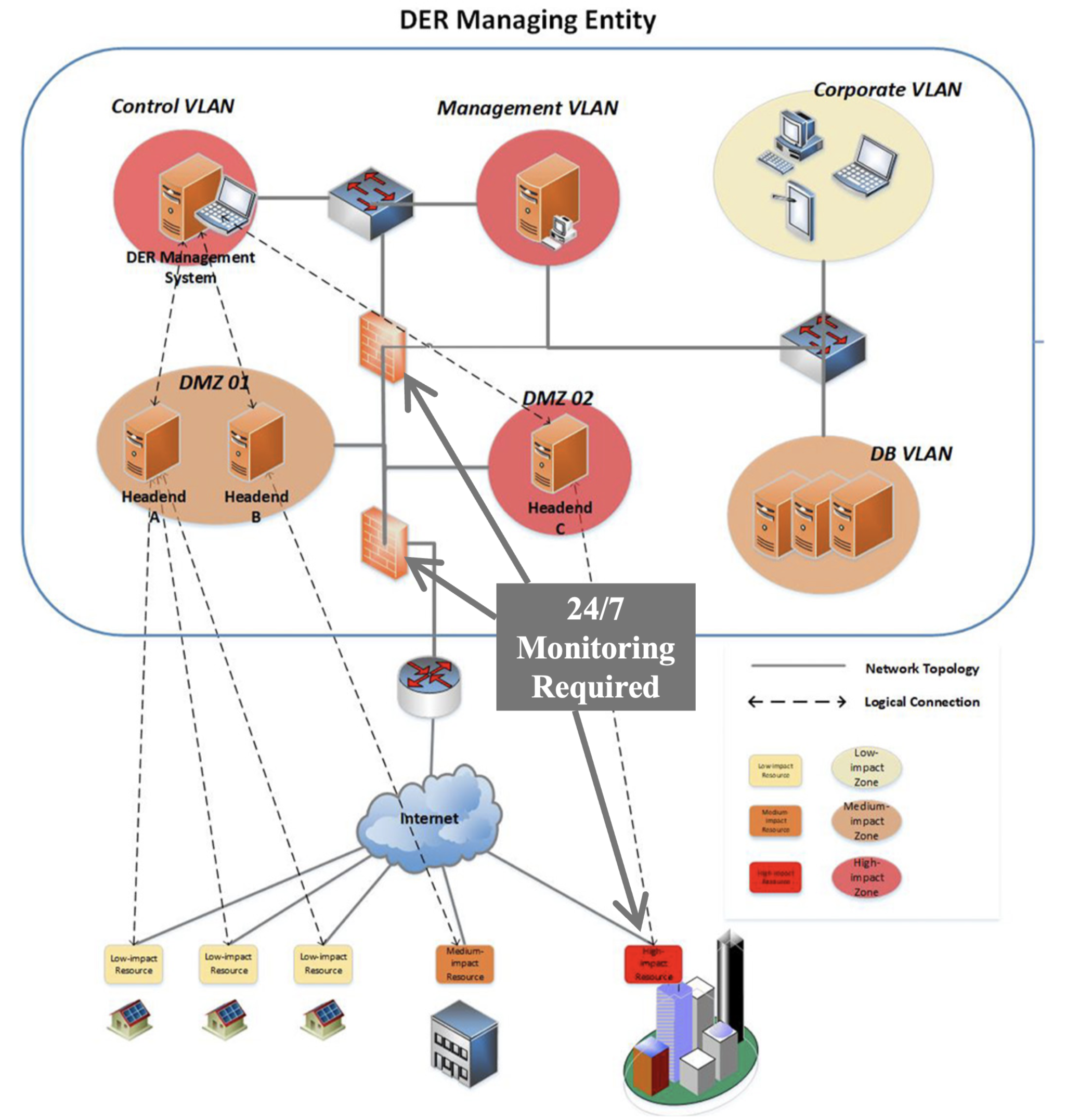}
    \caption{\color{black}Reference network architecture for the \acrshort{der}-based smart grid \cite{EPRI_NetworkDesignGuidelines}.}
    \label{fig:network segmentation}
\end{figure}

{\color{black}{\textit{\textbf{Network Architecture Guidelines:}}
A practical set of cybersecurity requirements pertaining to the network components supporting \acrshort{der} communications has been provided to minimize the likelihood, duration, or impact of a successful cyberattack \cite{EPRI_NetworkDesignGuidelines}. 
{This set of requirements does not make any assumption to the communication protocols, particular functional standards, or certain ownership/business models in terms of their effectiveness in cybersecurity.} 
Rather, it aims to provide a holistic view of the interconnected \acrshort{der}-based smart grid, and it suggests how they can be protected from cyberattacks. Four aspects of requirements and their implementation guidelines are detailed with a demonstration provided in Fig. \ref{fig:network segmentation}. 

   - \textit{Resource Criticality Level}: Each \acrshort{der} or supporting system in \acrshort{der} communications must be categorized into one of three criticality levels—high , medium or low impact as shown in Fig. \ref{fig:network segmentation}. {\color{black} The resource's criticality level is determined by the impact of any misuse of that resource to grid reliability, public safety, finances, and privacy. Different headends are allocated to different critical groups to accomplish separate control paths.} 
    
    - \textit{Network Segmentation}: Resources with different criticality levels must be located in different security zones. 
    {\color{black} As shown in Fig. \ref{fig:network segmentation}, the central management systems will typically consist of multiple zones containing headends of various criticality levels and a zone containing the core managing system, which will have the highest criticality of all the zones.
    Moreover, communications between two different security zones must be routed through the security gateways with access controls like a firewall. 
    In particular, communications between a system/resource in the high-impact zone and a system/resources in the low-impact zone must be routed through a \acrshort{dmz}\footnote{\color{black}The \acrshort{dmz} is a separate network zone where the traffic entering and exiting the \acrshort{dmz} is controlled by the relevant security gateways, but an additional level of control/traffic filtering is exerted by the devices inside the \acrshort{dmz}.} like the managing system in Fig. \ref{fig:network segmentation} communicating with low-impact residential \acrshort{der}s.}
    
    - \textit{Boundary Protection}: Access control in security gateways should be configured to deny a connection request  to a higher security zone by default. 
    {\color{black} In Fig. \ref{fig:network segmentation}, traffic should be blocked from the Internet to the \acrshort{dmz}s and from the \acrshort{dmz}s to the managing system, and should only be allowed in the opposite direction.} 
    Security gateways at the boundary of high-impact zones and interfacing with external networks must be monitored on a 24/7 basis to detect security events negatively impacting the operation of systems or resources in the security zone. 
    
    - \textit{Communications Partitioning}: \acrshort{der} communications to/from must be physically or logically partitioned from other types of communication. 
    {\color{black} In Fig. \ref{fig:network segmentation}, a shared switch uses \acrshort{vlan}s to segregate the corporate \acrshort{vlan} from the data base \acrshort{vlan}.} 
    Communications required for the administration of network infrastructure must be physically or logically partitioned from other types of communication. 
    {\color{black} The reference architecture in Fig. \ref{fig:network segmentation} shows a management \acrshort{vlan} for several switches and firewalls.}}}


\textbf{\textit{Access Control:}}
With multiple entities needing differing levels of access to \acrshort{der} data and control modes, there is a need to establish robust access control security policies and technologies. Access control restricts access to resource functionality unless the user is authorized, preventing unauthorized users from changing power system control settings. The \acrshort{rbac} is a natural choice for \acrshort{der} communication environments because there are clear roles for subjects based on their company of employment, job position, and responsibilities \cite{johnson2021recommendations}. Establishing an \acrshort{rbac} mechanism for the \acrshort{der}-based smart grid requires detailed information on the hardware and software. Based on the IEC 62351-8 \acrshort{rbac} implementation, these requirements are covered below.

    - \textit{User Authentication}: Users must provide one or more proofs of identity to ensure they are who they claim to be. Some options for user authentication includes Challenge-Response, Kerberos \cite{neuman1994kerberos}, and Digital Signatures. 
    
    - \textit{User Authorisation}: Users are permitted to access data, services, resources, or objects granted by the security policy. Two authorization mechanisms are structured query language \cite{halvorsen2016structured} and \acrshort{ldap} \cite{yeong1995lightweight}.

When selecting these mechanisms for the \acrshort{der} access control implementation, the administrative overhead and ease of implementing administration delegation are important. For instance, it is common to use Kerberos for authentication and \acrshort{ldap} for authorization. 


\textbf{\textit{Communication Requirements:}}
In the IEEE 1547-2018 interconnection and interoperability standard \cite{IEEStd1547_2018}, standardized information exchange interfaces between associated \acrshort{der} entities like IEEE 2030.5, IEEE 1815, SunSpec Modbus, and IEC 61850-7-420 have been identified to improve the interoperability. To ensure the security of information flows over public or private networks, \acrshort{der} communications and their corresponding security measures must be standardized, to prevent malicious control or misuse. 
{\color{black} For instance, some protocols lack authentication and authorization, allowing unauthorized control of \acrshort{der} equipment by individuals with network access and knowledge of the \acrshort{der}'s address. 
Moreover, implementing cryptographic methods in protocols lacking inherent security features may require a bump-in-the-wire approach, which does not provide application layer security and can introduce latency. 
In particular, to ensure the security of data-in-transit for \acrshort{der} equipment, it is crucial to address the security requirements to
In particular, the three basic requirements are to: 1) assure the data-in-transit authenticity, 2) verify the device identity, 3) confirm that encryption keys are securely managed, and 4) provide access control. }
Based on the analysis of the security strengths and weakness of communication technologies, a set of security recommendations for \acrshort{der} application protocols has been proposed \cite{onunkwo2020recommendations}:

    - {\textit{Data Encryption and Authentication for Bulk Traffic}}: Adopt \acrshort{tls} v1.3 with encryption like advanced encryption standard Galois Counter Mode \cite{rogaway2002authenticated}.
    
    - {\textit{Device Authentication}}: Use X.509v3 digital certificates with mutual client/server authentication \cite{X509}.
    
    - {\textit{Key Management}}: Align with \acrshort{tls} v1.3, adopt Elliptic Curve for ephemeral symmetric key exchange and rivest–shamir–adlema based node authentication \cite{hankerson2006guide}.

Conflicts exist between these security requirements and the processing limitations of \acrshort{der} equipment. For example, \acrshort{der} equipment without cryptographic hardware relies heavily on standard software libraries to support encryption, authentication, and hashing operations executed on the CPU, which may induce unacceptable latency for communication-based control of devices for grid supports. Nevertheless, some preliminary case studies indicate that the proper implementation of these security features will not impact \acrshort{der}-based grid control systems (well below the IEEE 1547-2018 limits for \acrshort{der} latency) but improved the security posture of the devices and networked system \cite{osti_1761846}. The change in roundtrip time due to addition of encryption is on the order of \textit{milliseconds}. 
To meet the stringent latency and messaging throughput requirements while retaining the benefits of public key cryptography, less-online/more-offline signatures model was proposed to allow the verification to be divided into online/offline phases such that online verification does not perform any expensive operations \cite{esiner2022lomos}.

\textbf{\textit{Code-Signing Software Patching:}}
Since the \acrshort{der} equipment is expected to operate in the field for 25 or more years, there will undoubtedly be newly discovered vulnerabilities in software packages or custom code that is running on the equipment during this period. 
The primary technology used for secure patching in the \acrshort{der} environment is the code-signing scheme \cite{osti_1825358}, which uses a digital signature mechanism to verify the identity of the data source and a checksum/hash to verify the data has not been altered in transit. Basically, the code-signing scheme mainly includes three actors:

    - The \textit{developer} of the code or data who submits the code to the signer.
    
    - The \textit{signer} entity that is responsible for managing the signing keys. The signer securely generates the private/public key pair and then provides the public key to a certification authority through a certificate signing request to tie their identity to the public key.
    
    - The \textit{verifier} that is responsible for validating the signed code signature.

There are multiple threats to the code-signed firmware. For example, it is possible that software developed by an organization has malicious firmware embedded in the signed version. This could be perpetrated by an \textit{insider} or through compromise of the firmware development environment, as was the case in the well-known SolarWinds attack \cite{willett2023lessons}. Awareness of this type of risk and application of appropriate mitigation methods are critical for \acrshort{der} vendors. 
A list of suggested firmware updates for \acrshort{der} equipment, product suppliers, aggregators, and owners is provided in \cite{osti_1825358}.

{\color{black}\textit{\textbf{Blockchain:}}
{Blockchain is a digital data structure comprised of a shared, decentralised, and distributed database or ledger with a continuous log of chronological transactions. Each block contains transaction data, a timestamp, and a hash point which is linked to the previous block. The hash values are crucial to its tamper-proof capability as the compromise of the block content requires to alter all subsequent blocks, which is practically impossible \cite{yap2023blockchain}. 
The blockchain technology can be introduced to establish a trustworthy network for multiple stakeholders comprising \acrshort{der} owners, \acrshort{der} aggregators, and utility operator without requiring a trusted third party.} 
Due to the decentralised data sharing/management scheme and transparent and immutable transaction for security, the potential of implementing \acrshort{der}-involved applications such as \acrshort{p2p} energy trading \cite{yildizbasi2021blockchain}, smart contract \cite{oprea2020two}, energy management \cite{luo2021blockchain}, competitive pricing \cite{wang2022two}, and secure control \cite{9681332,9709201,9931989} using blockchain has been widely investigated. {The investment costs and technological infrastructure are the greatest obstacles in integrating the blockchain into the \acrshort{der}-based smart grid.}

\textit{\textbf{Moving Target Defense:}}
\acrshort{mtd} is a proactive defense mechanism aiming to enhance security by dynamically modifying the controlling the attack surface through system configuration manipulation, rather than eliminating all vulnerabilities of system components \cite{osti_1761846}. The goals of \acrshort{mtd} include: i) Increase uncertainty and complexity for any adversary of the system; ii) Decrease the opportunities for the attacker to identify vulnerable system components; iii) Introduce higher cost in launching attacks or scans \cite{cho2020toward}. The \acrshort{mtd} technologies can be thought of as additional layers of defense to help protect a system from an adversary attempting to gain an understanding of a system in the early stages of an attack. The application of a \acrshort{mtd} tool that leverages the \acrshort{sdn} to randomize application port numbers, IP addresses, and communication paths in a \acrshort{ics} communication network was verified in \cite{osti_1812205}.}

{\color{black}\subsection{Physical Prevention Methods}
This subsection presents two representative physical preventative methods as below:

\textbf{\textit{Robust Control:}} The robust control based preventive method treats injected bounded biases as unknown uncertainties and the robust controller is designed to ensure that the tracking error under attacks could be bounded, which 
typically requires no other investments besides inducing some extra computation burdens \cite{zhou1998essentials}. Sadbadai \textit{et al.} designed a series of distributed cyber-resilient controllers for (parallel) DC and AC micorgrids (focusing on frequency regulation and active power sharing) to mitigate the adverse impact resulted from the bounded \acrshort{fdi} attacks against secondary communication links and actuator signals \cite{sadabadi2021distributed}. Once several key resiliency-related indices are designed to be large enough, the system states can converge to expect values with arbitrary small errors.

\textbf{\textit{Meter Protection:}} By strategically protecting a set of meters like smart meters from being compromised by the adversary, the attack-induced impact region can be bounded. From this perspective, Deng \textit{et al.} focused on designing the least-budget defense strategy to protect power systems against FDI attacks, which was then extended to investigate choosing which meters to be protected and determining how much defense budget to be deployed on each of these meters \cite{deng2015defending}.

{\color{black}\textbf{\textit{Operating Point Dispatch:}}
By extending the conventional security-constrained \acrshort{opf} analysis to incorporate the risk induced by cyberattacks, Xiang \textit{et al.} developed a holistic robustness framework to improve the power system operation's robustness under significant bias injections \cite{7817901}. With special focus on dummy data attacks, Du \textit{et al.} designed a robust mitigation strategy to thwart the construction of such highly stealthy attack scenarios \cite{10005238}. Regarding the resource's uncertainty possessed by the adversary and the attack's multiple periodicity, Xiang \textit{et al.} \cite{8286941} and Du \textit{et al.} \cite{9667096} proposed corresponding defender-attacker-defender models that consider the security personnel at the top-level, the attacker at the middle-level, and the power system operator at the bottom-level to develop cost-efficient and robust defense strategies.}

}

{\color{black}\textit{\textbf{Virtual \acrshort{der} Devices:}} 
As one of the typical deception technologies, the virtual \acrshort{der} device can offer multiple cybersecurity defense functionalities to capture adversary tactics and techniques to expand our understanding of the threat landscape and \acrshort{der} vulnerabilities. In particular, the virtualised \acrshort{der} device will be configured to provide protection by directing adversary's focus away from critical assets and detection by sending alerts when the adversary interacts with the artificial equipment. Virtual \acrshort{der} devices are usually deployed in the forms of i) Honeypots--internet-connected applicants to capture adversary actions, and ii) Canaries--virtualized device alongside real \acrshort{der} units.
A Laboratory Directed Research and Development project was conducted to design high-fidelity \acrshort{der} honeypot/canary prototypes \cite{osti_1821540}, providing informative references for further development.
Besides virtual DER devices, the methodology of creating virtual IEDs named as DecIED that imitates the device characteristics and communication models of IEC 61850-compliant IEDs was proposed, which can realize k-anonymous smokescreen by virtually showing $k-1$ indistinguishable decoy devices \cite{yang2020decied}.}


\textit{Lessons Learned}: 
{\color{black}As indicated by Fig. \ref{fig:prevention technologies}, the cyber prevention technologies play a leading role in the prevention phase, and basic implementation guidance and recommendations have been detailed to pave the way towards a resilient \acrshort{der}-based smart gird. The further prevention improvement resulted from physical prevention methods is usually not mandatory and depends on the vulnerability level and security demand of the specific scenario. For example, in the \acrshort{scada} centre, the meter protection strategy is recommended to ensure its functionality under extreme cyberattack events \cite{deng2015defending}. The cost-benefit tradeoff of adopting these advanced technologies and methods should be clearly analyzed to guarantee the cost-efficiency. Moreover, one critical perception is that there is no combination of cyber and physical prevention technologies/methods that can ensure 100\% security, i.e., all potential adversaries are prevented. Intuitive explanations to this kind of dilemma include zero-day vulnerabilities and insiders. Besides, the prevention capability improvement of physical methods can induce unacceptable control and performance degradation. It is not recommended to reach an extreme high security level while not considering the security budget or seriously degrading the system performance. Instead, the integration of a timely and effective monitoring and response framework has received much recognition recently and is much more recommended, which will be covered in the following sections.}

\section{Defense-in-Depth Strategies: Intrusion Detection System}\label{Section VI}
\begin{figure*}[]
  \centering
  \includegraphics[width=18cm]{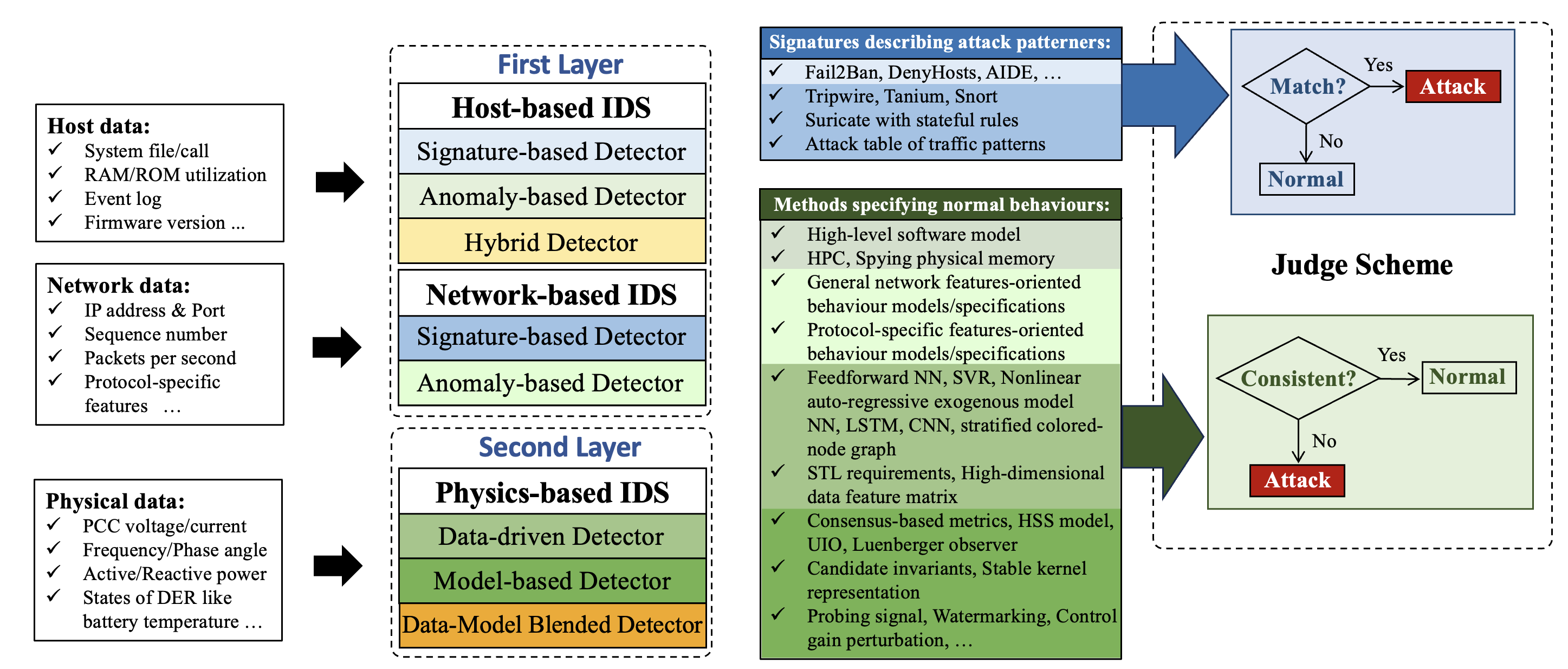}
  \caption{\color{black}Summary and classification of IDSs.}\label{fig:IDSClassification}
\end{figure*}

{\color{black}The \acrshort{ids} is responsible for detecting malicious activities by monitoring and analyzing the behaviour features originated from hosts, network devices, or physical-side sensors. According to Fig. \ref{fig:IDSClassification}, \acrshort{ids}s can be classified into three classes according to the origination of data: i) The \acrshort{hids} is to inspect the integrity of the host itself by examining the host-based features like system files, system calls, processes, RAM/ROM utilization, and firmware version. 
ii) The \acrshort{nids} aims to monitor and analyze network related attributes like IP addresses, service ports, traffic volumes, and protocol attributes. 
iii) The \acrshort{pids} is to detect the anomaly of physical measurements like PCC voltages/currents, frequency, and active/reactive power. 
Depending on the type of analysis carried out, each \acrshort{ids} can be further classified as signature-based or anomaly-based \cite{garcia2009anomaly}. The signature-based \acrshort{ids} aims to seek predefined \textit{patterns/signatures} of cyberattacks within the analyzed data. The anomaly-based \acrshort{ids} attempts to estimate the \emph{normal} behaviour of the system to be monitored using metrics, specifications, rules, observers, \acrshort{ml} training models, etc., 
and generates an anomaly whenever the deviation between the actual system and the \emph{normal} system exceeds a predefined threshold. Different from \acrshort{hids}s and \acrshort{nids}s, the attack signatures cannot be easily extracted from physical states, and thereby majority of \acrshort{pids}s are anomaly-based. According to the type of knowledge used to describe normal behaviours, \acrshort{pids}s are further classified as data-driven, model-based, and data-model blended. The data-driven \acrshort{pids} captures data-oriented characteristics of normal behaviours like \acrshort{ml} models, while the model-based \acrshort{pids} extract model-oriented properties of normal operations such as observers. The data-model blended \acrshort{pids} uses both data and model knowledge to feature the normal behaviours.}


\subsection{Host-based IDS}
The \acrshort{hids} is usually deployed at critical and vulnerable hosts like servers and workstations and \acrshort{ied}s to detect cyber intrusion. There are many signature-based \acrshort{hids} software available that can be directly installed into the upper hosts. Lai \emph{et al.} \cite{lai2021review} comprehensively reviewed these \acrshort{hids}s including Fail2Ban, DenyHosts, AIDE, Tripwire, OSSEC, Samhain, etc., analyzed their application scenarios, and highlight their features. As an integral component of \acrshort{ami} used in modern power systems, the security of smart meter has attracted great attention. Tabrizi \emph{et al.} \cite{tabrizi2014model} proposed an anomaly-based \acrshort{hids} based on the high-level model of the smart meter software, imposing little performance overhead, even under severe memory constraints, and effectively detecting both known and unknown attacks. {\color{black}To identify malicious instructions and counterfeit firmware within the inverter controller, Zografopoulos \emph{et al.} \cite{kuruvila2021hardware,zografopoulos2022time} developed an anomaly-based \acrshort{hids} utilizing custom-built \acrshort{hpc}s and time series classifiers.}
To further improve the detection performance, Liu \textit{et al.} designed a hybrid and collaborative \acrshort{hids} for smart meters by setting spying domain randomly in physical memory in combination with using secret information and event log, under which illegal reading and writing is identified once the spying domain is modified \cite{liu2015collaborative}. 

{\color{black}\textit{Lessons learned} from these HIDSs:

    - Current research status regarding \acrshort{hids}s mainly focuses on the upper hosts and smart meters.
    
    - As the most basic components that interfaces renewable sources with power grid, the energy conversion devices like converters have not obtained enough attention.
    - It is challenging to attain comparable performance using strictly limited resources on these energy conversion devices.}


\subsection{Network-based IDS}
The \acrshort{nids} is usually deployed at strategic points in the \acrshort{der} communication network, and careful considerations of the hardware and network components are needed to ensure effective security monitoring. The \acrshort{nids} using Snort equipped with default rules has been verified to be effective in detecting malevolent traffic in-between an aggregator and a single \acrshort{pv} inverter induced by naive cyberattacks \cite{jones2020implementation}. The collaboration among multiple \acrshort{nids}s placed at field device and control center levels are investigated in \cite{singh2020distributed}, where field device \acrshort{nids}s monitor Modbus-related traffic and control center \acrshort{nids}s monitor DNP3- and IEEE 2030.5-related traffic. To incorporate the physical characteristics into the design of \acrshort{nids}, Kang \textit{et al.} proposed a novel framework allowing stateful analysis methods to define its stateful rules that can be run on Suricata \cite{kang2016towards}. 
{\color{black}To relief the reliance on \acrshort{ids} software, Sun \textit{et al.} developed a signature-based \acrshort{nids} by establishing an attack table compromising the information of attack patterns in terms of attack types and time sequence of anomaly events based on the temporal failure propagation graph technique \cite{sun2019cyber}.}

The anomaly-based \acrshort{nids}s are further classified into three groups according to the feature types adopted to develop normal behaviour models. \textit{The first \acrshort{nids} group} uses general network features regardless of the protocol types. Based on the length and number of packets, the inverter behaviour model is learned using the adaptive resonance theory artificial neural network algorithm with online update capability \cite{jones2020implementation,jones2021unsupervised}. 
A distributed \acrshort{nids} framework is developed for \acrshort{ami}, where intelligent modules are deployed at three layers to perceive malicious network traffic collaboratively \cite{zhang2011distributed}. 
{\color{black}To strategically trade the false positives for a high detection probability, lightweight specification-based behavior rules are defined for critical devices of a modern electrical grid \cite{mitchell2013behavior}. }
\textit{The second \acrshort{nids} group} adopts protocol-specific features. Based on the semantics of GOOSE and SV messages, the specifications that define the normal behaviours of \acrshort{ied}s are developed and embedded in the built-in \acrshort{nids} inside \acrshort{ied}s to detect the GOOSE and SV related intrusions \cite{hong2017intelligent}. 
{\color{black}A finite state machine model for network communication was defined to detect the GOOSE-based poisoning attacks \cite{9303015}.} 
Through incorporating substation configuration description language and normal IEC 61850 traffic contents, the normal and correct behaviour models using in-depth protocol analysis are defined \cite{yang2016multidimensional}. For ZigBee-based HAN, a normal behaviour model is established according to SEP 2.0 and IEEE 802.15.4 standards \cite{jokar2016intrusion}. 
The third group concerns both general network and protocol-specific features. Using both statistical analysis of traditional network features and specification-based metrics of GOOSE and MMS, Kwon \textit{et al.} proposed a novel behavior-based \acrshort{nids} \cite{kwon2015behavior}. {\color{black}By monitoring the traffic data characteristics of transport, operation, and content levels in \acrshort{scada} network, Ren \textit{et al.} developed a edge-based multi-level anomaly detection framework \cite{8587533}.}

{\color{black}
\textit{Lessons learned} from these NIDSs:

    - Extra communication components like switches and network taps are usually required to ensure that NIDSs can access required network traffic for monitoring, and thus achieve expected detection performance. Thus, the deployment cost of NIDSs has to be concerned in the planning phase with numerous geographically dispersed terminal devices in the \acrshort{der}-based smart grid. 
    
    - The signature-based NIDS can generate a highly reliable result regarding known attacks, but is not capable of handling unknown attacks.
    On the contrary, the anomaly-based NIDS can handle unknown attacks such as zero-day attacks, while its rate of false positive alarms is higher than that of signature-based NIDS. The combination of the basic principles of signature- and anomaly-based methods to enhance NIDS's detection performance is still not clear. 
    
    - The NIDS based on general network features can be easily applied to various scenarios regardless of the communication protocol and communication architecture, while the NIDS using specific protocol-specific features can lead to better detection performance in terms of accuracy and response time. To meet the increasing applicability and performance requirements, more efforts should be devoted to the design of NIDSs incorporating both general network and protocol-specific features.}


\subsection{Physics-based IDS}
The \acrshort{pids} is usually deployed near the field devices, regarded as the last detection line, to directly interact with sensors or controllers for the sake of real-time measurement acquisition. The principal part of data-driven \acrshort{pids} is to train a \acrshort{ml} model using normal physical data, formulate specifications, or extract data features from normal physical data such that data-oriented characteristics of normal behaviours can be captured. 
{\color{black}After taking inputs of monitored data comprising of multi-interval \acrshort{der} dispatch signals and corresponding network status including nodal voltage magnitudes and phase angles, a kernel \acrshort{svr} model is adopted to predict the system margin of the time of interest \cite{kim2022identification}.}
{\color{black}By employing the Isolation Forest algorithm, which is trained on features determined from local current measurements, Saber \textit{et al.} proposed an anomaly-based scheme for detecting false-tripping attacks against line current differential relays, in the form of replay attacks, general FDI attacks, and time-synchronization attacks \cite{9804786}.} 

When it involves complex and fast-varying control dynamics, the prediction of system states would be even more challenging. Habibi \textit{et al.} tried to address this issue by adopting a nonlinear auto-regressive exogenous model neural network for the real-time estimation of voltages and currents in DC microgrids \cite{8963972}. The usage of \textit{electrical waveform data} has been verified to be powerful in the root cause diagnosis of anomalous events. Based on time-domain mean current vector-based features originated from raw waveform data, the \acrshort{lstm} and \acrshort{cnn} classifiers are able to distinguish between normal conditions, component failures, and \acrshort{fdi} attacks in \acrshort{ev}s and \acrshort{pv} farms \cite{guo2020cyberattack, guo2021data}. 
{\color{black}Besides attack detection and identification, the raw waveform data can also be used in the location of attack sources \cite{li2022adaptive,li2020detection}. 
To reduce the amount of required training data, transfer learning was incorporated into the cyberattack detection framework \cite{li2022data}. }
The specifications and data features extracted from physical data are also used to construct \acrshort{pids}s, which is training-free compared with the \acrshort{ml} methods. The \acrshort{stl} requirements, which are formalisms to monitor the output voltages and currents of DC microgrids against predefined specifications, were employed for anomaly detection \cite{beg2018signal} {If the ML model's input data is subject to adversarial perturbations, the detection accuracy can be degraded significantly. By exploring targeted and stealthy FDI attacks via adversarial machine learning, Tian \textit{et al} illustrated that the attack success rate can be as high as 80\% with small-scale attack targets \cite{9695995,9622117}. Besides, the success rate can be improved as the increase of attack scale.}

The key part of model-based \acrshort{pids}s is to develop consensus-based metrics, establish predictors/observers, or identify invariant based on the underlying model dynamics derived from physical structures and control algorithms such that the model-oriented properties of normal operations can be extracted. {\color{black}Based on on-the-fly power system dynamics simulation results, command authentication schemes were proposed to evaluate the legitimacy and validity of remote control commands near the edge of smart grid infrastructure (e.g., in substations), which can enhance the attack detection capability compared to the traditional schemes solely using steady-state information \cite{mashima2018securing,meliopoulos2016command}.} Due to the widespread adoption of consensus based secondary control in microgrids, various consensus-oriented detection metrics such as \acrshort{cvf} \cite{sahoo2018stealth} were derived to detect anomalous sensor measurements and communicated data in DC microgrids. 
{\color{black}When utilizing the primal-dual algorithm to solve the consensus optimization problem in isolated microgrids, dual variable-related detection metrics could be designed to detect \acrshort{fdi} attacks \cite{lu2019intrusion}. }

To further improve the detection accuracy, the physical dynamics obtained from Kirchhoff circuit laws were incorporated into the design of attack detectors. 
{\color{black}The \acrshort{hss} model was developed to predict current measurements of \acrshort{pv} farms, which were then used for integrity verification \cite{9580468}. }
By synthesising a Luenberger observer and a bank of \acrshort{uio}s, a distributed monitoring scheme was established for each \acrshort{der} unit to verify the integrity of neighbors' data \cite{gallo2020distributed}. 
{\color{black}Considering the robustness against unknown disturbances and parameter variations, a multi-objective optimization problem was formulated to design the generation scheme of detection residuals \cite{tan2022false}. }
The system properties that do not vary over time under normal operations are also adopted as indicators for the anomaly induced by cyberattacks. By identifying the variation of inferred candidate invariants that are extracted from both physical plant and controller software, Beg \textit{et al.} proposed a \acrshort{fdi} attack detection scheme for DC microgrids \cite{beg2017detection}.
{\color{black}With the small-signal model of islanded microgrids, Zografopoulos \textit{et al.} adopted the subspace method to identify its stable kernel representation in the attack-free situation such that any violation could be perceived \cite{zografopoulos2021detection}. }

Besides the passive anomaly perception principle, the proactive incentive-based detection scheme has also attracted great attention, which proactively adds secret perturbations to system dynamics or signals, for stealthy \acrshort{fdi} attack detection \cite{zhang2019analysis,zhang2020hiddenness,9497752,10400059,9721122}. 
After generating specified small probing signals and then injecting them into controllers, the output signals are compared with pre-determined values to locate the infracted controller components in microgrids \cite{7581101}. 
Further, by adding watermarks to communicated data between \acrshort{der}s, the replay attack could be detected by testing the existence of statistical properties of watermarks \cite{huang2020detection,10107611}. Considering the system dynamics involved in DC microgrids, the primary control gain was perturbed in a specific manner to uncover the inconsistency between original data and injected one \cite{9621221,9815319}. 

{\color{black}The data and model blended PIDS has also attracted increasing attention recently due to its benefits in performance enhancement and data requirement reduction. By incorporating physical dynamics into the data recovery algorithm, Xu \textit{et al.} proposed a blending data-driven and physics-based approach to improve the detection accuracy while reduce the operational cost resulted from MTD \cite{9998121}. Based on the combination of prior knowledge of physics and system metrics, a physics-informed context-based anomaly detection method was proposed to counter the stealthy attacks against \acrshort{agc} \cite{9961034}. 
{\color{black}To alleviate the data reliance on system topology and line parameters, a physically-inspired data-driven model was proposed for electricity theft detection with merely smart meter data comprising power consumption and voltage magnitudes \cite{8637772}. 
Given that cyberattacks can be strategically counterfeited to replicate grid faults, a physics-informed spline learning approach-based anomaly diagnosis mechanism was designed in \cite{9791853}, which not only provides compelling accuracy with limited data, but also reduces the training and computational resources signiﬁcantly.} To achieve timely and accurate attack localization and also output explainable detection results, Peng \textit{et al.} incorporated the nodal admittance matrix and physical property of power grid into the graph convolutional network \cite{10081329}.
}

{\color{black}\textit{Lessons learned} from these PIDSs:
    
    - Generally speaking, the HIDS and NIDS can perceive the anomalous traces on host and network related features resulted from malicious intruders, with a quicker rate, than the PIDS as the adversary will not disrupt the physical functionalities immediately after intruding the \acrshort{der} communication network. But the PIDS works as the last detection layer by observing the induced physical impacts when the HIDS and PIDS are both invalidated.
    
    - The data-driven and model-based PIDSs have their own cons and pros. The data-driven PIDS can achieve satisfactory detection performance against a wide varieties of cyberattacks without requiring any model knowledge. But it relies heavily on the diversity of training data and will induce substantial computation overhead, and the inexplainable output also limits its widespread application. The model-based PIDS can detect known types of cyberattacks in a timely and reliable manner with explainable detection results and acceptable computation burden. However, the detection performance can degrade significantly when the system parameters vary and it only works under limited types of cyberattacks.
    
    - The data and physics blended PIDS has became a prevailing topics as it is particularly suitable for the \acrshort{der}-based smart grid with massive measurement data and well-known physical dynamics.
    
    - The proactive detection strategy by perturbing system parameters can enhance the detection capability against powerful adversaries with acceptable sacrifice on either control or operation performance.}


{\color{black}The summary of reviewed IDSs, focusing on applied scenarios, utilized tools/methods, and evaluation metrics, is given in TABLE \ref{Table:IDSsummary1}. From a high-level perspective, a set of evaluation metrics regarding \acrshort{ids}s is refined: 1) Performance-related metrics: detected attack types, detection accuracy, and detection latency; 2) Cost-related metrics: memory and computation overhead, hardware investment, and control and operation performance sacrifice. The design of IDS should at least consider one type of performance- and cost-related metrics and address the trade-off between them.}

\begin{table*}[]
{\color{black}
\footnotesize
\centering
\begin{threeparttable}
\caption{\color{black}Summary of IDSs}\label{Table:IDSsummary1}
\begin{tabular}{m{2cm}m{1cm}m{2.5cm}m{3.5cm}m{7cm}}
\Xhline{1.2pt}
\multicolumn{5}{c}{\textbf{Host-based IDSs}} \\ \Xhline{1.2pt}
  \textbf{Type} & \textbf{Lit.} & \textbf{Scenario} & \textbf{Tools/Methods} & \textbf{Evaluation Metrics} \\ \Xhline{1.2pt}
   \multirow{1}{2cm}{Signature-based} & \cite{lai2021review} & Upper host & Tripwire, OSSEC, etc. & 
Attack: Known attacks; Detection latency: Timely \\ \cline{1-5}
   \multirow{3}{2cm}{Anomaly-based} & \cite{tabrizi2014model} & Smart meter & Abstract model based verification of core system calls & Attack: Known and unknown; Coverage: 100\% known and 69.9\% unknown; Latency: 10s; Memory overhead: 4.15\%
    \\ \cline{2-5}
   & \cite{kuruvila2021hardware,zografopoulos2022time} & Inverter controller & Custom-built HPCs and time series analysis & Attack: Firmware modification; Accuracy: 97.22\%, Latency: Not tested \\ \cline{1-5}
    \multirow{1}{1cm}{Hybrid} & \cite{liu2015collaborative} & Smart meter & Collaborative signature and anomaly combined detection & Attack: Known and unknown; Accuracy: $>$80\%; Memory overhead: 0.8\%  \\ \Xhline{1.2pt}
\multicolumn{5}{c}{\textbf{Network-based IDSs}} \\ \Xhline{1.2pt}
\textbf{Type} & \textbf{Lit.} & \textbf{Scenario} & \textbf{Tools/Methods} & \textbf{Evaluation Metrics} \\ \Xhline{1.2pt}
\multirow{7}{2cm}{Signature-based} & \cite{jones2020implementation} & DER comm. network & Snort with default rules & Attack: 5 scenarios; Detection coverage: 60\%; Memory overhead: 31.25\% \\ \cline{2-5}
 & \cite{singh2020distributed} & DER comm. network & Cross-level Snort-based detection with tailored rules & Attack: DoS attack; Accuracy: 100\%; Latency: $\le$500ms \\ \cline{2-5}
 & \cite{kang2016towards} & DER comm. network & Suricate with stateful rules & Attack: FDI attack; Accuracy: 100\%; Latency: Not tested  \\ \cline{2-5}
 & \cite{sun2019cyber} &DER comm. network & Attack table, Temporal failure propagation graph & Attack: DoS and FDI attacks; Accuracy: 100\%; Latency: Not tested  \\ \cline{1-5}
\multirow{20}{2cm}{Anomaly-based} & \cite{jones2020implementation,jones2021unsupervised} & DER comm. network & Adaptive resonance theory artificial neural network & Attack: 5 scenarios; Detection coverage: 80\%; Memory overhead: 45\%; Train/Test time: 33ms/14ms  \\ \cline{2-5}
 & \cite{zhang2011distributed,mitchell2013behavior} & SCADA network & Support vector machine and Artificial immune systems & Attack: FDI, DoS, and eavesdropping attacks; Accuracy: 99.33\%; Latency: Not tested  \\ \cline{2-5}
 & \cite{hong2017intelligent} & Substation network & Collaborative and distributed intrusion detection with normal behaviour model & Attack: FDI attack; Accuracy: N.A.; Latency: 2ms; Memory overhead: 2\%  \\ \cline{2-5}
 & \cite{9303015} & Substation network & Finite state machine model & Attack: FDI attack; Accuracy: 95\%; Latency: 0.06ms;  \\ \cline{2-5}
 & \cite{yang2016multidimensional} & Substation network & Access control, Protocol whitelisting, and Multiparameter-based detection & Attack: DoS and FDI attacks; Accuracy: 100\%; Latency: $<$0.3ms  \\ \cline{2-5}
 & \cite{jokar2016intrusion} & ZigBee network & Normal behaviour model established referring to SEP 2.0 and IEEE 802.15.4 standards & Attack: FDI, replay, and DoS attacks; Accuracy: $\ge$92.5\%; Latency: Not tested  \\ \cline{2-5}
 & \cite{kwon2015behavior} & Substation network & Statistical traffic features and Specification-based metrics & Attack: 27 scenarios; Accuracy: 98.89\%; Latency: Not tested  \\ \cline{2-5}
 & \cite{8587533} & SCADA network & Traffic data characteristics of transport, operation, and content levels & Attack: 12 scenarios; Accuracy: 100\%; Latency: 423ms \\ \Xhline{1.2pt}

 \multicolumn{5}{c}{\textbf{Physics-based IDSs}} \\ \Xhline{1.2pt}
 \textbf{Type} & \textbf{Lit.} & \textbf{Scenario} & \textbf{Tools/Methods} & \textbf{Evaluation Metrics} \\ \Xhline{1.2pt}
\multirow{11}{2cm}{Data-driven} & 
\cite{kim2022identification} & SCADA network & Kernel SVR & Attack: FDI attack; Accuracy: 100\%; Latency: 2 hours \\ \cline{2-5}
& \cite{9804786} & Substation network & Isolation forest algorithm & Attack: Replay attack, FDI attack, and Time-synchronization attack; Accuracy: 99.89\%; Latency: 8ms \\ \cline{2-5}
& \cite{8963972} & Microgrid & Auto-regressive exogenous model neural network & Attack: FDI attack; Accuracy: 100\%; Latency: Not tested \\ \cline{2-5}
& \cite{guo2021data,guo2020cyberattack} & Grid-tied DER & LSTM and CNN classifiers and Physics-guided features & Attack: Replay attack and FDI attack; Accuracy: $\ge$98.44\%; Latency: Not tested   \\ \cline{2-5}
& \cite{li2022adaptive,li2020detection,li2022data} & Grid-tied DER & LSTM and CNN classifiers and Transfer learning & Attack: FDI attack; Accuracy: $\ge$95.23\%; Latency: Not tested  \\ \cline{2-5}
& \cite{beg2018signal} & Microgrid & STL requirements based specifications & Attack: DoS attack and FDI attack; Accuracy: 100\%; Latency: $<$1s  \\ \cline{1-5}

\multirow{11}{2cm}{Model-based} & \cite{mashima2018securing,meliopoulos2016command} & Substation network & On-the-fly power system dynamics simulation & Attack: FDI attack; Accuracy: 83\%; Latency: 859ms \\ \cline{2-5}

& \cite{sahoo2018stealth} & Isolated Microgrid & Consensus-oriented metric CVF & Attack: FDI attack; Accuracy: 100\%; Latency: $<$1s  \\ \cline{2-5}

& \cite{lu2019intrusion} & Isolated Microgrid & Dual variable-related detection metrics & Attack: FDI attack; Accuracy: 100\%; Latency: $<$1s  \\ \cline{2-5} 

& \cite{gallo2020distributed,tan2022false} & Isolated Microgrid & Luenberger observer, UIO & Attack: FDI attack; Accuracy: 100\%; Latency: $<$1s  \\ \cline{2-5}

& \cite{beg2017detection} & Microgrid & Candidate invariant & Attack: FDI attack; Accuracy: 100\%; Latency: $<$1s \\ \cline{2-5}

& \cite{zografopoulos2021detection} & Isolated Microgrid & Stable kernel representation & Attack: FDI attack; Accuracy: 100\%; Latency: $<$1s  \\ \cline{2-5}

& \cite{7581101} & Microgrid & Probing-based proactive attack detection & Attack: FDI attack; Accuracy: 100\%; Latency: $<$1s; 
Cost: No  \\ \cline{2-5}

& \cite{9621221,10107611,huang2020detection,9815319} & Microgrid & Watermarking and Primary control gain perturbation & Attack: FDI attack and Replay attack; Accuracy: 100\%; Latency: $<$1s; Cost: Neglectable  \\ \cline{1-5}

\multirow{8}{2cm}{Data and model blended} & \cite{9998121} & SCADA network & LSTM and Event-triggered MTD & Attack: FDI attack; Accuracy: 98.16\%; Operational cost: 0.5\% \\ \cline{2-5}

& \cite{9961034} & Automatic generation control & CNN, LSTM, and Physical knowledge & Attack: FDI atatck; Accuracy: 93.2\%; Latency: Not tested \\ \cline{2-5}

& \cite{8637772} & Distribution network & Linear regression and Power flow dynamics & Attack: FDI attack; Accuracy: 94\%; Latency: Not tested \\ \cline{2-5}

& \cite{9791853} & Grid-tied DER & Spline learning and Power electronics dynamics & Attack: FDI attack; Accuracy: 98.23\%; Latency: 25ms \\ \cline{2-5}

& \cite{10081329} & SCADA network & Graph convolutional network & Attack: FDI attack; Accuracy: 99.25\%; Latency: Not tested \\ \Xhline{1.2pt}

\end{tabular}
\end{threeparttable}
}\end{table*}

\section{Defense-in-Depth Strategies: Impact Mitigation System}\label{section: VII}
\begin{figure*}[!ht]
  \centering
  \includegraphics[width=18cm]{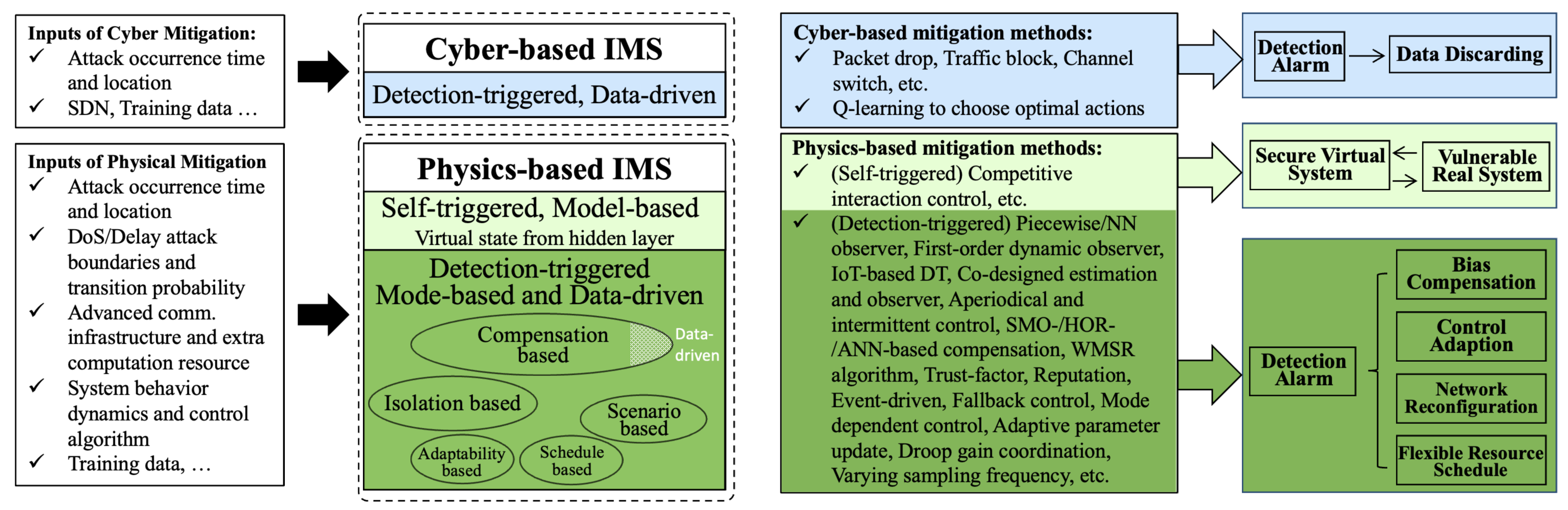}
  \caption{\color{black}Summary and classification of IMSs.}\label{fig:IMSClassification}
\end{figure*}

{\color{black}The IMS aims to restrict the impacts caused by cyberattacks and tries to restore the system performance. According to the basic knowledge domain of adopted mitigation actions, IMSs are classified as cyber-based and physics-based: The cyber-based IMS uses intuitive cyber-side actions like packet drop to exclude the malicious components from the remaining network; The physics-based IMS adopts local control capability or global resource schedule flexibility to compensate for the data integrity/availability loss (FDI/DoS). Furthermore, based on the activation scheme of the mitigation action, IMSs are divided into self-triggered and detection-triggered. The detection-triggered IMS can be only activated when the IDS alarms while the self-triggered IMS can work autonomously regardless of IDS's outputs. As shown in Fig. \ref{fig:IMSClassification}, the cyber-based IMSs are all detection-triggered as the cyber-involved mitigation actions can only work after taking inputs of attack occurrence time and location. The physics-based IMSs consist of both self-triggered and detection-triggered, and some concepts like bias compensation, control adaption, network reconfiguration, and flexible resource schedule are further utilized to distinguish the specific methods adopted in IMSs. The self-triggered physics-based IMSs heavily rely on a secure virtual system enabled by the advanced blockchain and \acrshort{sdn} technologies that can interact with vulnerable real systems to correct any misbehaving actions caused by data disruption/compromise without the need for an in dependent detection module. {In parallel with the detection- and self-triggered classification metrics, IMSs can be also classified as data-driven and model-based. All cyber-based IMSs are data-driven, and most physics-based IMSs are model-based with only a small portion of detection-triggered IMSs being data-driven as illustrated by Fig. \ref{fig:IMSClassification}.} }

\subsection{Cyber-based Detection-triggered IMS} This type of IMS adopts the most intuitive cyber-side mitigation actions encompassing packet drop, traffic block, channel switch to thwart the cyber-side propagation of attack impacts. The simplest mitigation decision is to block the associated malicious network traffic regardless of the anomaly type. When a \acrshort{der} unit is found subject to DoS attacks, it will enter the protective mode where only outgoing network traffic is permitted \cite{appiah2020decentralized,hong2017intelligent}. In the worst case, if anomaly is detected twice while the compromised unit cannot be located, then all \acrshort{der} units will enter protective mode to gain time for the control center to attend to the aroused security issues. For the microgrid enabled by \acrshort{sdn} technologies, the SDN controller is designed to block the network traffic from/to the malicious \acrshort{der} unit to guarantee the normal operation of the remaining units when anomaly is perceived \cite{li2018sdn}. 
{\color{black}To achieve the cost-benefit trade-off, Jokar \textit{et al.} presented a Q-learning based intrusion prevention system for the ZigBee-based home area network to automatically adjust the mitigation strategies facing a wide varieties of cyberattacks \cite{jokar2016intrusion}.}

{\color{black}\textit{Lessons learned from cyber-based IMSs:} The cyber-based mitigation strategies are suitable for the pure IT system where the data availability is not the primary concern. However, when involving the closed-loop control functionalities that require real-time interaction with the physical plant, these cyber-side strategies can be too aggressive as the data availability loss may induce severe stability issues. Moreover, it is not enough to thwart the propagation of attack impacts by merely excluding the cyber-side malicious sources as the physical couplings could also be exploited for impact propagation. Hence, the cyber-side actions are usually not regarded as the primary choice for impact mitigation in the \acrshort{der}-based smart grid.}

\subsection{Physics-based Self-triggered IMS}
Since the DoS attack can be easily detected and the subsequent mitigation actions will be activated accordingly, the self-triggered IMS mainly focuses on the FDI attack. 
This type of IMS relies on the construction of a compensation term, which can be just a variable with no physical meaning such that the attack impact can be mitigated to a certain extent after incorporating the compensation term into the controller. 
With the assistance of a \textit{hidden and secure} network layer enabled by advanced SDN technologies, a series of virtual states are established to interact with the original control layer such that the anomalous activities could be corrected in an autonomous manner. Liu \textit{et al.} designed resilient secondary controllers for micorgrids such that the frequency synchronization and active power sharing can be regulated to an arbitrarily small region around the expected point under bounded FDI attacks \cite{liu2021robust}. 
{\color{black}To handle \textit{unbounded} FDI attacks, Zuo \textit{et al.} proposed a novel attack-resilient control framework to assure the uniformly ultimately bounded voltage containment and frequency regulation \cite{zuo2020resilient}.}

{\color{black}\textit{Lessons learned from physics-based self-triggered IMSs}:

    - Majority of physics-based IMSs are detection-triggered, and only a small portion are self-triggered.
    
    - Although the self-triggered does not require the inputs from IDSs, which can avoid potential false positive alarms, two limitations also come along with this cons: i) A hidden secure network layer independent from the original control layer should run all the time, inducing extra computation and communication overheads; ii) The introduction of hidden layer can expose larger attack surfaces if not equipped with appropriate security strategies. The two limitations hinder the further investigation of self-triggered IMSs.}

\subsection{Physics-based Detection-triggered IMS}
According to the adopted mitigation methods, the detection-triggered IMSs are further classified as compensation-based, isolation-based, scenario-based, adaptability-based, and schedule-based. The compensation-based methods involved in this type of IMS are to estimate/observe the unavailable data (DoS attack) or injected bias (FDI attack) after the IDS perceives anomaly. The mitigation strategies against DoS attacks are mainly detection-triggered. 
{\color{black}Given the duration-restricted DoS attacks in the \textit{centralized} \acrshort{lfc} of islanded AC microgrids, a piecewise observer was established to provide real-time estimates of unavailable system states \cite{hu2022resilient}. }
To guarantee the tracking performance of variable-speed WTs when the rotor velocity measurement is unavailable under DoS attacks, Zhao \textit{et al.} proposed a dual-triggered adaptive control strategy \cite{zhao2022dual}. 
In addition, considering the {\color{black}\textit{distributed} secondary control in multi-bus DC microgrids subject to DoS attacks, a first-order dynamic observer is adopted to estimate the unavailable load information \cite{li2022distributed}. 
Similar to the idea of hidden network layer, Saad \textit{et al.} established a IoT-based \acrshort{dt} by emulating the dynamics of cyber-physical networked microgrids to help estimate the unavailable data induced by DoS attacks \cite{saad2020implementation}.}

In terms of FDI attacks, the same idea also works by estimating/observing injected biases and healthy states. The estimation/observer can be accomplished using the corrupted signal together with some extra securely communicated data. Jiang \textit{et al.} designed distributed \acrshort{smo} and \acrshort{hod} based resilient secondary controllers for DC microgrids to compensate for the adverse impact of bounded FDI attacks \cite{jiang2021high}. 
{\color{black}Taking inputs of legitimate voltage and frequency information, a distributed observer was established to observe the healthy reactive and active power measurements, respectively, guaranteeing $L_2$-gain performance under FDI attacks \cite{shi2021observer}. 
To guarantee the uniformly ultimately bounded voltage regulation and proportional load sharing under \textit{unbounded} FDI attacks, an adaptive observer is employed to estimate the aggregated term induced by attacks on the secondary control input \cite{zuo2020distributed}. }

In addition, an \acrshort{ann} based decentralized cyberattack mitigation framework was proposed to relief the reliance on model accuracy \cite{habibi2021decentralized}. The incorporation of physical circuit dynamics can benefit the estimation of injected biases or healthy states. {\color{black} By utilising the adversarial \acrshort{drl}, Wang \textit{et al.} designed a robust mitigation method to find the optimal combination of droop gains such that the impact of destabilising attacks can be eliminated \cite{10089185}. Based on the nonlinear \acrshort{der} \textit{circuit dynamics} along with constant power loads, distributed nonlinear adaptive observer and high-order SMOs were established to jointly track the current variation, which may be corrupted by cyberattacks \cite{cecilia2021addressing}. Based on the information (voltage varying slope) observed from attack impacts, a distributed estimator was designed as per explicit impact analysis results to obtain the injected bias \cite{jin2022distributed}. To guarantee the tracking performance of variable-speed WTs in the presence of the FDI attacks tampering with velocity measurements, Zhao \textit{et al.} co-designed the estimator and observer to estimate the impact induced by cyberattacks and observe the injected biases simultaneously \cite{zhao2022adaptive}.}

The isolation-based IMSs aim to isolate the malicious components from the remaining parts to restrict the attack impact with acceptable performance degradation. Different from directly blocking network traffic in the cyber-based IMS, the isolation-based strategy will not only involve the cyber-side traffic block but also incorporate the knowledge of system dynamics and control algorithms to further enhance the mitigation performance. By switching the data exchange mode among \acrshort{der}s and master controllers in an aperiodical and intermittent manner, FDI attacks resulting in unexpected data transmission modes can be easily detected and \textit{both the communication links and associated \acrshort{der}s} will be isolated \cite{zhou2020cyber}. 

{\color{black} For the consensus-based economic dispatch and secondary frequency/voltage regulation in microgrids, Zhang and Yassaie \textit{et al.} employed the {\acrshort{wmsr}} algorithm to discard the the extreme values among the data received from neighbors \cite{zhang2021resilient,yassaie2021resilient}. 
Moreover, based on the consensus objectives from either deterministic or statistical perspectives, the \textit{trust-factors} implying the trust level of its own observation and the data received from neighbors are incorporated into the secondary control to eliminate the adverse impact and isolate suspected malicious components \cite{abhinav2017synchrony,mustafa2019detection,lu2020distributed}.} 
Besides simply discarding the corrupted data, some further actions can be adopted to mitigate the impact of data loss like replacing the transmitted anomalous data with a local calculated safe but not accurate one. 
The idea of \textit{reputation} was integrated into the consensus-based ED in microgrids to thwart non-colluding and colluding FDI attacks\cite{huang2021distributed, cheng2020resilient}. If the reputations of half of its neighbors are lower than a predefined threshold, the malicious information will be replaced with locally calculated one. 
Sahoo \textit{et al.} proposed a event-driven impact mitigation scheme against the FDI attacks in islanded DC/AC microgrids \cite{zhang2021mitigating,sahoo2020resilient1}. The event, defined as the attack detection, will trigger the mitigation strategy to replace the compromised data with the one received from trustworthy neighbors.

The scenario-based IMS will adjust the control algorithm to adapt to different attack scenarios (the number and location of malicious components), which can largely reduce the performance degradation induced by control conservativeness but only work under a number-limited attack scenarios. Considering the DoS attack targeting at the communication link connecting the \acrshort{ess} and energy management system in microgrids, Chlela \textit{et al.} designed a rule-based fallback control strategy to mitigate its impact. When the \acrshort{ess} cannot receive dispatch signals from the EMS, it will enter the decentralized control mode and manage the state of charge in a standalone manner \cite{chlela2017fallback}. 
{\color{black} To handle the excessive latency and damaged cyber connectivity under DoS attacks in islanded microgrids, an event-triggered network reconfiguration scheme was proposed \cite{yao2022cyber}. 
By modeling random DoS attacks as markovian jumps, Liu \textit{et al.} proposed a mode-dependent resilient controller to restore the control performance of centralized islanded microgrids \cite{liu2018stochastic}. The chosen of control parameters under different DoS attacks scenarios (namely different modes) is explicitly investigated to guarantee the stochastic stability of microgrids. }

The adaptability-based IMS is to adjust the control algorithm in an adaptive manner without knowing the specific attack scenarios. Obviously, this type of mitigation strategy may be subject to the problem of excessive performance degradation when a over-conservative control parameters are chosen. A self-adaptive resilient control algorithm was proposed to preserve secondary consensus in hierarchical networked microgrids under multi-layer DoS attacks \cite{ge2022cyber}. 
{\color{black}For the centralized event-triggered control framework of DC microgrids subject to DoS attacks, Hu \textit{et al.} developed an adaptive parameter update scheme to mitigate the attack impact \cite{hu2019attack}. }
The schedule-based IMS tries to schedule flexible resources like \acrshort{der}s to mitigate the impacts of cyberattacks. By adjusting the droop gains of \acrshort{der}s, the destabilizing effect of load alteration attacks (a type of FDI attack) could be effectively mitigated \cite{chu2022mitigating}. Moreover, the sampling scheme with time-varying frequency was proposed to restore the communication as soon as the DoS attack terminates \cite{lian2021distributed,deng2021distributed}.


{\color{black}\textit{Lessons learned from physics-based detection-triggered IMSs:}

   - After incorporating the inputs of IDSs, more mitigation strategies like isolation- and scenario-based appear for the detection-triggered IMSs. The most common compensation-based strategy can work for both FDI and DoS attacks, and usually needs to integrate adaptive control and NN methods to estimate the injected bias/healthy data. The isolation-based IMS can be regarded as the simplest strategy, but it only works under FDI attacks and is subject to the number of attacks. The scenario- and adaptability-based IMSs are usually used to counter DoS attacks, where the former is customized for attack scenarios (less conservativeness, limited attack scenarios) and the latter adapts automatically without requiring specific attack information (more conservativeness, unlimited attack scenarios). The schedule-based IMS can mitigate both FDI and DoS attacks by utilizing extra flexible resources.
     
     - Actually, the performance of each IMS can be guaranteed only when the adversary's capability is restricted like bounded FDI attacks. In particular, the compensation-/isolation-/scenario-/adaptability-based IMSs try to enhance the tolerance of control algorithms against cyberattacks and can work immediately once perceiving anomaly. When the adversary's capability exceeds control algorithms' tolerance, the schedule-based IMS is expected to alleviate the severe consequence by adopting available flexible resources. Hence, the cooperative design of control-enabled and schedule-driven mitigation strategies should lead to a more general and effective mitigation scheme.

     - The investigation of data-driven physics-based IMSs is rare, and most of them are model-based. This phenomenon is caused by the inherent difficulty of recovering control-acceptable healthy data from compromised data using purely data-driven methods, since it is difficult to train the model covering all attack forms.
}

{\color{black}
The summary of reviewed IMSs, focusing on applied scenarios, utilized tools/methods, and evaluation metrics, can be found in TABLE \ref{Table:Mitigation}. From a high-level perspective, a set of evaluation metrics regarding IMS is refined: 1) Performance-related metrics: mitigated attack types and mitigation effect; 2) Cost-related metrics: computation and communication overhead and hardware investment. It is recommended to appropriately balance the trade-off between these metrics when designing IMSs. However, it is indeed difficult to give a comparative study regarding all detection and mitgation methods in the literature due to the lack of a set of benchmark testbeds or datasets. On one hand, it is unrealistic to establish a high-fidelity smart grid testbed without considering space and budget limitation. On the other hand, the sensitivity information contained in the real-world power system data hinders its disclosure for research purpose. Many efforts are still required from academic and industry as well as governments to address this critical issue and pave the way towards the cyber-resilient smart grid under highly penetrated \acrshort{der}s.}

\begin{table*}[]
\footnotesize
\centering
{\color{black}
\begin{threeparttable}
\caption{\color{black}Summary of IMSs}\label{Table:Mitigation}
\begin{tabular}{m{2.5cm}m{1cm}m{2.5cm}m{2.5cm}m{7.5cm}}
\Xhline{1.2pt}
\multicolumn{5}{c}{\textbf{Cyber-based IMSs}} \\ \Xhline{1.2pt}
\textbf{Types} & \textbf{Lit.} & \textbf{Scenarios} & \textbf{Methods/Ideas} & \textbf{Evaluation Metrics} 
\\ \Xhline{1.2pt}
    \multirow{5}{3cm}{Detection-triggered, \\Data-driven} & \cite{appiah2020decentralized,hong2017intelligent} & Grid-tied DER & Block network traffic & Attack: DoS and FDI attacks; Effect: Isolation; Extra cost: No 
    \\ \cline{2-5}
    
    &  \cite{li2018sdn} & Microgrid & SDN enabled traffic block & Attack: FDI and replay attacks; Effect: Isolation; Extra cost: SDN  \\ \cline{2-5}
    & \cite{jokar2016intrusion} & ZigBee HAN  & Q-learning & Attack: FDI and replay attacks; Accuracy: 93.46\%; Latency: Neglectable  \\ \Xhline{1.2pt}
    
    \multicolumn{5}{c}{\textbf{Physics-based IMSs}} \\ \Xhline{1.2pt}
    \textbf{Types} & \textbf{Lit.} & \textbf{Scenarios} & \textbf{Methods/Ideas} & \textbf{Evaluation Metrics} 
\\ \Xhline{1.2pt}
    \multirow{-1.5}{3cm}{Self-triggered,\\ Compensation-based} & \cite{liu2021robust,zuo2020resilient} & Isolated Microgrid control & Competitive interaction control & Attack: Bounded and unbounded FDI attacks; Effect: Input-to-state and uniformly ultimately bounded stability; Extra cost: SDN-based secure hidden communication layer \\ \cline{1-5}

     \multirow{24}{3cm}{Detection-triggered,\\ Compensation-based} & \cite{hu2022resilient} & Load frequency control & Piecewise observer based robust control & Attack: Resources constrained FDI and DoS attacks; Effect: $H_\infty$ performance guarantee; Extra cost: No \\ \cline{2-5}

     & \cite{zhao2022dual} & Variable-speed WT & NN observer, Dual-triggered control & 
     Attack: Resources constrained DoS attack; Effect: Exponential convergence guarantee; Extra cost: No
      \\ \cline{2-5}

    & \cite{li2022distributed} & Isolated Microgrid control& First-order dynamic observer & Attack: Resource constrained DoS attack; Effect: Exponential convergence; Extra cost: No  \\ \cline{2-5}

    & \cite{saad2020implementation} & Isolated Microgrid control& IoT-based DT, Luenberger observer & Attack: DoS and FDI attacks; Latency: Timely; Extra cost: DT and Cloud service  \\ \cline{2-5}

    & \cite{jiang2021high} & Isolated Microgrid control& SMO and HOD & Attack: Bounded FDI attack; Effect: Lyapunov stable; Extra cost: No  \\ \cline{2-5}

    & \cite{shi2021observer} & Isolated Microgrid control& Robust output feedback control & Attack: Bounded FDI attacks; Effect: $L_2$-gain boundedness; Extra cost: No  \\ \cline{2-5}

    & \cite{zuo2020distributed} & Isolated Microgrid control& Adaptive observer & Attack: Bounded and Unbounded FDI attacks; Effect: Uniformly ultimately bounded and Asymptotic stability; Extra cost: No \\ \cline{2-5}

    & \cite{habibi2021decentralized} & Isolated Microgrid control& ANN, PI-based controller & Attack: FDI attack; Effect: Compensation error $\le0.02\%$; Latency: $<$0.15s; Extra cost: No \\ \cline{2-5}

    & \cite{10089185} & Isolated Microgrid control& Adversarial DRL & Attack: Destabilising FDI attack; Effect: Asymptotic stability; Extra cost: No \\ \cline{2-5}

    & \cite{cecilia2021addressing} & Isolated Microgrid control& Nonlinear adaptive observer & Attack: FDI attack; Effect: Input-to-state stability; Extra cost: No  \\ \cline{2-5}

    & \cite{jin2022distributed} & Isolated Microgrid control& Impact-oriented compensation & Attack: Constant FDI attack; Latency: 2s; Extra cost: No \\ \cline{2-5}

    & \cite{zhao2022adaptive} & Variable-speed WT & Adaptive resilient control & 
    Attack: Bounded FDI attack; Effect:  uniformly ultimately bounded stability; Extra cost: No  \\ \cline{1-5}

    \multirow{10}{3cm}{Detection-triggered,\\ Isolation-based} & \cite{zhou2020cyber} & Isolated Microgrid control & Aperiodically intermittent control & Attack: Quantitatively limited FDI attack; Effect: Asymptotic stability; Extra cost: No \\ \cline{2-5}

    & \cite{zhang2021resilient,yassaie2021resilient} & Economic dispatch, Microgrid & Weighted mean subsequence reduced algorithm & Attack: Quantitatively limited FDI attack; Effect: Optimal dispatch, Asymptotic stability; Extra cost: No  \\ \cline{2-5}

    & \cite{abhinav2017synchrony, mustafa2019detection,lu2020distributed} & Isolated Microgrid control & Trust-factor based control & Attack: Quantitatively limited FDI attack; Effect: Asymptotically stability; Extra cost: No  \\ \cline{2-5}

    & \cite{huang2021distributed,cheng2020resilient} & Economic dispatch & Reputation-driven bad data replacement & Attack: Quantitatively limited FDI attack; Effect: Optimal dispatch; Extra cost: Multiple-hop communications  \\ \cline{2-5}

    & \cite{zhang2021mitigating,sahoo2020resilient1} & Isolated Microgrid control & Event-driven bad data replacement & Attack: Quantitatively limited FDI attack; Effect: Successful mitigation: Extra cost: No  \\ \cline{1-5}

    \multirow{6}{3cm}{Detection-triggered, \\Scenario-based} & \cite{chlela2017fallback} & ESS management & Rule-based fallback control & Attack: DoS attack; Effect: Maintain ESS's SOC within allowable range; Extra cost: No \\ \cline{2-5}

    & \cite{yao2022cyber} & Isolated Microgrid control & Adaptive control, network reconfiguration & Attack: Resource constrained DoS attack; Effect: Stochastic stability; Extra cost: Network reconfiguration  \\ \cline{2-5}

    & \cite{liu2018stochastic} & Isolated Microgrid control & Mode dependent control & Attack: Resource constrained Markovian DoS attack; Effect: Stochastic stability; Extra cost: No   \\ \cline{1-5}

    \multirow{4}{3cm}{Detection-triggered, Adaptability-based} & \cite{ge2022cyber} & Isolated Microgrid control & Adaptive control & Attack: Resource constrained DoS attack; Effect: Secure consensus; Extra cost: No \\ \cline{2-5}

    & \cite{hu2019attack} & Isolated Microgrid control & Adaptive event-triggered control & Attack: Resource constrained DoS attack; Effect: Global asymptotically stability; Extra cost: No \\ \cline{1-5}

    \multirow{4}{3cm}{Detection-triggered, \\Schedule-based} & \cite{chu2022mitigating} & Load frequency control & DER droop gain adjustment & Attack: IoT Botnet attack; Effect: Exponential stability; Extra cost: Generation cost \\ \cline{2-5}

    & \cite{lian2021distributed,deng2021distributed} & Isolated Microgrid control & Sampling frequency adjustment & Attack: Resource constrained DoS attack; Effect: Asymptotic stability; Extra cost: Communication overhead \\ \Xhline{1.2pt}

\end{tabular}
\end{threeparttable}
}\end{table*}

\section{Defense-in-Depth Strategies: Recovery}\label{section: VIII}
The recovery scheduling is to recover the degraded system states after mitigation to the normal states. It is vital as after the response of IMSs the blackout/isolated areas cannot be reconnected, the malicious payloads inserted by adversaries still exist, and the damaged electrical devices need repair/replacement.


\begin{figure*}
    \centering
    \includegraphics[width=16cm]{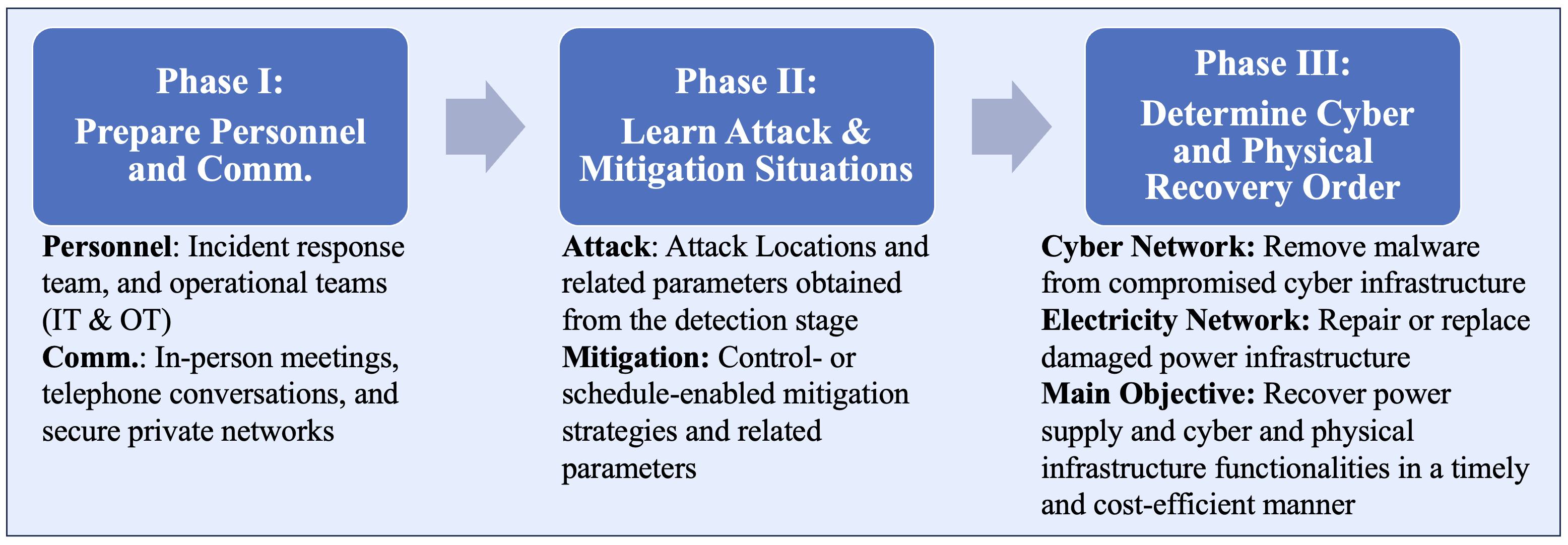}
    \caption{\color{black}Phases included in the cyber-physical interdependent recovery under cyberattacks.}
    \label{fig:RecoveryPhases}
\end{figure*}

{\color{black}{According to NIST's guide for cybersecurity event recovery \cite{bartock2016guide}, the recovery schedule comprises Phase I: Prepare personnel and communication, Phase II: Learn attack and mitigation situations, and Phase III: Determine recovery the order as shown in Fig. \ref{fig:RecoveryPhases}. Phases I and II are more like preparation steps based on the information from the previous detection and mitigation steps, and Phase III is the core part that determines the recovery order of blackout areas, compromised cyber components, and damaged physical equipment to achieve the restoration of power supply and infrastructure functionalities. In particular, the power supply restoration is given the first priority and should be completed timely (hours) \cite{9913670}, where both cyber and physical recovery actions will be involved. The cyber-related restoration actions aim to reconnect the communication network using flexible emergency communication vehicles. The physics-related restoration actions try to restore the electricity supply to the blackout area in transmission and isolated areas in distribution via emergency generators like mobile power supply vehicles or other black-start-capable local generators. }

After the restoration of power supply, the infrastructure recovery will be activated to repair/replace the compromised/damaged software and hardware facilities to enable properties of $N-1$ security and loss-efficiency, as well as economical dispatch. Compared to the power supply restoration, the full restoration of infrastructure functionalities requires a much longer period (days/weeks). The cyber-related recovery actions include the removal of virus, malware, and other malicious payloads from the computation and communication environment, generally completed through software reinstall and antivirus tools. The physics-related recovery actions aim to repair the damaged power lines and transformers, synchronize the grid islands to return to interconnected operation, and replace backup and emergency systems with components used in normal operation. 
The essential challenge here is to schedule the recovery actions in multiple time scales to achieve power supply and infrastructure restoration in a cost-efficient manner considering resource constraints and performance requirements. }
Forensic analysis should be conducted to summarize and learn lessons from the pre-, during, and post-event phases, providing guidelines for better prevention, detection, and mitigation capabilities.

{\color{black}
Many efforts have been devoted to designing optimal black-start strategies following blackouts and planning cost-efficient physical recovery actions under natural disasters like extreme weather events \cite{9121323}. The black-start service aims to energize power grid without requiring external power supplies in the event of partial or total shutdown, and the generator providing this service is called as black-start capable generator like diesel generators. Considering the complex practical constraints that vary with time, Sun \textit{et al.} \cite{sun2010optimal} reformulated the generator start-up sequencing problem as a tractable \acrshort{milp} problem such that an optimal solution that outperforms conventional heuristic or enumerative techniques \cite{perez2008optimal,toune2002comparative} in terms of result performance and computational speed can be obtained to maximise the system generation capability. The feasibility of using DERs to provide a “bottom-up” black start approach has been investigated \cite{yan2021feasibility}, which has potential advantages of reduced restoration time and more flexible recovery procedure compared with the conventional large thermal plants. These black-start capable DER units can also provide ancillary services like reactive power support to assist the voltage control during synchronization and grid re-connection \cite{ganesh2022ancillary}.

In the presence of extreme weather events, Qin \textit{et al.} developed a hierarchical dispatch model, where the three dispatch stages including preventive dispatch, emergency dispatch, and restorative dispatch for transmission systems are coordinated within a robust optimisation problem, to improve the short-term power supply restoration capability \cite{qin2021coordination}. Given the grid-support services provided by DERs, Sun \textit{et al.} proposed a distributed restoration strategy for integrated distribution and transmission systems by the utilisation of limited sharing information of boundary buses \cite{8727510}. Considering the simultaneous disaster damages on cyber and physical networks, Wang \textit{et al.} developed a post-disaster restoration scheduling model for active distribution networks to coordinately dispatch the repair and operation crews, reconfigure power and communication networks, and adjust DER operations \cite{wang2022cyber}. Nevertheless, merely limited attention has been paid to developing the cyber-recovery problem under cyberattacks. }

{\color{black} The recent pioneering work \cite{liu2023cyber} about the cyber recovery from the cyber-resilient mode that is activated to mitigate the destabilizing consequence resulted from dynamics load alternating attacks has been comprehensively studied. Considering the links of electricity, transportation, and
cyber networks, a \acrshort{milp} problem is formulated to determine the optimal repair crew route and adaptive adjustment of system operation such that the extra cost of operating the cyber-resilient mode and the malicious manipulation of IoT malware on bus loads can be both eliminated as soon as possible. The possible attack movements in each recovery step that aim to maximize the attack impacts of remaining compromised resources are considered to obtain the robust recovery strategy. In addition to this preliminary study, the incorporation of more realistic scenarios like stochastic travel/repair time and more complex interactions between the attacker and defender still require nontrivial further efforts.

{For clarity purpose, TABLE \ref{Table:Recovery comparsions} is given to summarise the differences between the focused cyber-recovery and conventional black-start and physical-recovery. The black-start capable units are adopted in both the physical- and cyber-recovery processes to restore the power supply service in the physical side. Besides that, the reestablish of communication network is also involved when the extreme event damages/compromises cyber components and induces network disconnection. The difference between physical- and cyber-recovery processes in restoring the power supply service lies in their focuses. The physical-recovery \textit{mainly} focuses on the power grid energization in the \textit{physical} side since natural disasters usually first damage physical power lines and generators \cite{wang2022cyber}. While the cyber-recovery needs to concern both the physical side's grid energization and cyber side's communication network reestablishment as cyberattacks will first invalidate cyber components and then affect the power supply service. In restoring the infrastructure functionality, the cyber-recovery needs to additionally schedule recovery crews to remove the cyber malware compared with the physical-recovery. Two key challenges of cyber-recovery are thereby identified: i) Particular attention should be paid to the cyber-side modeling and the resultant strong cyber-physical coupling may complicate the recovery schedule problem; ii) Additional attack movements may occur when the adversary perceives the recovery actions. }

It is also vital to discuss the  compatibility between defense-in-depth strategies, i.e., how much each defense mechanism supports/conflict with one another to achieve defense-in-depth protection \cite{tan2019tabulating}. Most of the compatibility relations belong to neutral, while several relations are regarded as dependent or conflicted. In particular, the effectiveness of access control, secure communication protocols, and code-signing software update relies on the network architecture's capabilities like network segmentation and communications partitioning. Moreover, the dependency relation also exists between IDSs and detection-triggered IMSs, as well as between cyber-recovery and IDSs/IMSs. The conflict relation mainly comes from the computation burden resulted from encryption-enabled secure communication protocols, which may degrade the subsequent detection, mitigation, and recovery performance and even invalidate these functionalities.
}

{\color{black}Pairwise comparisons of the capability between defense-in-depth strategies are summarized in TABLE \ref{Table:Compability comparsions}, i.e., how much each defense mechanism supports one another to achieve defense-in-depth protection \cite{tan2019tabulating}. In particular, the network segmentation is the most basic cyber prevention technology, and it does not requires dependencies from other technologies. Based on a well-segmented network architecture, an appropriate access control mechanism is developed to grant participators' accesses to resources with different criticality. Then, secure communication protocols are designed to allow entities to transmit information in a secure manner. On top of these three prevention technologies, the code-signing software update scheme is established to guarantee the integrity of installed software. The conflict mainly comes from the computation burden resulted from encryption-enabled secure communication protocols, which may degrade the subsequent detection, mitigation, and recovery performance and even invalidate these functionalities. Moreover, the dependency relation also exists between IDSs and detection-triggered IMSs as well as recovery and IDSs/IMSs.}

\begin{table*}[]
\footnotesize
\centering
\begin{threeparttable}
\caption{\color{black}Comparsions between black-start, physical-recovery, and cyber-recovery.}\label{Table:Recovery comparsions}
\begin{tabular}{m{1.5cm}m{3.5cm}m{6cm}m{6cm}}
\Xhline{1.2pt}
    \textbf{Tasks} & \textbf{Black-Start under Blackouts} & \textbf{Phyical-Recovery under Natural Disasters} & \textbf{Cyber-Recovery under Attacks} \\ \Xhline{1.2pt}
    \multirow{3}{1.5cm}{Power Supply Service} & \multirow{-1.5}{3.5cm}{\textit{Physical}: Energize power grid without requiring external power supplies in the event of partial or total shutdown} & \textit{Cyber}: Reestablish comm. network utilizing mobile comm. vehicles or other resources (\textit{Occasional}) & \textit{Cyber}: Reestablish comm. network utilizing mobile comm. vehicles or other resources (\textit{Main}) \\ 
     & & \textit{Physical}: Energize power grid utilizing mobile generation vehicles and black-start generators (\textit{Main}) & \textit{Physical}: Energize power grid utilizing mobile generation vehicles or black-start generators (\textit{Main})\\ \cline{1-4}
     
    \multirow{2}{1.5cm}{Infrastructure Function} & \multirow{2}{3.5cm}{N.A.} & \multirow{2}{6cm}{\textit{Cyber and Physical}: Repair or replace damaged components}  & \textit{Cyber}: Remove cyber malware \\
     & & & \textit{Physical}: Repair or replace damaged components \\
    \Xhline{1.2pt}
\end{tabular}
\end{threeparttable}
\end{table*}


\begin{table}[]
\footnotesize
\centering
{
\begin{threeparttable}
\caption{\color{black}Pairwise Comparisons of the Capability between Defense-in-Depth Strategies}\label{Table:Compability comparsions}
\begin{tabular}{m{2cm}|m{0.5cm}p{0.1cm}<{\centering}p{0.1cm}<{\centering}p{0.1cm}<{\centering}p{0.1cm}<{\centering}p{0.1cm}<{\centering}p{0.1cm}<{\centering}p{0.1cm}<{\centering}p{0.1cm}<{\centering}p{0.1cm}<{\centering}p{0.2cm}<{\centering}}
\Xhline{1.2pt}
\multicolumn{2}{c}{\textbf{Defense Strategies}} & (1) & (2) & (3) & (4) & (5) & (6) & (7) & (8) & (9) & (10) \\ \Xhline{1.2pt}
\multirow{4}{1.5cm}{Prevention} & (1) & & \cellcolor{gray1} N & \cellcolor{gray1} N & \cellcolor{gray1} N& \cellcolor{gray1} N& \cellcolor{gray1} N& \cellcolor{gray1} N& \cellcolor{gray1} N& \cellcolor{gray1} N& \cellcolor{gray1} N \\ 
& (2) & \cellcolor{blue1} D &  & \cellcolor{gray1} N  & \cellcolor{gray1} N & \cellcolor{gray1} N & \cellcolor{gray1} N & \cellcolor{gray1} N & \cellcolor{gray1} N & \cellcolor{gray1} N & \cellcolor{gray1} N \\
& (3) & \cellcolor{blue1} D & \cellcolor{blue1} D & & \cellcolor{gray1} N & \cellcolor{gray1} N & \cellcolor{red1} C & \cellcolor{red1} C & \cellcolor{red1} C & \cellcolor{red1} C & \cellcolor{red1} C \\
& (4) & \cellcolor{blue1} D & \cellcolor{blue1} D & \cellcolor{blue1} D  &  & \cellcolor{gray1} N& \cellcolor{gray1} N& \cellcolor{gray1} N& \cellcolor{gray1} N& \cellcolor{gray1} N& \cellcolor{gray1} N\\ \hline
\multirow{3}{1.5cm}{Detection} & (5) & \cellcolor{gray1} N & \cellcolor{gray1} N & \cellcolor{gray1} N & \cellcolor{gray1} N &  & \cellcolor{gray1} N & \cellcolor{gray1} N & \cellcolor{blue1} D & \cellcolor{gray1} N & \cellcolor{blue1} D \\
& (6) & \cellcolor{gray1} N & \cellcolor{gray1} N & \cellcolor{red1} C & \cellcolor{gray1} N & \cellcolor{gray1} N & &  \cellcolor{gray1} N & \cellcolor{blue1} D & \cellcolor{gray1} N&\cellcolor{blue1} D \\
& (7) & \cellcolor{gray1} N & \cellcolor{gray1} N& \cellcolor{red1} C& \cellcolor{gray1} N& \cellcolor{gray1} N& \cellcolor{gray1} N& & \cellcolor{blue1} D& \cellcolor{gray1} N&\cellcolor{blue1} D \\ \hline
\multirow{2}{1.5cm}{Mitigation} & (8) & \cellcolor{gray1} N &\cellcolor{gray1} N &\cellcolor{red1} C & \cellcolor{gray1} N& \cellcolor{blue1} D&\cellcolor{blue1} D&\cellcolor{blue1} D& &\cellcolor{gray1} N &\cellcolor{blue1} D \\
& (9) & \cellcolor{gray1} N &\cellcolor{gray1} N &\cellcolor{red1} C & \cellcolor{gray1} N& \cellcolor{gray1} N&\cellcolor{gray1} N&\cellcolor{gray1} N&\cellcolor{gray1} N & &\cellcolor{blue1} D \\ \hline
Cyber Recovery & (10) & \cellcolor{gray1} N &\cellcolor{gray1} N &\cellcolor{red1} C & \cellcolor{gray1} N& \cellcolor{blue1} D& \cellcolor{blue1} D& \cellcolor{blue1} D& \cellcolor{blue1} D& \cellcolor{blue1} D& \\ \Xhline{1.2pt}
\end{tabular}
\begin{tablenotes}
  \footnotesize
  \item (1): Network segmentation, (2) Access control
  \item (3): Secure communication protocol, (4) Software update verification
  \item (5): HIDS, (6): NIDS, (7): PIDS, (8): Detection-triggered IMS
  \item (9): Self-triggered IMS, (10): Cyber-physical interdependent recovery
  \item {\colorbox{blue1}D}: Dependency, {\colorbox{gray1}N}: Neural, {\colorbox{red1}C}: Conflict
\end{tablenotes}
\end{threeparttable}}
\end{table}

\section{Challenges and Future Directions}\label{Section IX}
In this section, challenges and future directions are discussed from six phases including identification, prevention, detection, mitigation, and recovery.

\subsection{Threat Identification}
In terms of threat identification, the potential vulnerabilities and corresponding attack impact have been extensively investigated. Standing on the perspective of attacker, a successful attack event requires to exploit multiple vulnerabilities and coordinate them appropriately to induce targeted and precise consequences. {\color{black}There still lack a \textbf{highly integrated and automatic} framework to identify vulnerability exploitation paths that can cause critical hazards given specific system configurations. This research direction is vital as its outputs can identify the most critical parts needed to be protected, but meanwhile is difficult as both the expert knowledge of IT and OT domains are required in the top-bottom design process. Moreover, the automation of the identification tool is challenging as the smart grid modeling involving the interoperability of various types of DER devices, the complicated couplings among the layers of hierarchical framework, and the strict real-time functionality requirements usually needs substantial manual interventions \cite{9961003}.} {\color{black}Besides, it is of great importance to establish a sharing platform of the security issues and cyber vulnerabilities where the participators' privacy will not be leaked by the disclosure of these critical information.}

{For the adversarial attacks on smart grid applications, current studies mainly focus on the model-oriented attack, while limited attentions have been paid to the privacy- and platform-oriented attacks. When the highly-valued commercial data and user's sensitive information are used for ML training, the well-tuned membership inference methods \cite{salem2018ml} can expose the economic condition and the user's preferences to the adversary. The platform-based attack remains another ML-related cyber threat, where the zero-day and unpatched vulnerabilities of hosts/servers that run the ML-based applications can be exploited to install backdoor to the ML code \cite{xiao2018security}. Future efforts are required to investigate the privacy- and platform-oriented adversarial attacks on the ML-based smart grid applications.}

\subsection{Prevention Technology}
Prevention technologies with high security levels have been standardized for the interaction between \acrshort{der}s and power systems. Nevertheless, the blockchain, MTD, and virtualized \acrshort{der} and cyber-physical integrated prevention technologies can be enhanced to enable the further security improvement.


{\color{black}1) \textbf{Grid-Edge Lightweight Blockchain}: To enable the implementation of blockchain in grid-edge \acrshort{der}s, the future efforts should focus on the optimisation of computation complexity, data handling, and number of transactions in blockchain to reduce its energy consumption and provide timely response while guaranteeing required security levels \cite{10310250}.}

2) \textbf{Cost-Efficient MTD}: The triggering scheme, cost, and performance of MTD can be systematically optimized. More adaptive MTD triggering schemes need to be developed, which requires the advanced detection or learning capabilities of the defender \cite{cho2020toward}. The key challenge is how to infer an adversary's action or learn system security condition to guide MTD deployment.

3) \textbf{Physics-Informed Virtualized \acrshort{der} equipment}: To make the emulated virtualized \acrshort{der}s indistinguishable from \acrshort{der} devices, the physical/plant dynamics should be deeply integrated to mimic the behaviours of \acrshort{der} devices instead of simply displaying the historically recorded inputs and outputs. The key challenge is how to emulate complex physical/plant dynamics using resource constrained computation and storage capabilities. 

{\color{black}4) \textbf{Cyber-Physical Integrated Prevention Technologies:} To prevent the adversary from intruding into critical networks and inducing hazardous consequences in a cost-efficient manner, it is vital to design cyber-physical integrated prevention technology such that conventional IT security technologies can be combined with OT's emerging robust control and dispatching methods. The key challenge lies in modelling and quantifying the interactions of cyber and physical domains' methods.}


\subsection{Intrusion Detection System}
Existing IDSs can perceive anomalous activities with satisfactory performance using single-domain features (host, network, or physical) but require add-on detection hardware. The next step needs to integrate IDSs into embedded hardware like inverters, where the computation and memory resource is highly restricted, and investigate the possibility of improving the detection performance by fusing multi-domain features, coordinating multi-layer resources, and blending data and model knowledge.

{\color{black}1) \textbf{Lightweight HIDS in Grid-edge Devices}: Tailoring HIDSs for inverters can greatly help counter against the threats arouse from Trojan, firmware manipulation, supply-chain, etc. The primary challenging is that the HIDS's detection overhead on computation and memory cannot significantly decrease/affect the original control performance of grid-edge converters and inverters.}

{\color{black}2) \textbf{Cyber-Physical Integrated IDSs}: 
The deep integration of multiple features from cyber and physical domains including the host data, network traffic, and physical measurements can improve the detection performance, but the investigation of an appropriate fusion scheme of multiple domain data is particularly challenging. Pan \textit{et al.} made their attempts towards this direction and successfully applied a data mining technique called common path mining to automatically and accurately learn patterns for scenarios from a fusion of synchrophasor measurement data, and information from relay, network security logs, and EMS logs \cite{7063234,6939262}. Nevertheless, research efforts are still needed to address the issues of limited amount of available data, real-time decision-making requirement, and increasingly complicated system dynamics with the penetration of \acrshort{der}s.}

3) \textbf{Local-Centralized Coordinated PIDS}: Co-designing model-based and data-driven PIDSs in a local-centralized collaboration manner can help incorporate their respective advantages. Specifically, the data-driven PIDS is employed in the control centre to perceive the existence of anomaly, while the model-based PIDS is adopted in each distributed entity to reveal the malicious component location.

{\color{black}4) \textbf{Physics-Awareness Data-driven IDSs} By incorporating the well-established physical knowledge into the design of data-driven IDSs, the interpretability of making unexpected false/missed alarms can be potentially improved while being able to address high uncertainties of renewable resources and adversary behaviours. Several challenges exist in the establishment of feasible and efficient integration schemes of model and data, the utilisation of highly heterogeneous data, and the consideration of ML model's robustness against data poisoning and evasion attacks \cite{zhang2023vulnerability}.
}

{5) \textbf{Data-driven IDSs with High Robustness to Adversarial Attacks}: The robustness to adversarial attacks needs to be considered when designing data-driven IDSs in the smart grid. From the perspective of ML model, improving the ML model's resilience through adversarial training \cite{10015583} and detecting adversarial samples in the training dataset and model inputs \cite{8787888} can be possible solutions. From the perspective of smart grid, the idea of robust control \cite{zhou1998essentials} that is able to deal with unknown disturbances may be promising to tolerate the intractable adversarial perturbations.
}

\subsection{Impact Mitigation System}
Although numerous IMSs have been proposed to quickly respond to cyberattacks, their design phases ignore the cost-efficiency, cross-level coordination, and adaptability of IMSs, leading to possible future directions: 

{\color{black}1) \textbf{Data and Model Blended Bad Data Reconstruction}: Fusing data characteristics of multiple domains can help improve the bad data recovery performance while reducing the cost of extra hardware. The statistical and spatio-temporal correlation properties can be employed to predict the \textit{data interval} of the next time slot, while the semantic model knowledge can be utilised to construct estimators to quickly and accurately find the value within that interval. The key challenges are to develop stability and efficiency guaranteed integration schemes of data and model as well as to train robust ML models

2) \textbf{Data-driven IMSs under Varying and Uncertain Environments}: To defend against the cyberattacks with uncertain and varying types, duration, and intensities, it is promising to incorporate advanced data-driven methods like the deep reinforcement learning methods, which can enable the optimal decision-making by continuously interact with and learn from changing environments. The main challenges here include the establishment of a high-fidelity DER environments that support real-time data interaction and the design of efficient and robust model training algorithms that are able to converge quickly to optimal decisions with high robustness to untimely and imperfect data.}

{\color{black}3) \textbf{Edge-Centre Coordinated Holistic IMSs}: The coordination of grid-edge devices' and SCADA centre system-level mitigation methods can respond holistically to multiple complicit and strategically evolving powerful adversaries. As shown in Fig. \ref{fig:HolisticIMSframework}, grid-edge mitigation methods are adopted as timely (seconds to minutes) responses to cyber contingencies to isolate bad data and replace it with recovered legitimate data such that the decentralized DERs' control performance can be guaranteed with appropriate control adaptions. When the grid-edge DERs' controllability cannot handle the attack severity, the SCADA system level safe-mode operations with a longer response time (minutes-hours) will be triggered to schedule flexible resources like \acrshort{der}s and reconfigure the communication network such that the grid-edge DERs' attack tolerance can be enhanced. The basic enabling techniques are from control theory and \acrshort{ml}. Key challenges exist in identifying the bound beyond which the system-level mitigation schedule should be triggered and developing an appropriate integration scheme accommodating for grid-edge and system-level mitigation.

\begin{figure}
  \centering
  \includegraphics[width=9cm]{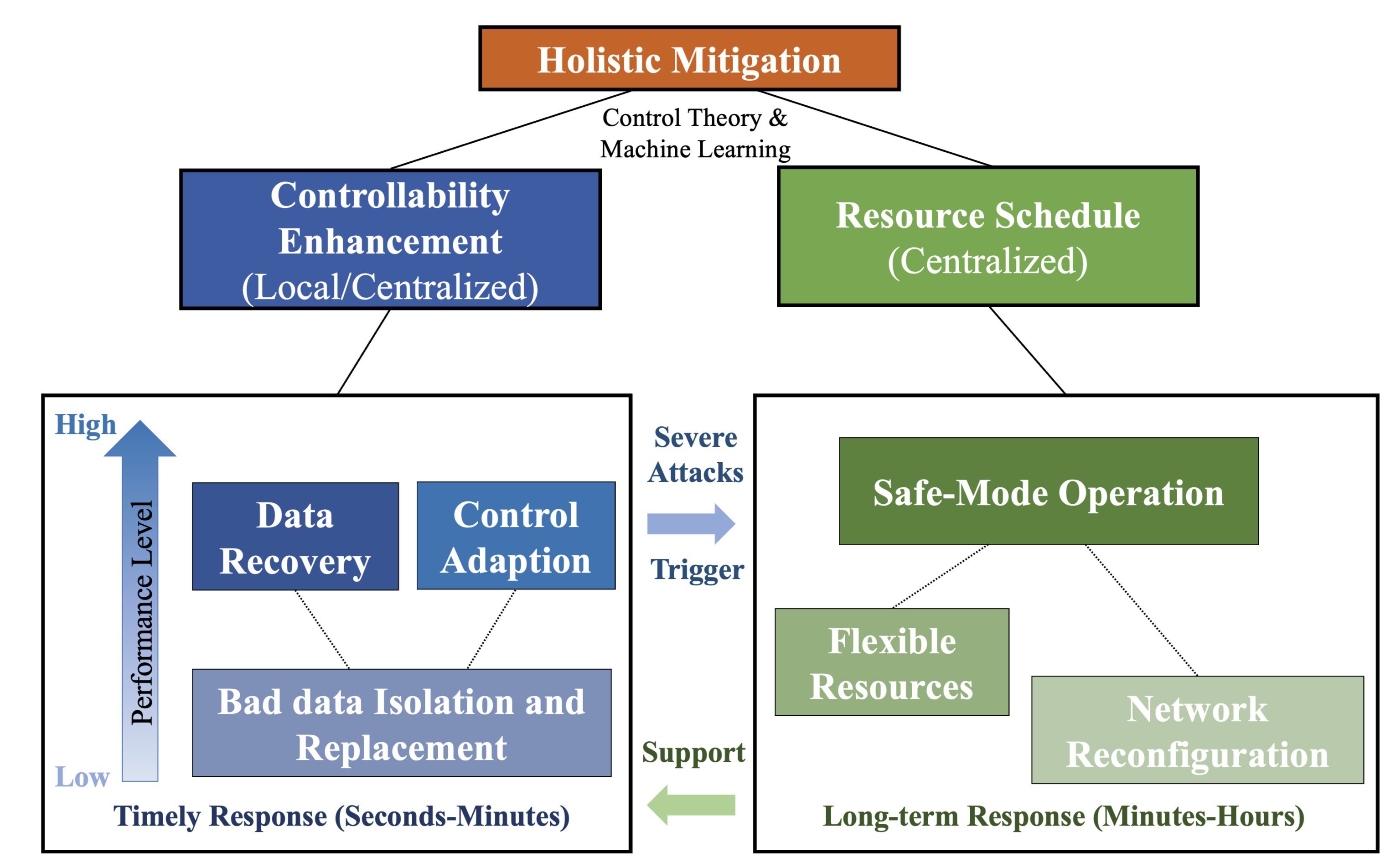}
  \caption{\color{black}A holistic impact mitigation framework.}\label{fig:HolisticIMSframework}
\end{figure}

4) \textbf{MTD-based IMS}: MTD-based IMSs are designed to improve the mitigation capability against the powerful adversaries, who can obtain or infer the principles of mitigation strategies and thus are able to invalidate them, by adding uncertainties to the mitigation actions in a periodical or triggered manner. The key challenge is to design appropriate perturbation schemes to balance the trade-off between the cost of introducing perturbations and the accordingly improved mitigation performance.}

{\color{black}\subsection{Cyber Recovery Schedule}
Although many efforts have been devoted to designing recovery schemes under natural disasters like extreme weathers \cite{wang2022cyber}, the \textbf{cyber-physical interdependent recovery schedule under HILP cyberattacks} has not been extensively investigated yet. 

1) \textbf{Cyber-Physical Interdependent Recovery Schedule:} Since the attack will first compromise cyber infrastructure and then affect the physical infrastructure and power supply service, the cyber recovery schedule needs to consider both cyber and physical components' repair and replace and the interdependence between them should be paid special attention. The restoration of cyber and physical infrastructure should be prioritized to swiftly restore power supply services or resume normal system operations. The key challenges include the formulation of the cyber recovery problem considering the adversary's movements when the recovery actions are perceived and the solving of the formalised bi- or tri-level problem in a fast and accurate manner.

2) \textbf{Data-driven Cyber Recovery Schedule Considering Uncertainties: }The uncertainties originated from the adversary, repair time, and crew travel time can exist in the cyber recovery process, which can be well addressed utilising the deep reinforcement learning approach. The agent will be trained to learn the optimal recovery sequence from the interactive uncertain and varying environment. Key challenges in establishing the real-time recovery-supported environment, incorporating the physical-domain knowledge, and designing learning algorithms that can achieve robust convergence in the presence of data compromise.

3) \textbf{Decentralized Cyber Recovery Schedule:} As the decentralisation of the control and operation in the DER-based smart grid, it is necessary to design a multi-agent deep reinforcement learning method to coordinately obtain the optimal recovery sequence based on each DER's or DER cluster's resources. The limited data availability of each agent and its resource-constrained memory and computation abilities pose new challenges to the design of learning algorithms.}




\section{Conclusion}\label{Section X}
In this paper, we provided a comprehensive survey regarding the CRE process in the \acrshort{der}-based smart grid, where threat modeling, risk assessment, and defense-in-depth strategies encompass the key enablers. First, a hierarchical architecture of the cyber-physical \acrshort{der}-based smart grid was presented to illustrate the actors and their functionalities. An integrated threat modeling methodology was tailored for the hierarchical \acrshort{der}-based smart grid with special emphasises on vulnerability identification and consequence investigation, based on which a general risk assessment matrix can be established to inform the system operator about attack scenarios' severity. Then, the state-of-the-art progresses made in prevention, detection, mitigation, and recovery technologies were comprehensively reviewed, systematically classified, and extensively summarized. It is observed that current CRE-related researches mainly focus on the improvement of security-oriented performance and utilization of local and single-domain resources while rarely consider the restriction of security cost and coordination of multi-layer and cross-domain resources. Based on this, challenges and future directions were highlighted and discussed in details.

\section*{Acknowledgement}
The authors would like to thank Professor Xin Zhang from Sheffield of University and anonymous reviewers very much for their fruitful and insightful suggestions during the revision of this manuscript.

\begin{spacing}{1}
\bibliographystyle{IEEEtran}
\small
\bibliography{root}
\end{spacing}

\begin{IEEEbiography}[{\includegraphics[width=1in,height=1.2in,clip,keepaspectratio]{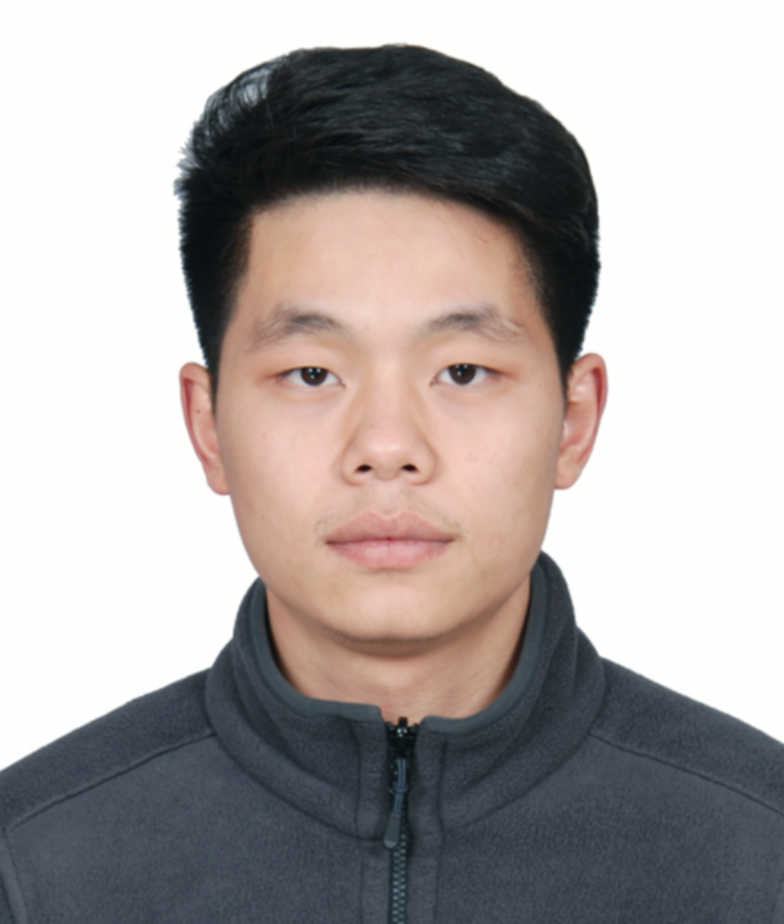}}]{Mengxiang Liu} received the B.Sc. degree in Automation from Tongji University, Shanghai, in 2017 and the Ph.D. degree in Cyberspace Security from Zhejiang University, Hangzhou, in 2022. He is currently a Research Assistant with the Department of Electrical and Electronic Engineering, Imperial College London, London, UK. His research interests include cyber resiliency, DER-based smart grid, active defense.\end{IEEEbiography}

\begin{IEEEbiography}[{\includegraphics[width=1in,height=1.2in,clip,keepaspectratio]{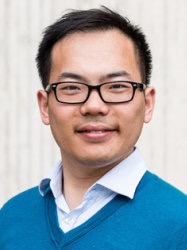}}]{Fei Teng} (Senior Member, IEEE) received the B.Eng. degree in electrical engineering from Beihang University, China, in 2009, and the M.Sc. and Ph.D. degrees in electrical engineering from Imperial College London, U.K., in 2010 and 2015, respectively, where he is currently a Senior Lecturer with the Department of Electrical and Electronic Engineering. His research focuses on the power system operation with high penetration of inverter based resources (IBRs) and the cyber-resilient and privacy-preserving cyber-physical power grid. He serves/served as an Associate Editor for IEEE Transactions on Power System, Control Engineering Practice, IEEE Open Access Journal of Power and Energy, and IEEE Power Engineering Letters and a Guest Editor for IEEE Transactions on Cloud Computing, IEEE Transactions on Industry Applications, Applied Energy, and IET Renewable Power Generation. 
\end{IEEEbiography}

\begin{IEEEbiography}[{\includegraphics[width=1in,height=1.2in,clip,keepaspectratio]{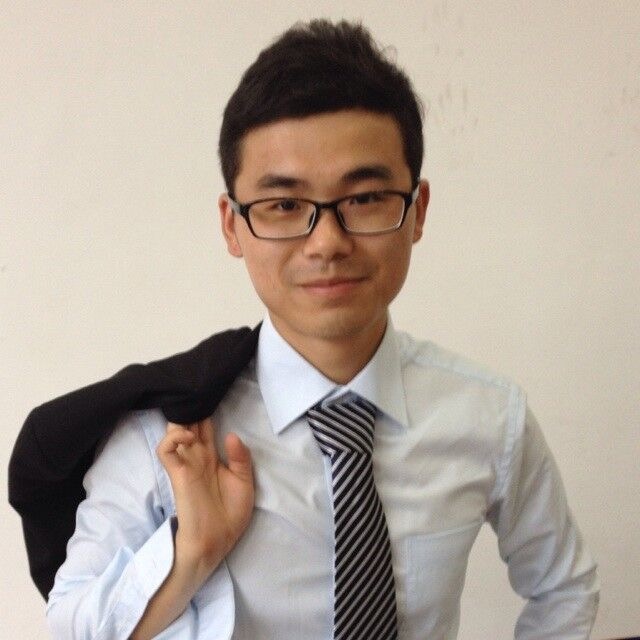}}]{Zhenyong Zhang} (Member, IEEE) received his Ph.D. degree from Zhejiang University, Hangzhou, China, in 2020, and bachelor degree from Central South University, Changsha, China, in 2015. He was a visiting scholar in Singapore University of Technology and Design, Singapore, from 2018 to 2019. Currently, he is a professor in the college of Computer Science and Technology, Guizhou University, Guiyang, China. His research interests include cyber-physical system security, applied cryptography and machine learning security.\end{IEEEbiography}

\begin{IEEEbiography}[{\includegraphics[width=1in,height=1.2in,clip,keepaspectratio]{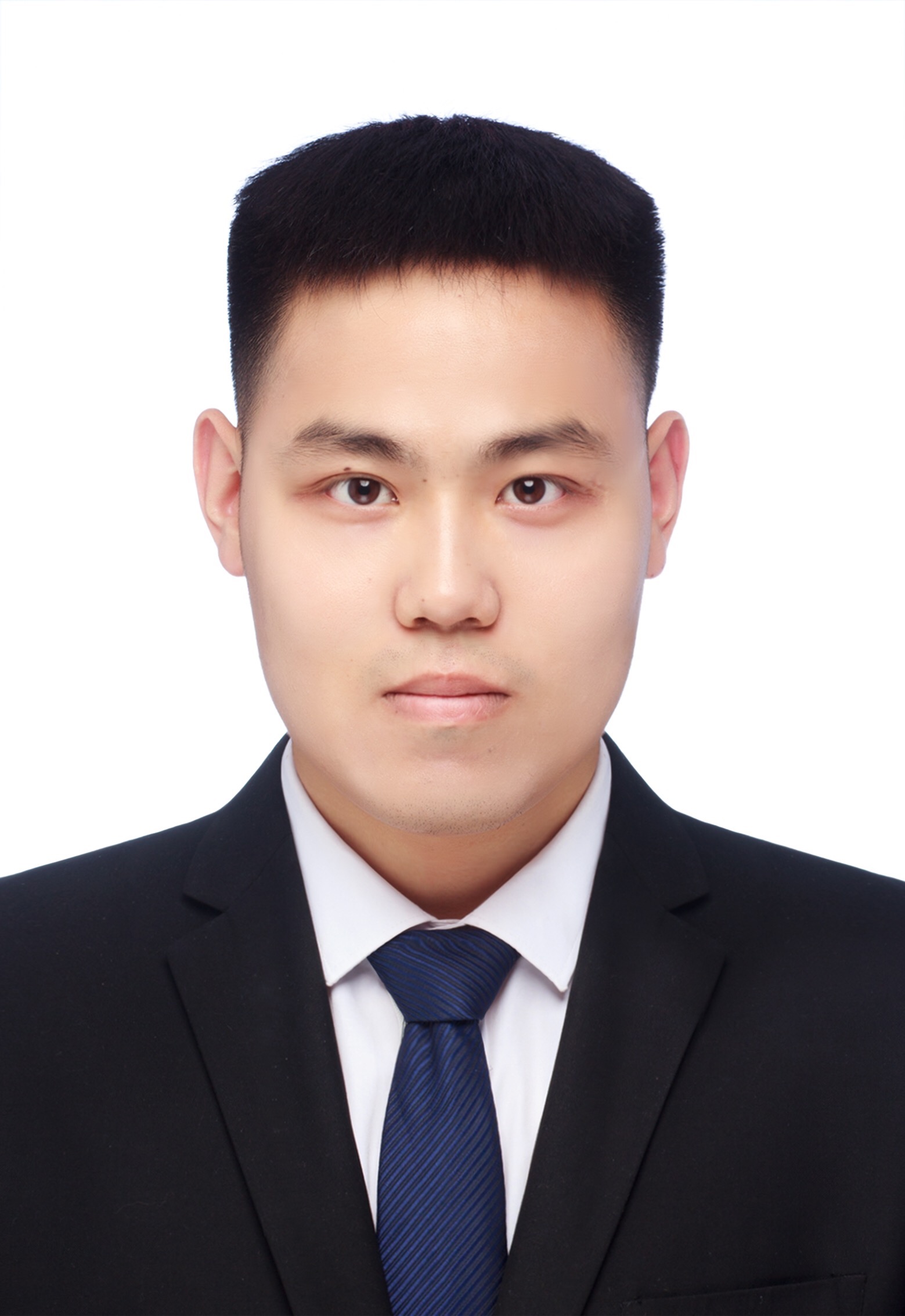}}]{Pudong Ge} (Student Member, IEEE) received the M.Sc degree in electrical engineering from Southeast University, Nanjing, China, in 2019, and he is currently a Ph.D Student at the Department of Electrical and Electronic Engineering, Imperial College London. His current research focuses on the distributed control of cyber-physical coupling microgrids, and cyber-resilient energy system operation and control.\end{IEEEbiography}

\begin{IEEEbiography}[{\includegraphics[width=1in,height=1.2in,clip,keepaspectratio]{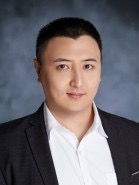}}]{Mingyang Sun} (Senior Member, IEEE) received the Ph.D. degree from the Department of Electrical and Electronic Engineering, Imperial College London, London, U.K., in 2017. From 2017 to 2019, he was a Research Associate and a DSI Affiliate Fellow with Imperial College London. He is currently a Professor of Control Science and Engineering under the Hundred Talents Program at Zhejiang University, Hangzhou, China. Also, he is an Honorary Lecturer at Imperial College London. His research interests include AI in energy systems and cyber-physical energy system security and control.\end{IEEEbiography}

\begin{IEEEbiography}[{\includegraphics[width=1in,height=1.2in,clip,keepaspectratio]{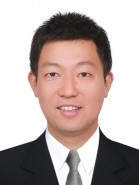}}]{Ruilong Deng} (Senior Member, IEEE) received the B.Sc. and Ph.D. degrees both in Control Science and Engineering from Zhejiang University, Hangzhou, Zhejiang, China, in 2009 and 2014, respectively. He was a Research Fellow with Nanyang Technological University, Singapore, from 2014 to 2015; an AITF Postdoctoral Fellow with the University of Alberta, Edmonton, AB, Canada, from 2015 to 2018; and an Assistant Professor with Nanyang Technological University, from 2018 to 2019. Currently, he is a Professor with the College of Control Science and Engineering, Zhejiang University, where he is also affiliated with the School of Cyber Science and Technology. His research interests include cyber security, smart grid, and communication networks. 
\end{IEEEbiography}

\begin{IEEEbiography}
[{\includegraphics[width=1in,height=1.2in,clip,keepaspectratio]{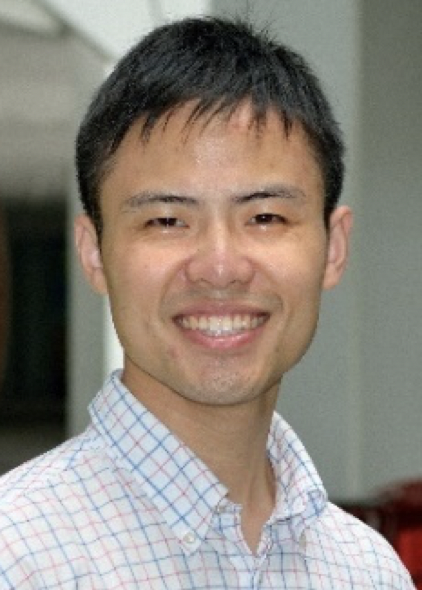}}]{Peng Cheng}
(M'10) received the B.Sc. and Ph.D. degrees in control science and engineering from Zhejiang University, Hang Zhou, China, in 2004 and 2009, respectively. He is currently a Professor and an Associate Dean of the College of Control Science and Engineering, Zhejiang University.  He has been awarded the 2020 Changjiang Scholars Chair Professor. He serves as Associate Editors for the IEEE Transactions on Control of Network Systems. He also serves/served as Guest Editors for IEEE Transactions on Automatic Control and IEEE Transactions on Signal and Information Processing over Networks. His research interests include networked sensing and control, cyber-physical systems, and control system security.\end{IEEEbiography}

\begin{IEEEbiography}[{\includegraphics[width=1in,height=1.2in,clip,keepaspectratio]{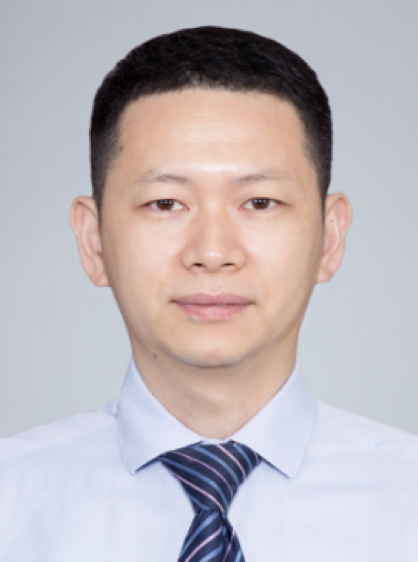}}]{Jiming Chen} (Fellow, IEEE) received the Ph.D. degree in control science and engineering from Zhejiang University, Hangzhou, China, in 2005. He is currently a Professor with the Department of Control Science and Engineering, Zhejiang University, where he is also the Vice Dean of the Faculty of Information Technology. His research interests include network optimization and control, cyber security, and internet of things (IoT) and big data for industry. He was a recipient of the 7th IEEE ComSoc Asia/Pacific Outstanding Paper Award, the JSPS Invitation Fellowship, and the IEEE ComSoc AP Outstanding Young Researcher Award. He serves as the general Co-Chairs for the IEEE RTCSA’19, the IEEE Datacom’19, and the IEEE PST’20. He is an IEEE VTS Distinguished Lecturer. He serves on the editorial boards of multiple IEEE TRANSACTIONS.\end{IEEEbiography}

\end{spacing}
\end{document}